\documentclass{pasj00}

\def\Oversim#1#2{\lower0.5ex\vbox{\baselineskip0pt\lineskip0pt%
            \lineskiplimit0pt\ialign{%
          $\mathsurround0pt #1\hfil##\hfil$\crcr#2\crcr\sim\crcr}}}

\begin{document}

\SetRunningHead{K. Omukai}
{Observational Characteristics of the First Protostellar Cores}
\Received{2006/1/1}
\Accepted{2006/1/1}
  
\title{Observational Characteristics of the First Protostellar Cores} 
\author{Kazuyuki \textsc{Omukai}} 
\affil{National Astronomical Observatory,  
Mitaka, Tokyo 181-8588, Japan; omukai@th.nao.ac.jp} 


\maketitle 

\begin{abstract}
First protostellar cores are young stellar objects in the 
earliest evolutionary stage.
They are hydrostatic objects formed soon after the central portions 
of star-forming cores become optically thick to dust emission. 
We consider their characteristics in the emitted radiation, and
discuss their evolution with increasing mass of the cores. 
Particular attention is paid to detailed radiative and chemical processes 
in the postshock relaxation layer located at the surface of the core, 
where the majority of radiation is emitted.
Most of the radiation is originally emitted in the dust continuum
in mid-infrared wavelength ($\sim 10-30 \mu$m), 
which reprocessed to far-infrared with $\sim 100-200\mu$m.
Although some fraction ($\sim 0.1$) of the radiation energy is 
emitted in the H$_2$O lines at the accretion shock, 
most is absorbed and reemitted in the dust continuum in the envelope.
The H$_2$O lines account for at most $\sim 1/100$ of the observed luminosity.
If a cavity is present in the envelope due to outflow or rotation, 
the dust and H$_2$O line emission in the mid-infrared wavelength 
from the shock can be observed directly, or as a reflection nebula.
Among forthcoming observational facillities, 
{\it SPace Infrared telescope for Cosmology and Astrophysics (SPICA)}
is the most suitable for detecting either direct or processed radiation 
from first-core objects.

\end{abstract}  
 
\section{Introduction} 
Evolution from the collapse of dense molecular cores to the formation of 
protostars has been discussed theoretically for several decades 
(e.g., Hayashi 1966; Shu, Adams, \& Lizano 1987).
Advances in radio and infrared observations have confirmed
many aspects of the evolution.  
The earliest phase of protostellar evolution, i.e., the first core, 
however, still eludes observational confirmation 
though a theoretical prediction was made almost forty years ago 
(Larson 1969).   
A first core is a hydrostatic object that forms after a star-forming core 
becomes optically thick to the dust continuum and the temperature evolution 
becomes adiabatic owing to inefficient radiative cooling.
Their roles in protostellar evolution include enabling 
binary formation and/or driving protostellar outflow (e.g., Inutsuka 2004).
The difficulty in finding first-core objects is due to the following 
reasons. 
First, first-core objects are short-lived.
After formation, they grow in mass by accretion of the envelope 
matter and the temperature inside increases.
When the mass of the first core reaches about $0.05M_{\odot}$ and 
simultaneously the temperature reaches 2000K, molecular hydrogen
begins to dissociate.
Owing to effective cooling by dissociation, the first core begins
collapse dynamically.
Using a typical accretion rate of $10^{-5}M_{\odot}/{\rm yr}$,  
the timescale in the first-core phase is only $5 \times 10^{3}$ yr, 
which is one or two orders of magnitude shorter than the Class 0 phase.
This means that the expected number of the first-core objects is only a few 
hundredth of Class 0 objects.
Besides, the fact that first-core objects are deeply embedded in their 
parental molecular cores prevents detection.

Such difficulties are being overcome with a rapidly increasing  
number of observed young stellar objects, as well as with 
dramatical improvement of the sensitivity in the infrared/submillimeter 
wavelengths.
Encouraged by these facts, the search for protostars in their 
earliest phase is being undertaken vigorously.
Among them, Onishi, Mizuno, \& Fukui (1999) found L1527F 
(also known as MC27) in Taurus by the ${\rm H^{13}CO^{+}}$ emission, which
is a candidate for an object in the earliest stage of a protostar 
owing to its very high density ($\sim 10^{6}{\rm cm^{-3}}$).
Recently, a central submillimeter source (L1521F-IRS) and near-mid infrared 
bipolar reflection nebulae with a luminosity 0.05$L_{\odot}$ was detected
in this object by {\it Spitzer Space Telescope} (Bourke et al. 2006).
Another more promissing object is Cha-MMS1, whose evolutionary stage
appears to lie between L1527F and a starless core L1544 
from the degree of ${\rm N_{2}H^{+}}$ fractionation and 
from the detection of a mid-infrared source that is several times less luminous 
than L1521F-IRS (Belloch et al. 2006).

Knowledge of expected observational appearence of first-core objects 
is crucial in finding such objects and confirming their identity.  
Some efforts have already been made along this line.
Boss \& Yorke (1995) presented the expected spectral energy distribution 
of an object in the first-core phase, using the result of a three dimensional 
simulation of star formation.
By performing a one-dimensional radiative hydrodynamical calculation, 
Masunaga, Miyama \& Inutsuka (1998) followed the formation and evolution 
of the first cores, and presented the evolution of their emission spectrum.
The above mentioned hydrodynamical studies, however, failed to follow 
the structure of the postshock flow of the accretion shock
due to its extreme thinness, although the majority of emission 
is created there (Stahler, Shu, \& Taam 1980).
In addition to this, those authors included only the dust thermal 
emission in their calculation.
In this paper, we take another approach. 
With a simpler assumption concerning the dynamics of the accreting envelope, 
we correctly solve the structure and emission process in the postshock flow.
Using the injection radiation from the shock, we solve radiative transfer 
in the protostellar envelope. 
In this way, we obtain the observable radiation spectrum from 
the first-core object.

The organization of this paper is as follows.
In Section 2, we describe the model for calculating the emission 
from the first-core objects.
In Section 3, we present the results. 
In Section 4, we summarize this paper, and present a discussion 
and the conclusion.

\section{Model}
\begin{figure}
  \begin{center}
    \FigureFile(80mm,80mm){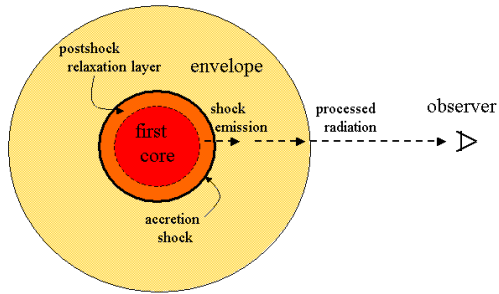}
  \end{center}
\caption{First core and its environment (schematic). 
This shows our model geometry of the first core and the envelope.
We calculated the processed radiation through the envelope, 
in addition to the emission from the accretion shock.}
\label{fig:schematic}
\end{figure}

In Figure \ref{fig:schematic}, we present a schematic view of
our model of a first core and its envelope. 
For the structure and evolution of the envelope, we adopt 
a self-similar solution.
Given the radius of the first core, which is the location 
of the accretion shock, the shock velocity and preshock density are
obtained from a self-similar solution. 
Chemical and radiative processes in the postshock relaxation 
layer are analyzed by assuming a steady state shock.
Finally, using the radiative energy injection obtained as mentioned above, 
we solve the radiative transfer in the envelope.
We describe specific procedures in the following.
 
\subsection{Envelope density distribution}
We adopt a similarity solution as an envelope density structure.
Two types of similarity solutions, (i) the Larson-Penston 
(LP; Larson 1969; Penston 1969) and 
(ii) Shu (Shu 1977) solutions, are studied.
In translating the similarity solutions to physical quantities, 
we assume isothermal evolution at 10K.
In Figure 2, we show the density distributions in the envelope 
both for LP and Shu similarity solutions.
The structure at three epochs, where the mass of the first core
$M_{\rm FC} = 0.0125, 0.025$ and $0.05M_{\odot}$, is illustrated 
for each model.
As the maximum mass of the first core, we take 
$M_{\rm FC, max}=0.05M_{\odot}$ following Masunaga \& Inutsuka (2000).
As the total mass of the core, i.e. the sum of the first-core mass and 
the envelope mass, we take $1M_{\odot}$.
In both solutions, the density low follows $\rho \propto r^{-3/2}$ for 
inner radius and becomes steeper, $\propto r^{-2}$, outside. 
The rarefaction wave, which separates the inner and outer parts, 
propagates outward at the sound speed.
In the outer region, the Shu solution coincides with the singular 
isothermal sphere, while the density of the LP solution is 4.4 times higher. 

\begin{figure}
  \begin{center}
    \FigureFile(80mm,80mm){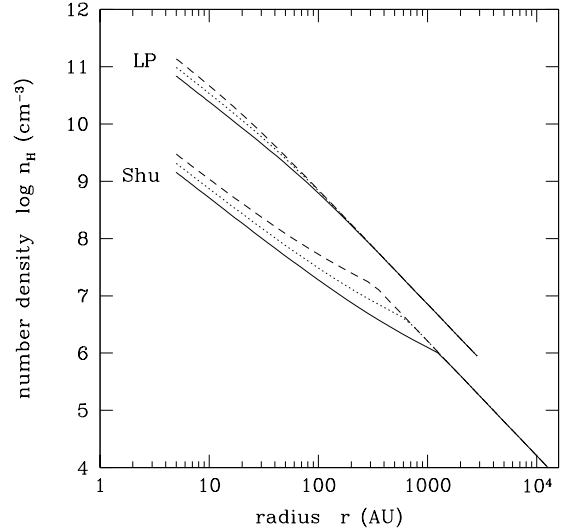}
  \end{center}
\caption{Number density distributions in the envelope.
Depicted are both of the LP and Shu solutions 
for three epochs of the first-core mass 
$M_{\rm FC}= 0.0125 M_{\odot}$ (dashed),  
$0.025 M_{\odot}$ (dotted) and $0.05 M_{\odot}$ (solid).}
\label{fig:r_rho}
\end{figure}

Note that even with the same $M_{\rm FC}$, 
the time which elapsed since the formation of the first core
is different for the two solutions.
Since the mass-accretion rate is $\dot{M}=46.9c_{s}^3/G
=7.5 \times 10^{-5}M_{\odot}/{\rm yr}$ 
($\dot{M}=0.975c_{s}^3/G=1.6 \times 10^{-6}M_{\odot}/{\rm yr}$) 
for the LP (Shu, respectively) solution, which is  
constant for isothermal similarity solutions, 
the age of the first core is $t_{\rm FC}=M_{\rm FC}/\dot{M}$, and 
its entire lifetime is $t_{\rm FC, max}=M_{\rm FC, max}/\dot{M}=670$yr 
for the LP solution, and $3.2 \times 10^4$yr for the Shu solution.
For the same first-core mass, the rarefaction wave propagates to 
about a 50-times larger radius in the Shu solution than in the LP one.
Thus, the LP solution is hardly modified, except at the innermost region, 
while the Shu solution evolves remarkably during the first-core stage 
(see Figure \ref{fig:r_rho}). 

We take the radius of the first core, $R_{\rm sh}$, as a free parameter.
For the fiducial case, we take $R_{\rm sh}=5$AU from the 
spherical symmetric model (Masunaga, Miyama, \& Inutsuka 1998).  
We fix the radius in time because it is found to be 
almost constant from hydrodynamical simulations (Masunaga et al. 1998).
Taking account of the possibility that rotating first cores 
can be more flattened and extended (Saigo \& Tomisaka 2006), 
we also study the case of $R_{\rm sh}=10$AU in order to 
mimic this effect.
We then consider four models that are combinations of either LP or Shu 
solutions, and either $R_{\rm sh}=$5, or 10 AU. 
These four models are called as LP05, LP10, S05, and S10, hereafter.
At a given time, the mass inside $R_{\rm sh}$ is used as 
the first core mass, $M_{\rm FC}$.
 
\subsection{Structure and radiative processes in the 
postshock relaxation layer}
The postshock flow is treated as a steady-state shock problem.
The preshock density is given by a self-similar solution once the 
shock radius is fixed.
The preshock temperature is given by solving radiative transfer in the 
envelope (Sec. 2.3).

Using the shock velocity $v_{s}$, Mach number ${\cal M}$, 
preshock pressure $p_1$ and number density $n_1$, 
the postshock initial values can be
obtained by the following Rankin-Hugoniot relations:
\begin{eqnarray}
\frac{p_{2}}{p_{1}}&=&\frac{2 \gamma_{\rm ad} {\cal M}^2 
-(\gamma_{\rm ad} -1)}{\gamma_{\rm ad}+1},\\
\frac{v_{2}}{v_{s}}&=&\frac{n_{1}}{n_{2}}=\frac{(\gamma_{\rm ad}-1) 
+ 2/{\cal M}^2}{\gamma_{\rm ad}+1},
\end{eqnarray}
where we use subscript 1 (2) for the preshock (postshock, respectively)
variables, and $\gamma_{\rm ad}$ is the ratio of specific heat.

In the postshock relaxation layer, since 
the gas kinetic temperature $T$ is higher than the radiation 
temperature $T_{\rm rad}$, the thermal energy is converted 
to the radiation energy. 
We follow the evolution of a postshock gas parcel 
until $T$ reaches $1.1T_{\rm rad}$ in the following way.
The equations of conservation of mass and momentum fluxes, 
\begin{eqnarray}
\rho v&=&\rho_{2} v_{2} \equiv j, \\
\rho v^{2} + p &=& \rho_{2} v_{2}^{2} + p_{2}, 
\end{eqnarray}
can be incorporated in a single equation,  
\begin{equation}
p+\frac{j^{2}}{\rho} = \rho_{2} v_{2}^{2} + p_{2}= const.
\label{eq:conserv}
\end{equation}
Coupling the above relation with the equation of state for an ideal gas
\begin{equation}
p=\frac{\rho k_{\rm B} T}{\mu m_{\rm H}},
\end{equation}
we solve the energy equation
\begin{equation}
\frac{d e}{d t}=-p\frac{d}{dt}(\frac{1}{\rho})-\frac{1}{\rho}\Lambda,
\label{eq:energy}
\end{equation}
where the energy density per unit mass is 
\begin{equation}
e=\frac{k_{\rm B} T}{(\gamma_{\rm ad}-1) \mu m_{\rm H}}, 
\end{equation}
and $\Lambda$ is the radiative cooling rate.
This determines the structure of the postshock flow.

The radiative cooling is by line emission and dust thermal
emission,
\begin{equation}
\Lambda=\Lambda_{\rm line}+\Lambda_{\rm gr}.
\label{eq:lambda}
\end{equation}
In the first term $\Lambda_{\rm line}$, we include 
the fine-structure and metastable
transitions of C, C$^{+}$ and O, and molecular rotational 
transitions of H$_2$, CO, OH and H$_2$O.
Those are treated as in Omukai et al. (2005), but
the related coefficients for CO and H$_2$O are updated using those 
in Leiden Atomic and Molecular Database (Sch\"{o}ier et al. 2005). 
The cooling rate per unit volume by a transition $i$ is 
\begin{equation}
\Lambda_{i}=h \nu_{i} n(i_u) A_{i} \beta_{i},
\end{equation}
where $n(i_u)$ is the population in the upper level $i_u$, and 
$A_{i}$ is the spontaneous radiative decay rate.
The effect of photon-trapping is accounted for by the 
escape probability (Neufeld \& Kaufman 1993), 
which is given by using the optical depth at the line centroid 
$\tau_{i}$
\begin{equation}
\beta_{i}=\frac{1}{1+3\tau_{i}},
\end{equation}
and 
\begin{equation}
\tau_{i}(s)
=\int \alpha_{i}(s') ds',
\end{equation}
where $s$ is the distance from the shock.
The above intergal is over flows with 
$|v(s')-v(s)|< [\Delta v_{D}(s')+\Delta v_{D}(s)]/2$, where
$\Delta v_{D}$ is the thermal Doppler width.
The second term in equation (\ref{eq:lambda}) is  
the cooling rate by the dust continuum
\begin{equation}
\Lambda_{\rm gr}=4 \pi \int \kappa_{\nu} 
\left[ B_{\nu}(T_{\rm dust})-B_{\nu}(T_{\rm rad}) \right] d \nu, 
\end{equation}
where $\kappa_{\nu}$ is the dust opacity and 
$T_{\rm dust}$ is the dust temperature.
The dust opacity is taken from the IPS (``iron-poor'' silicate) 
composite aggregates model 
of Semenov et al. (2003), who extended Pollack et al. (1994)'s calculation.
The dust temperature is calculated by the energy balance between 
radiative loss from the dust and collisional heating by gas 
(see Omukai et al. 2005 for details).
The postshock flow is optically thin to the dust continuum.
Among radiative cooling processes mentioned above, the H$_2$O emission 
and dust thermal emission are dominant.

Chemical processes among compounds of H, C and O are
included, as in Omukai (2000), where the reaction rate 
coefficients for the C and O chemistry are based on Millar et al. (1997).
Hydrogen is assumed to initially be fully molecular.
The initial abundances of other species in the gas phase are set to be 
$y({\rm CO})=1.10 \times 10^{-4}$, $y({\rm C})=1 \times 10^{-6}$,
$y({\rm O})=3.31 \times 10^{-4}$, and $y({\rm H_2O})=1.7 \times 10^{-5}$
following Pollack et al. (1994)'s ``Cloud Cores'' model in their Table 1D.
If the postshock temperature is higher than the evaporation 
temperature of ice, which is about 150K, 
molecules locked in the ice are released into the gas phase.
In this case, the amounts of H$_2$O, CO and CO$_2$ in the ice mantle
$y_{\rm ice}({\rm H_{2}O})=9.3 \times 10^{-5}$, 
$y_{\rm ice}({\rm CO})=2.8 \times 10^{-5}$, and 
$y_{\rm ice}({\rm CO_{2}})=5.9 \times 10^{-5}$ are added to the 
gas-phase chemistry.

\subsection{Radiative Transfer in the Envelope}
To solve the temperature structure in the envelope and the final 
processed radiation, we need to specify  
the injection radiation from the accretion shock, which is the 
sum of the emission from the postshock relaxation layer 
and radiation from interior of the first core.
In the previous section, we described the method to calculate the 
emission from the relaxation layer.
At the bottom of this layer, the radiation and matter are thermally 
well-coupled.
Thus the radiation intensity is given by the blackbody radiation
at temperature $T_{\rm rad}=T_{\rm post}$, where $T_{\rm post}$ 
is the temperature there.
To find the post relaxation temperature $T_{\rm post}$, 
we follow Stahler, Shu, \& Taam (1981).
For the LP solution, the envelope is optically thick both 
to the shock emission and to dust re-emission in the envelope. 
In this case, both the preshock temperature $T_{\rm pre}$ 
and the temperature at the bottom of the postshock relaxation layer 
$T_{\rm post}$ are equal to the radiation temperature in the postshock 
layer, which is optically thin.
Thus, the condition 
\begin{equation}
T_{\rm post}=T_{\rm pre}
\label{eq:shockBC1}
\end{equation}
needs to be satisfied.
For the Shu solution, the envelope is optically thick to 
high energy shock emission, but optically thin to 
dust re-emission.
In this case, the condition for $T_{\rm post}$ becomes
\begin{equation}
T_{\rm post}
=\left[\frac{1}{\sigma}
(\frac{3}{4}F_{\rm pre}-\frac{1}{4}F_{\rm post})\right]^{1/4}, 
\label{eq:shockBC2}
\end{equation}
where $F_{\rm pre}$ ($F_{\rm post}$) is the radiation energy 
flux in the preshock flow (at the bottom of the relaxation layer, 
respectively).

Using the postshock relaxation temperature $T_{\rm post}$ obtained above, 
the injection radiation to the envelope from the accretion shock is
\begin{equation}
I_{\rm \nu, in}=I_{\rm \nu, shock}+I_{\rm \nu, core} 
=\frac{1}{2}\dot{E}_{\rm \nu, sh}/\pi+B_{\nu}(T_{\rm post}),
\label{eq:I_in}
\end{equation}
where
$\dot{E}_{\rm \nu, sh}$ is the energy emission rate in the postshock
layer per unit area, and can be obtained by integrating the radiative 
cooling rate along the postshock flow,
\begin{equation}
\dot{E}_{\rm \nu, sh}=\int \Lambda_{\nu} ds =\int \Lambda_{\nu} v dt,
\end{equation}
where $\Lambda_{\nu}$ is the radiative cooling rate per unit frequency.
The factor 1/2 in the first term on the right hand side of 
equation (\ref{eq:I_in}) comes from the fact 
that the other half of the radiation goes into the first core.

The radiation field in the envelope is determined by 
solving the radiative tranfer under the input radiation from the inner
boundary.
We used the variable Eddington factor method (Mihalas \& Mihalas 1984).
The temperature $T$ is determined by the radiative equilibrium
\begin{equation}
\int \kappa_{\nu} B_{\nu}(T) d\nu 
= \int \kappa_{\nu} J_{\nu, {\rm dust}} + \sum_{l} \kappa_{\nu_{l}} J_{l},
\end{equation}
where $J_{\nu, {\rm dust}}$ is the monochromatic 
mean intensity in the continuum, 
and $J_{l}$ is the mean intensity integrated over frequency in 
a line $l$.
Since the gas and dust are thermally coupled in the envelope, we do not 
distinguish their temperatures.
In solving radiative transfer, we only consider the dust continuum
as a source of opacity in the envelope.
In all calculated mass ranges in the LP05 model and some in the LP10 model, 
the dust ice mantle sublimates 
and H$_2$O vapor exists in the innermost portion of the core.
However, this part is also optically very thick to dust continuum, 
and the H$_2$O-line luminosity is significantly attenuated by the dust 
absorption in the surrounding region. 
Since we are not interested in such weak line emission, 
we do not include the H$_2$O-line opacity in our model.

\subsection{Method of Model Construction}
The model is constructed as follows:

(i) Given the model parameters, namely, 
the shock radius $R_{\rm sh}$ and the type of 
similarity solution (LP or Shu).

(ii) Calculate the shock velocity $v_{s}$ and 
the preshock density $\rho_{1}$ for a given $R_{\rm sh}$ (\S 2.1). 
Then, assume the preshock temperature $T_{\rm pre}$ and
the radiation temperature in the postshock layer $T_{\rm rad}$.
 
(iii) Solve the postshock flow (\S 2.2) and 
obtain the injection radiation field $I_{\nu, \rm in}$ from the shock.

(iv) Solve radiative transfer in the envelope (\S 2.3) and obtain 
$T_{\rm pre}$.
 
(v) If the assumed and calculated $T_{\rm pre}$ coincide and the shock 
boundary condition (eq. \ref{eq:shockBC1} or \ref{eq:shockBC2}) is satisfied, 
the model is constructed. If not, go back to (ii) with
modified $T_{\rm pre}$ and $T_{\rm rad}$, and repeat this procedure 
until convergence.

\section{Results} 
\subsection{Emission in the postshock relaxation layer}
\begin{figure}
  \begin{center}
\FigureFile(80mm,80mm){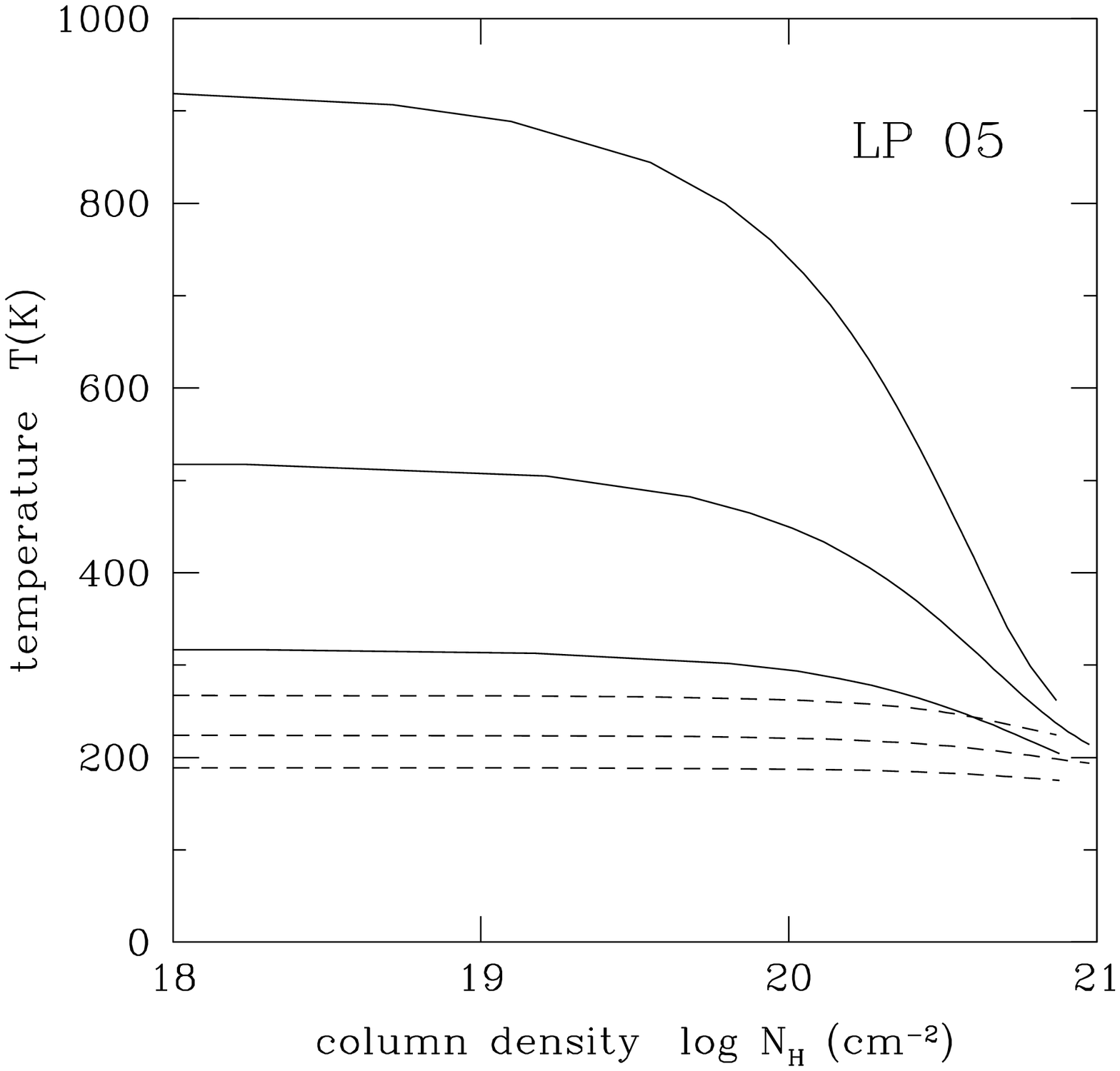}
\FigureFile(80mm,80mm){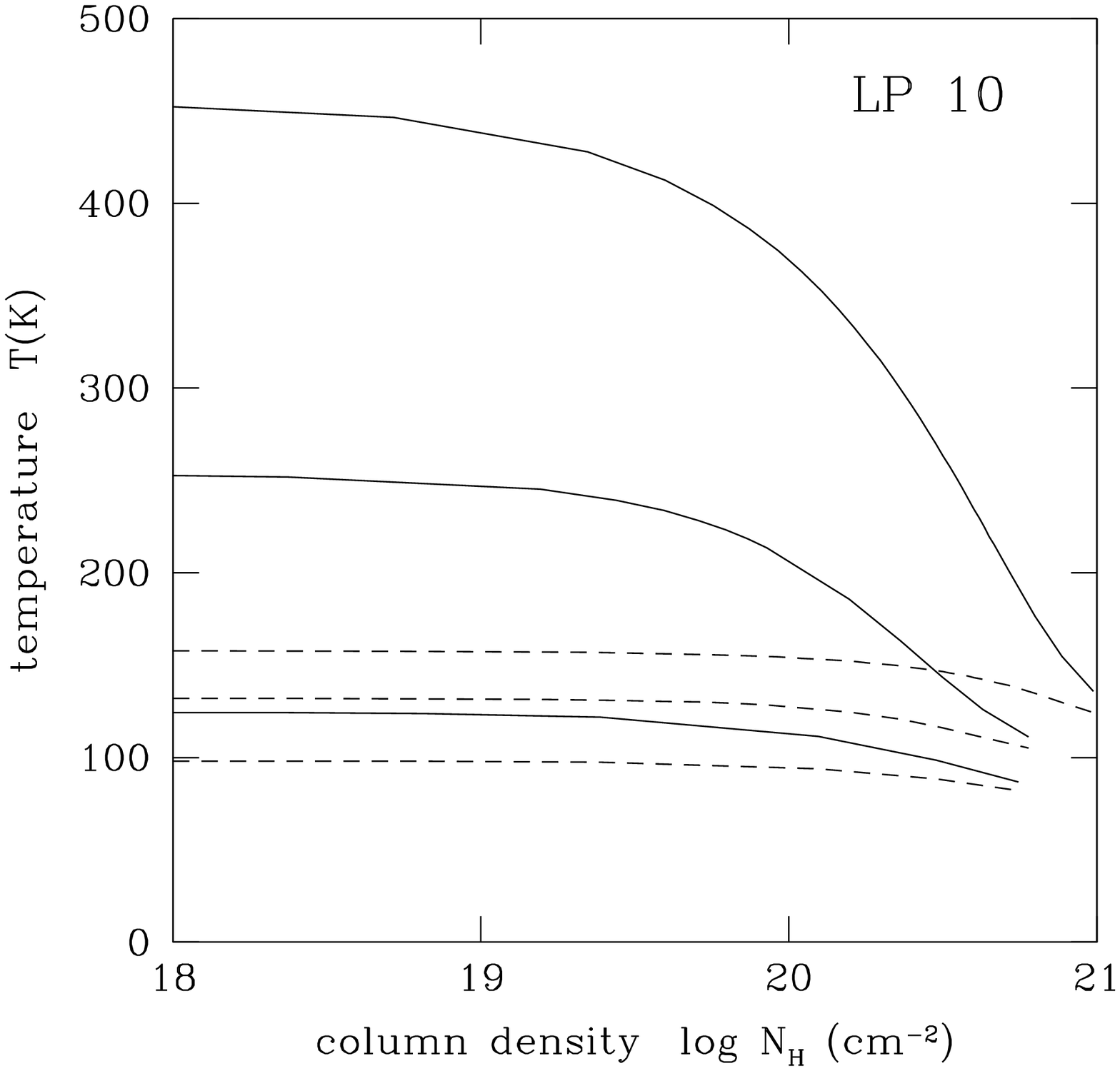}
\FigureFile(80mm,80mm){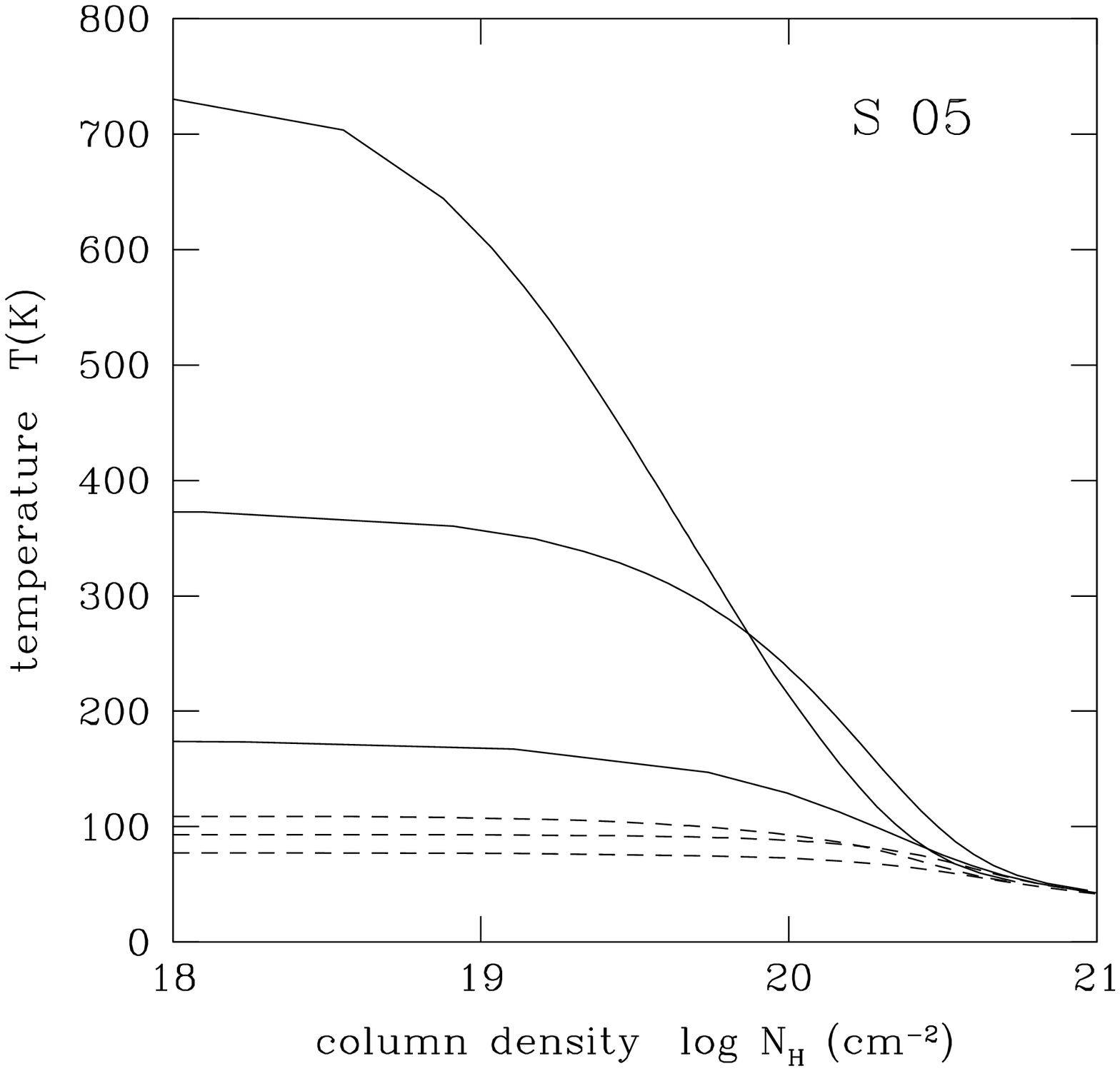}
\FigureFile(80mm,80mm){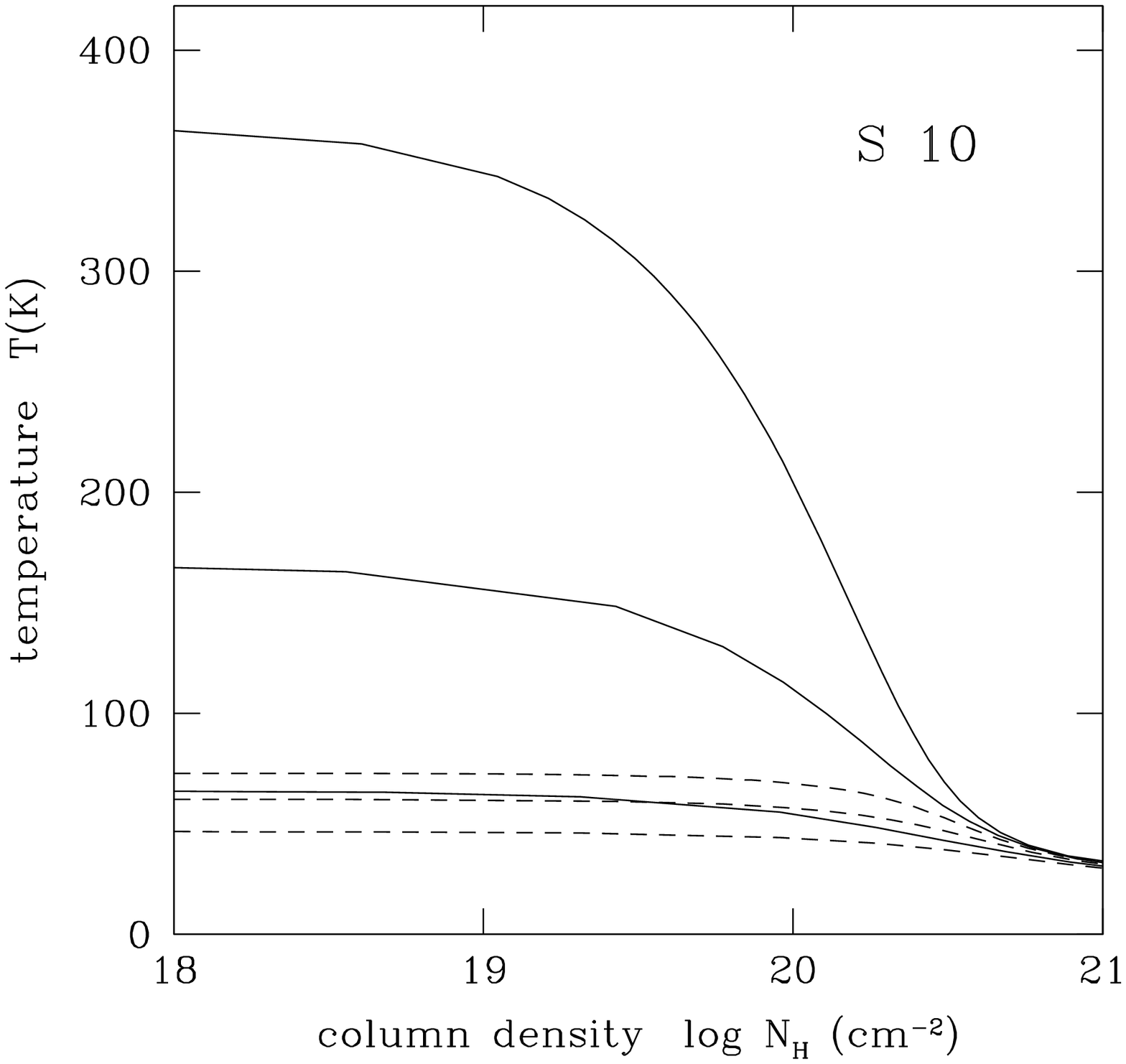}
  \end{center}
\caption{Temperature distributions in the postshock relaxation layer.
The panels are for (a) LP05, (b) LP10, (c) S05, (d) S10 models. 
For each model, those for three epochs of 
$M_{\rm FC}= 0.0125 M_{\odot}$ (bottom),  
$0.025 M_{\odot}$ (middle) and $0.05 M_{\odot}$ (top) are shown.
The solid curves indicate the gas temperature, while the dashed ones are 
for the dust temperature.}
\label{fig:shock}
\end{figure}

\begin{figure}
  \begin{center}
\FigureFile(80mm,80mm){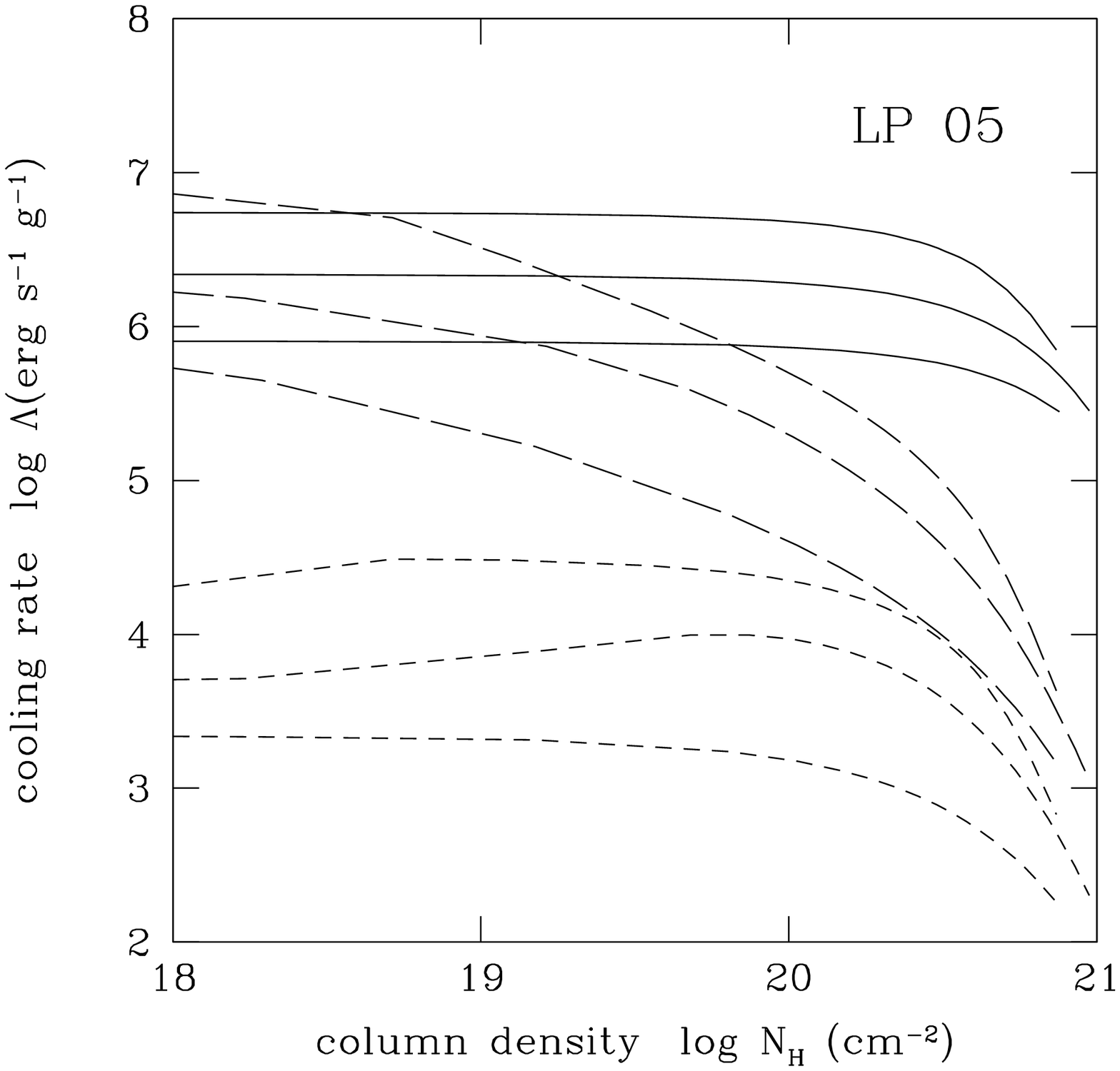}
\FigureFile(80mm,80mm){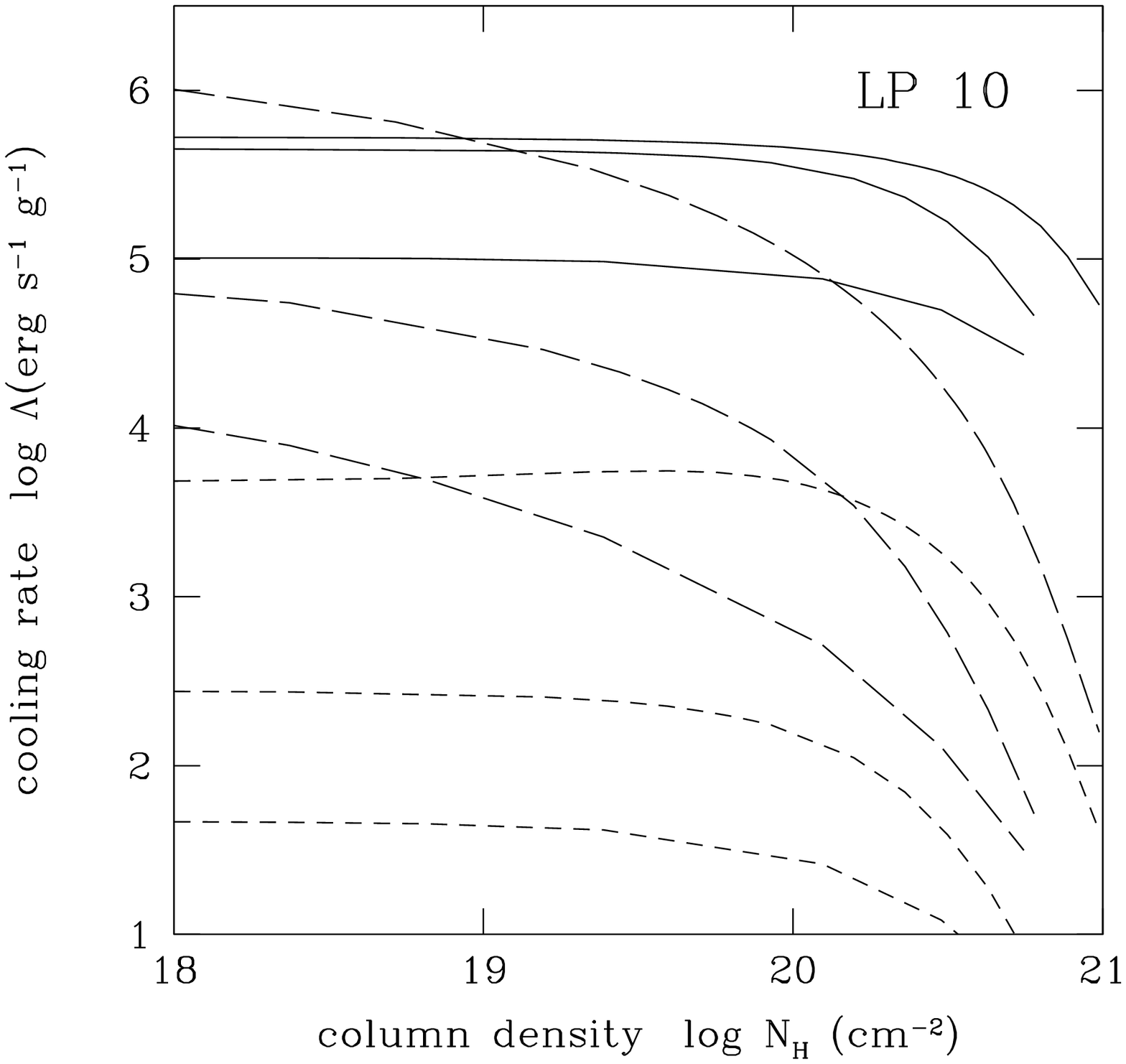}
\FigureFile(80mm,80mm){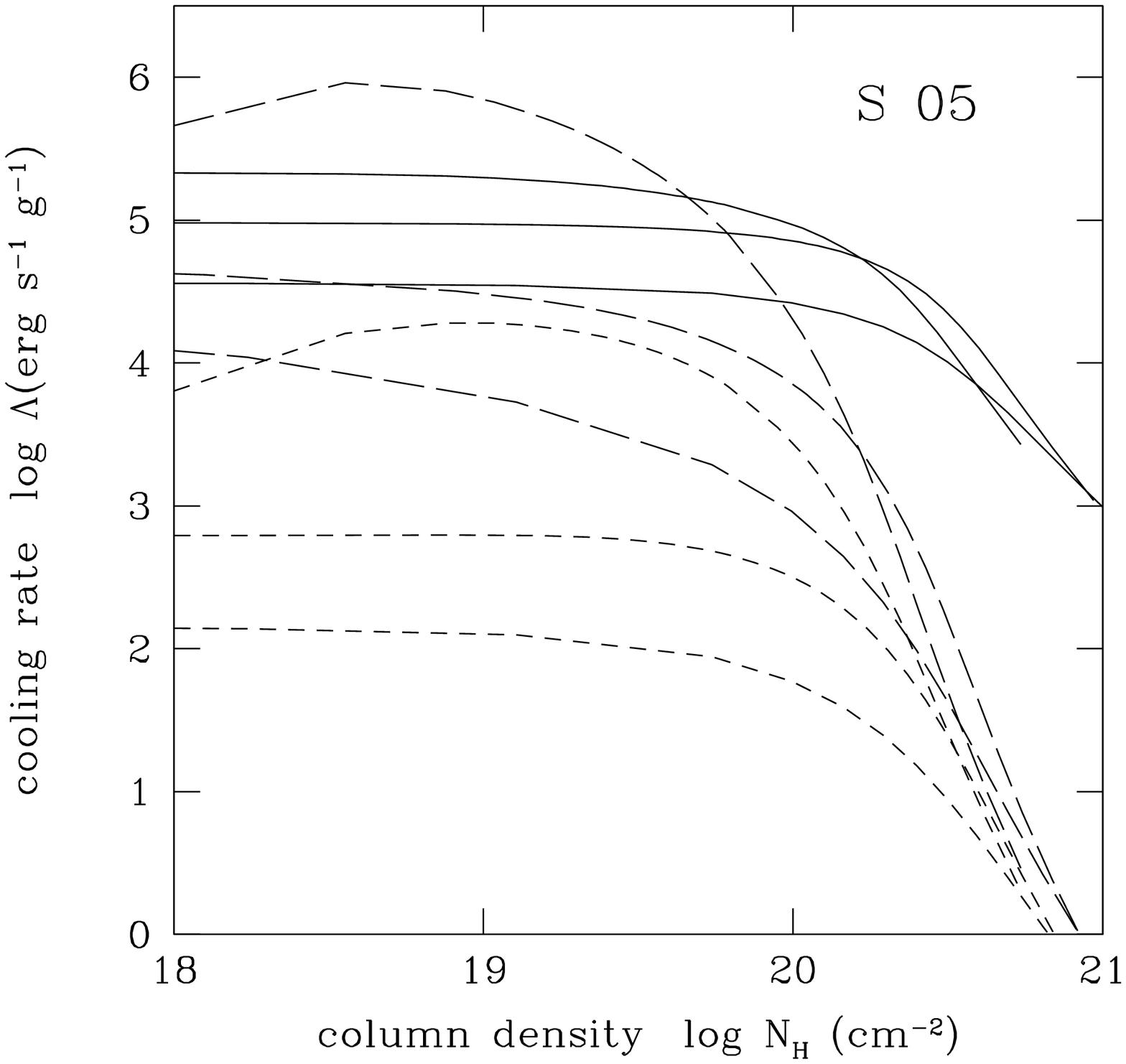}
\FigureFile(80mm,80mm){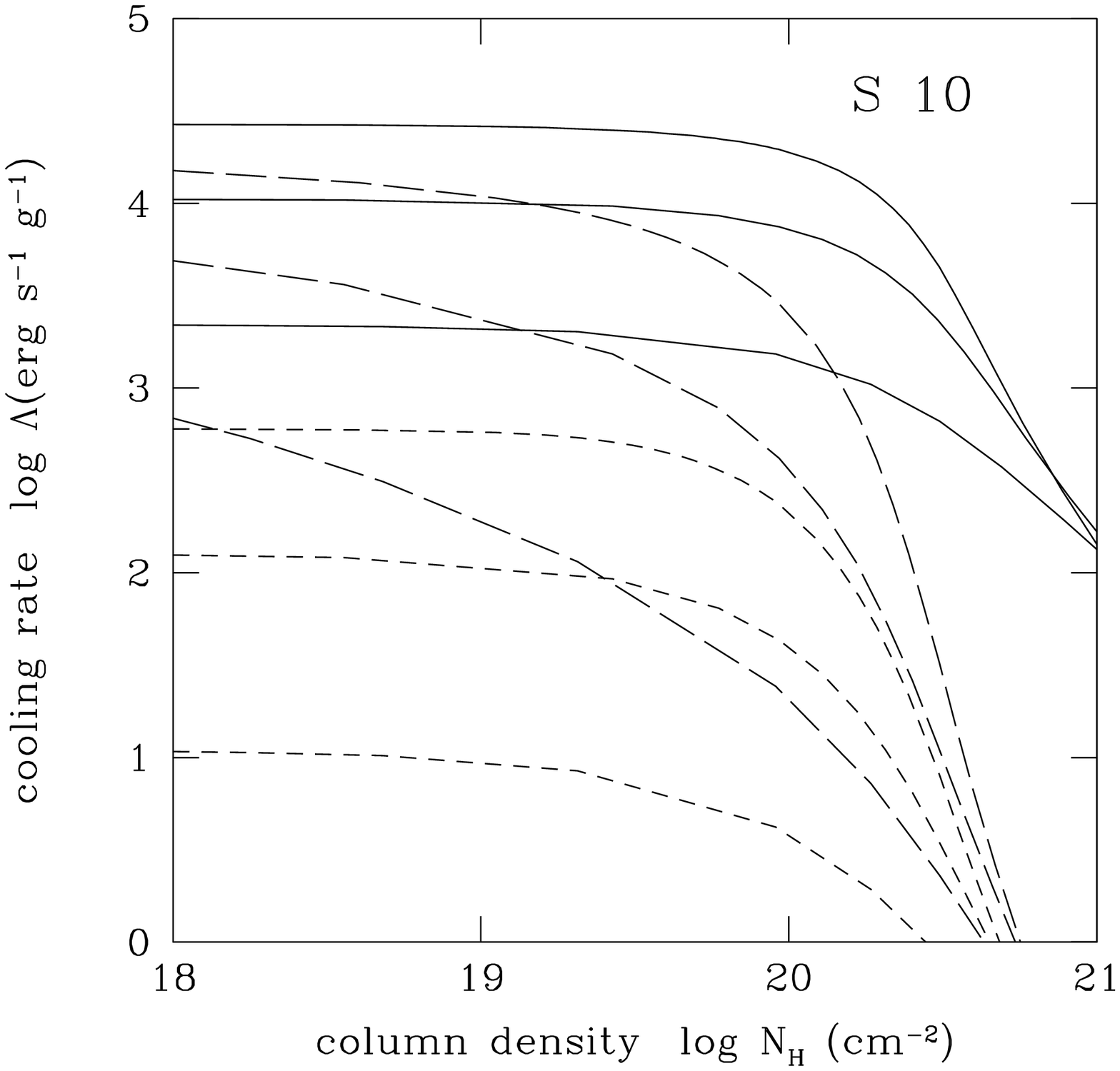}
  \end{center}
\caption{Cooling rates in the postshock relaxation layer.
The panels are for (a) LP05, (b) LP10, (c) S05, (d) S10 models. 
For each model, three epochs of $M_{\rm FC}= 0.0125 
M_{\odot}$ (bottom), $0.025 M_{\odot}$ (middle) and $0.05 M_{\odot}$ (top)
are shown.
The solid, dashed, and dotted curves indicate the cooling rates by
dust, H$_2$O, and ${\rm H_2^{18}O}$, respectively.}
\label{fig:shock2}
\end{figure}

In Figure \ref{fig:shock}, we present the temperature 
distribution in the postshock relaxation layer 
for three epochs with the first core mass being 0.0125, 0.025, and 
0.05 $M_{\odot}$ (i.e., 0.25, 0.5, and 1.0 $M_{\rm FC, max}$, respectively). 
The distribution is shown as a function of the column density 
from the accretion shock: a higher column density corresponds 
to more downstream.   
The cooling rate by each cooling agent, namely, the dust thermal emission
and the H$_2$O and ${\rm H_2^{18}O}$ line emission, is shown individually 
in Figure \ref{fig:shock2} for the same flows.

As can be seen in Figure \ref{fig:shock}, the postshock temperature 
$T_{2}$ is more dependent on the assumed first-core radius 
$R_{\rm sh}$, rather than the type of similarity solution.
It is $700-900$K for the cases of 
$R_{\rm sh}=5$ AU, while $T_{\rm 2} \simeq 400$K 
for those of $R_{\rm sh}=10$ AU. 
This can be understood as follows.
By assuming $\rho_{1}v_{s}^{2} \gg p_{1}$ in the preshock
and  $\rho_{2}v_{2}^{2} \ll p_{2}$ in the postshock, 
\begin{equation}
\frac{1}{2}v_{s}^2 
\simeq \frac{\gamma_{\rm ad}}{\gamma_{\rm ad}-1} \frac{p_{2}}{\rho_{2}}
= \frac{\gamma_{\rm ad}}{\gamma_{\rm ad}-1} \frac{k_{\rm B} T_{2}}{\mu m_{\rm H}}.
\end{equation} 
Since the flow velocity is approximately given by 
free-fall
\begin{equation}
\frac{1}{2} v_{s}^{2} \simeq \frac{G M_{\rm FC}}{R_{\rm sh}},
\end{equation}
the postshock temperature can be written as
\begin{equation}
T_{2} \simeq \frac{\gamma_{\rm ad}-1}{\gamma_{\rm ad}} \frac{\mu m}{k_{\rm B}} 
\frac{G M_{\rm FC}}
{R_{\rm sh}}
\simeq 600K \left( \frac{M_{\rm FC}}{0.05M_{\odot}} \right) 
\left( \frac{R_{\rm sh}}{5{\rm AU}} \right)^{-1}.
\label{eq:T2}
\end{equation}
Since the equation above does not
contain the type of similarity solution, the values of postshock temperature 
are similar for the two solutions with the same $R_{\rm FC}$ and $M_{\rm FC}$.
On the other hand, the dust temperature, which is  
close to the radiation temperature, is higher for the LP solutions, 
because of higher luminosity and higher optical depth in the envelope.
Assuming the radiative diffusion equation
\begin{equation}
\frac{\partial T}{\partial r}=-\frac{3}{16 \pi a c} 
\frac{\kappa_{\rm R} \rho L}{r^{2}T^{3}}
\end{equation}
in the envelope and approximating the left hand side by 
$T_{\rm 1}/R_{\rm env}$, where $R_{\rm env}$ is the outer 
radius of the envelope, we obtain 
\begin{equation}
T_{1}^4 \propto \frac{\tau_{\rm env} L}{R_{\rm sh}^{2}},
\end{equation}
where $\tau_{\rm env}$ is the envelope optical depth. 
This shows that the higher luminosity and higher envelope optical depth 
result in a higher radiation temperature ($\simeq T_{1}$) 
at the first-core surface.

Despite the smaller cooling rates (see Fig. \ref{fig:shock2}), 
the shocked gas cools at a lower column density 
in the Shu models ($\lesssim 10^{20} {\rm cm^{-2}}$) 
than in the LP ones ($\gtrsim 10^{20} {\rm cm^{-2}}$).
This is because of the longer timescale in the Shu cases:
since the accretion rates of the LP and Shu solutions 
are different by a factor of about fifty, 
for the same column density the time which elapsed after 
the shock is also different by the same factor.

In all models, for low column densities, 
thermal energy in the postshock flow is dissipated mostly 
in dust continuum emission (see Fig.\ref{fig:shock2}). 
The postshock temperature is several hundred degree K (Fig.\ref{fig:shock}), 
and increases with the first-core mass, 
reaching 900K (LP05) for $M_{\rm FC, max}=0.05M_{\odot}$.
In this temperature range, the most efficient molecular coolant 
is H$_2$O, and its cooling rate is as large as that by dust continuum 
for small column densities (see Fig.\ref{fig:shock2}).
However, for $N_H \gtrsim 10^{19} {\rm cm^{-2}}$, 
the H$_2$O lines become optically thick and the cooling rate decreases.
The optical depth effect is apparent when we compare the H$_2$O cooling rate 
with that by its isotope H$_{2}^{18}$O, which remains optically thin.

\begin{figure}
  \begin{center}
\FigureFile(80mm,80mm){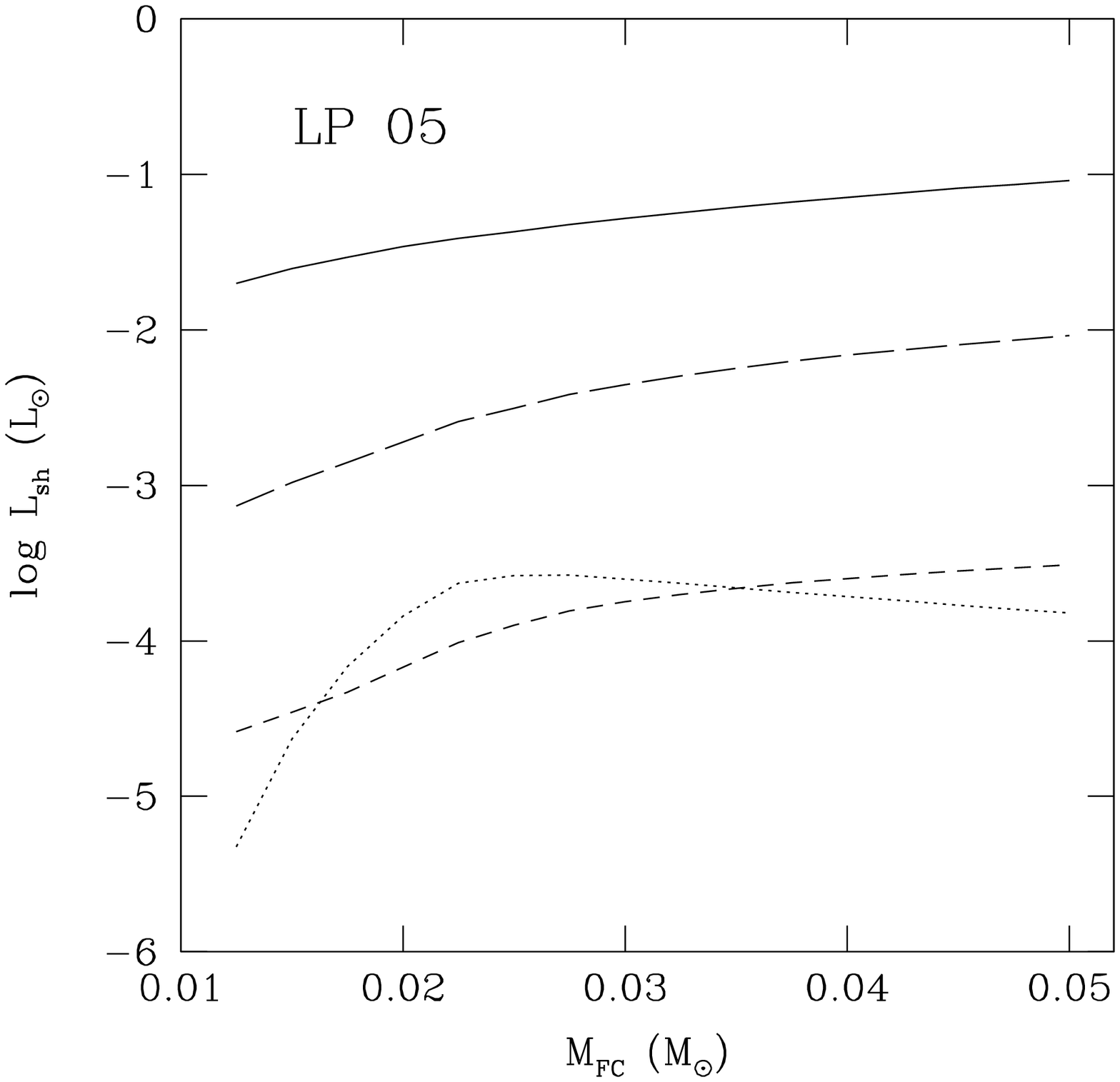}
\FigureFile(80mm,80mm){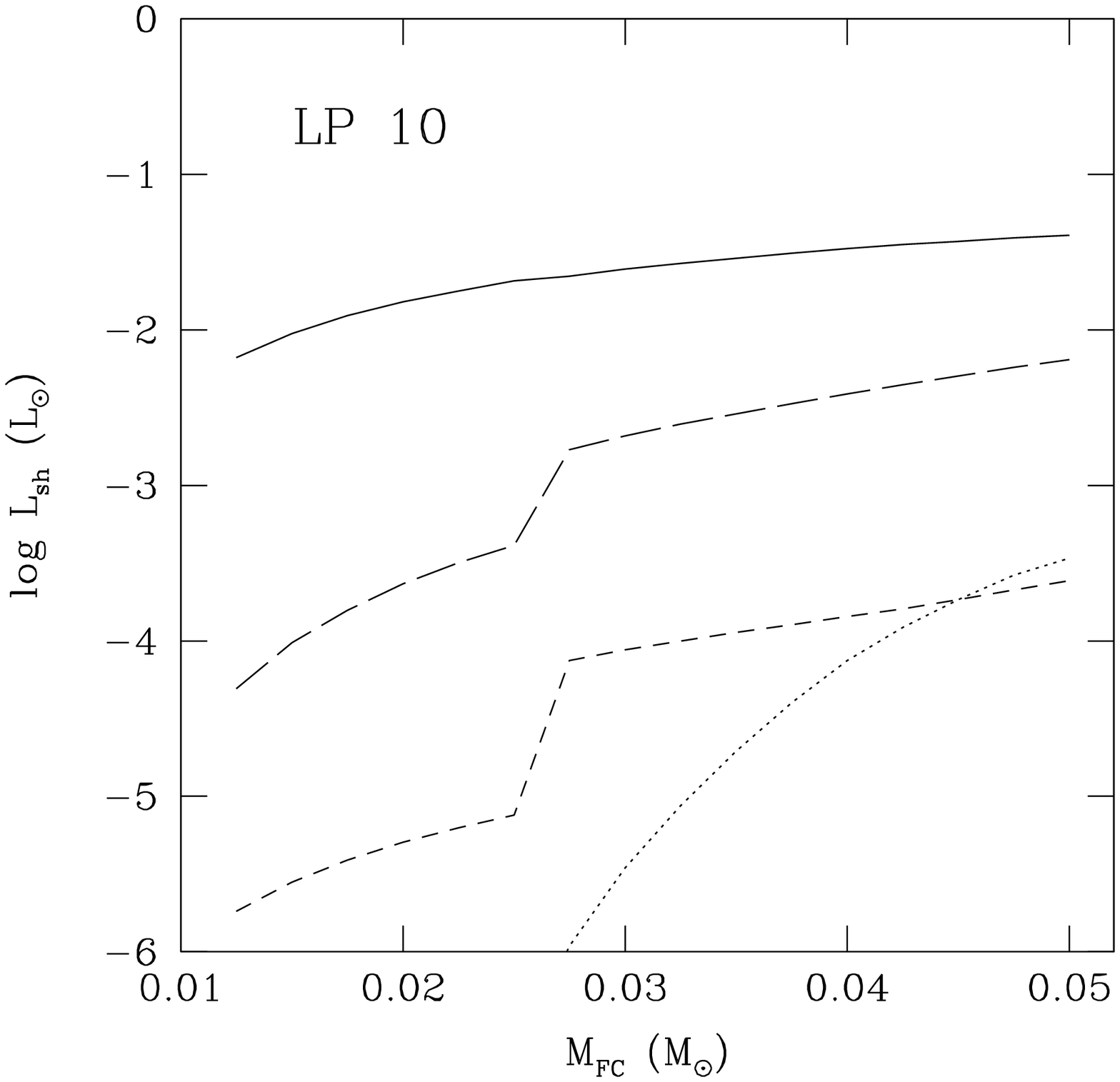}
\FigureFile(80mm,80mm){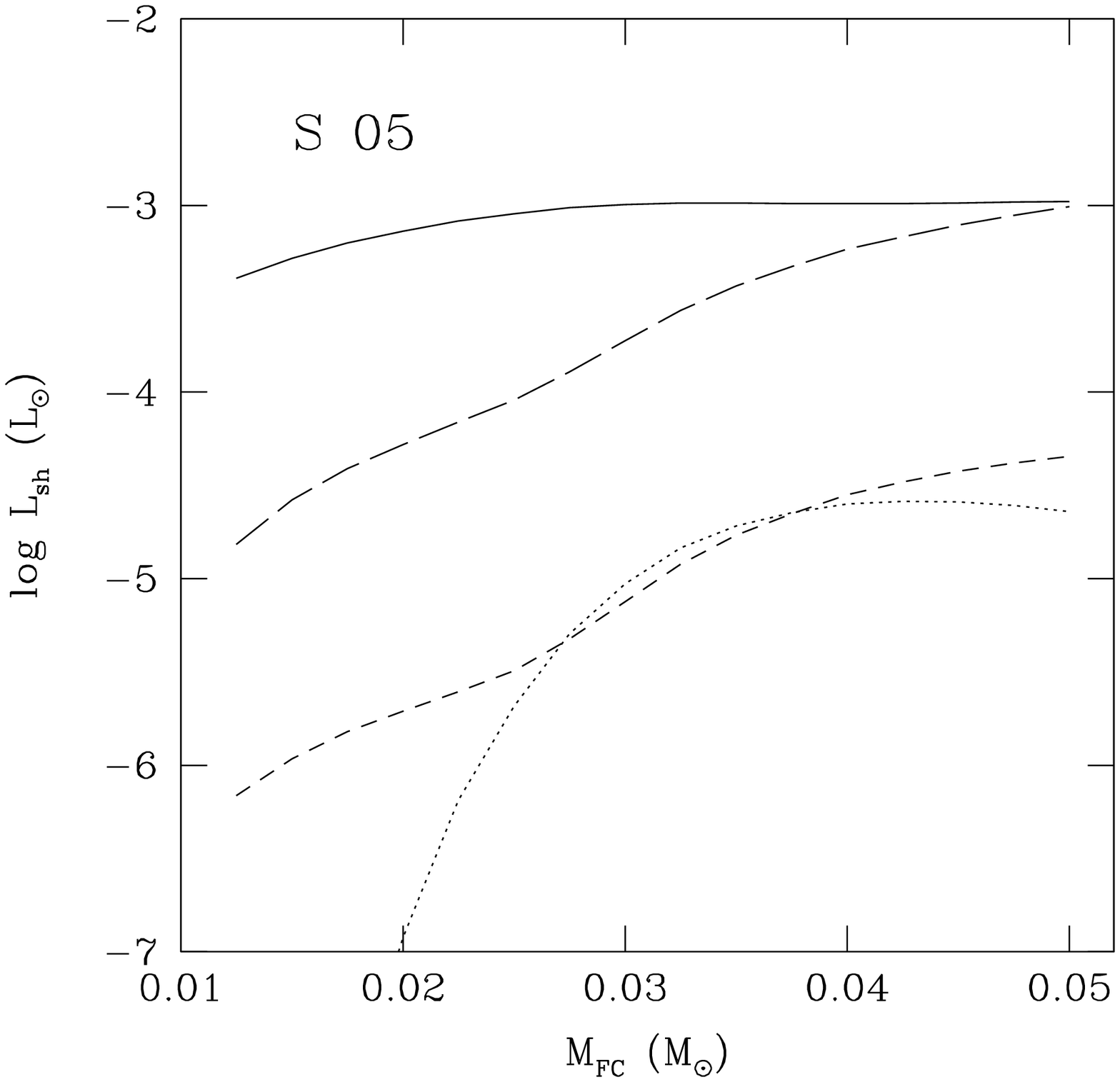}
\FigureFile(80mm,80mm){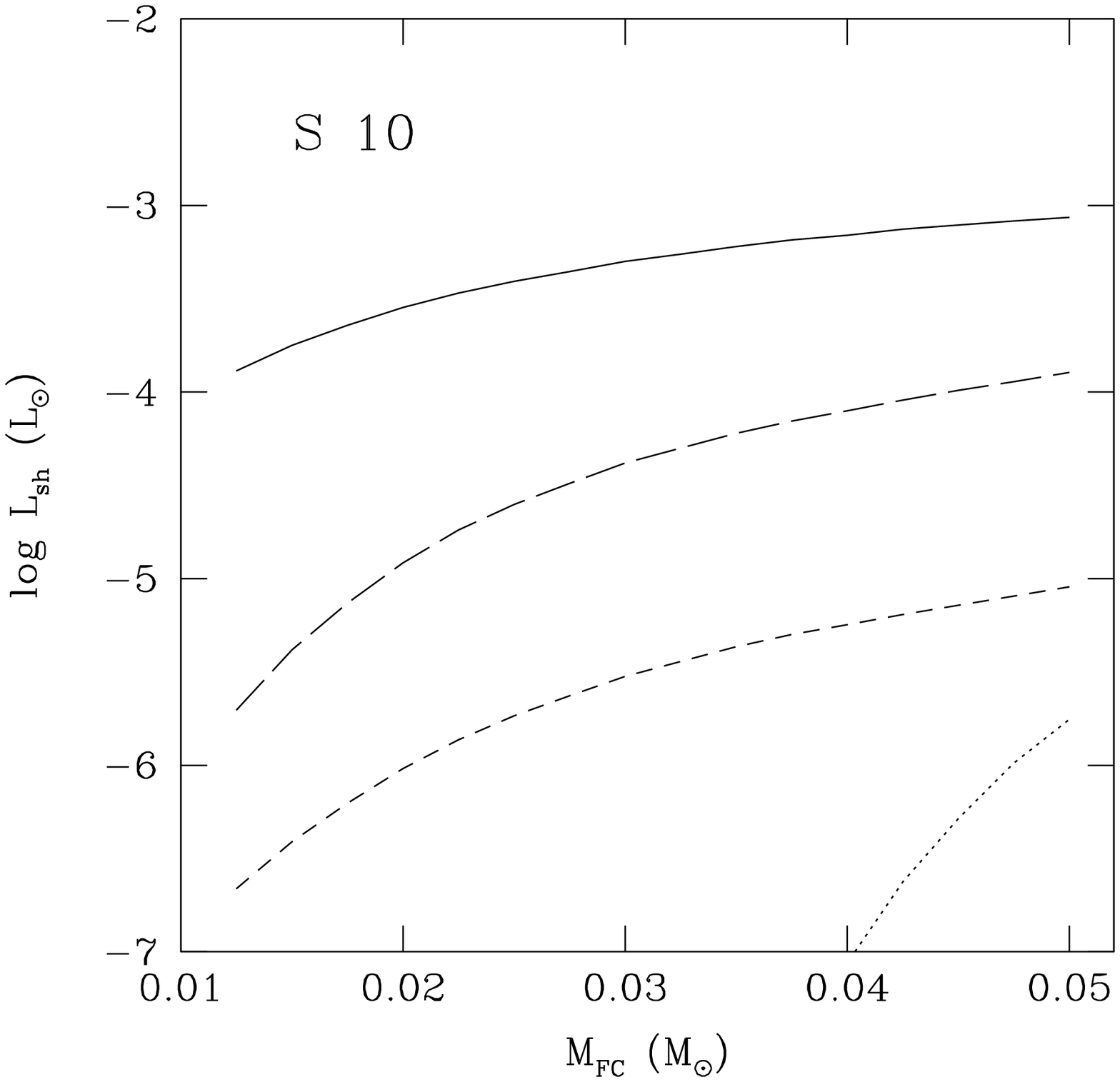}
  \end{center}
\caption{
Emitted luminosities in the postshock layer in the dust continuum (solid), 
H$_2$O lines (long-dashed), 
H$_2 ^{18}$O lines (short-dashed), and OH lines (dotted) as a function of 
the first-core mass. 
The panels are for (a) LP05, (b) LP10, (c) S05, (d) S10 models. }
\label{fig:L_sh}
\end{figure}

In Figure \ref{fig:L_sh}, we show the contribution to the luminosity emitted 
in the postshock layer by the individual components; 
dust continuum, H$_2$O lines, H$_2^{18}$O lines, and OH lines. 
Although the contribution of the H$_2$O-line cooling to the shock 
dissipation increases with the first-core mass, 
the maximum fraction of radiated energy in the H$_2$O lines 
is at most about 10\%, except for the S05 model.
In S05 model, the H$_2$O-line cooling is as important as the dust 
continuum at the end of the first-core phase $M_{\rm FC}=0.05M_{\odot}$.
In this model, since the postshock gas temperature is high (700K) 
and it takes a long time for the layer to become optically thick, 
the H$_2$O lines contribute to cooling as much as the dust emission.

The contribution of the water isotope ${\rm H_2^{18}O}$ is very small 
owing to its small abundance.
The second important species as a coolant is OH.
However, its role is only temporary:
since OH is intermediate species in chemical reactions leading to H$_2$O 
formation, its cooling rate decreases as OH is being converted to H$_2$O.

\begin{figure}
  \begin{center}
\FigureFile(80mm,80mm){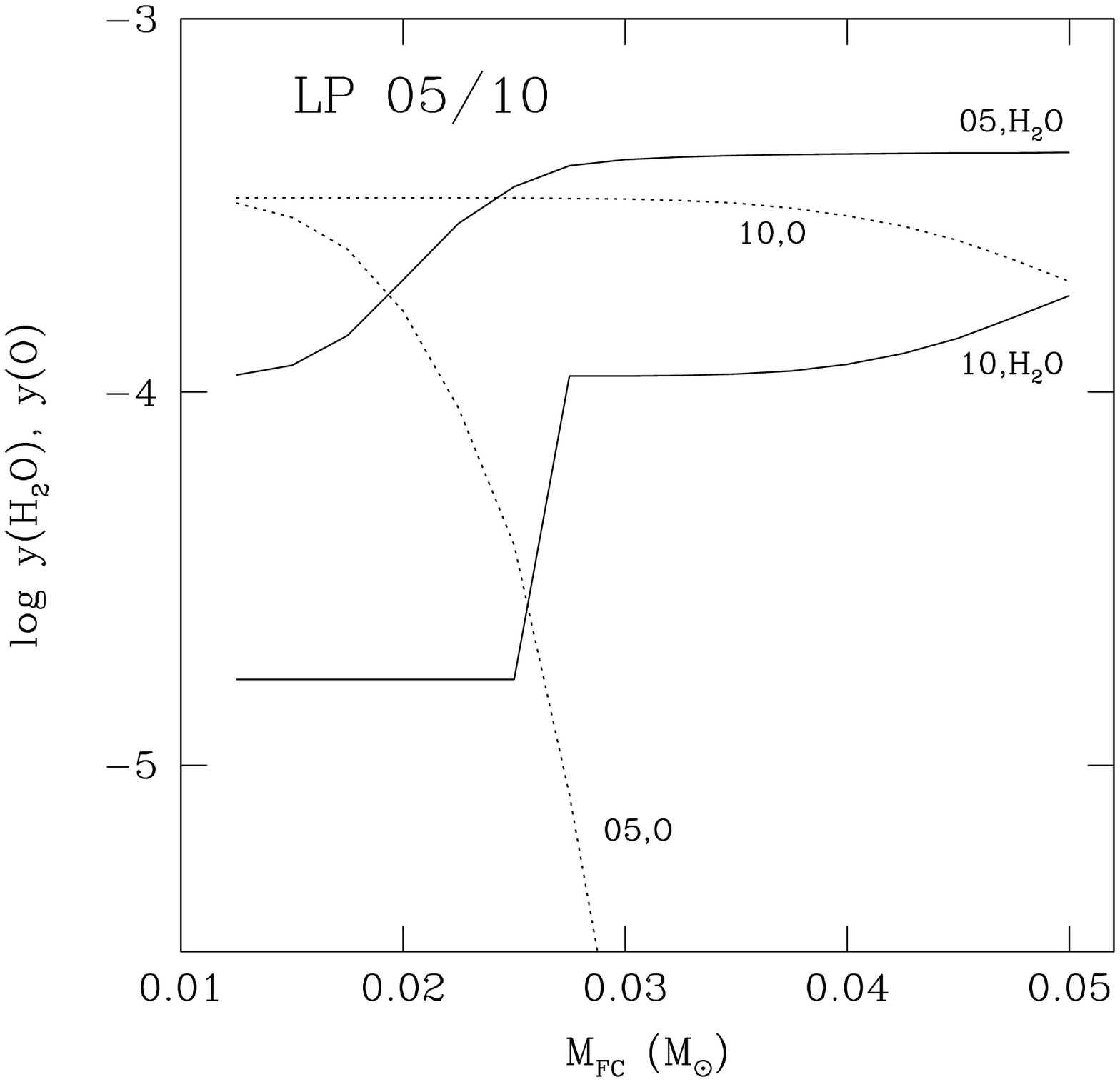}
\FigureFile(80mm,80mm){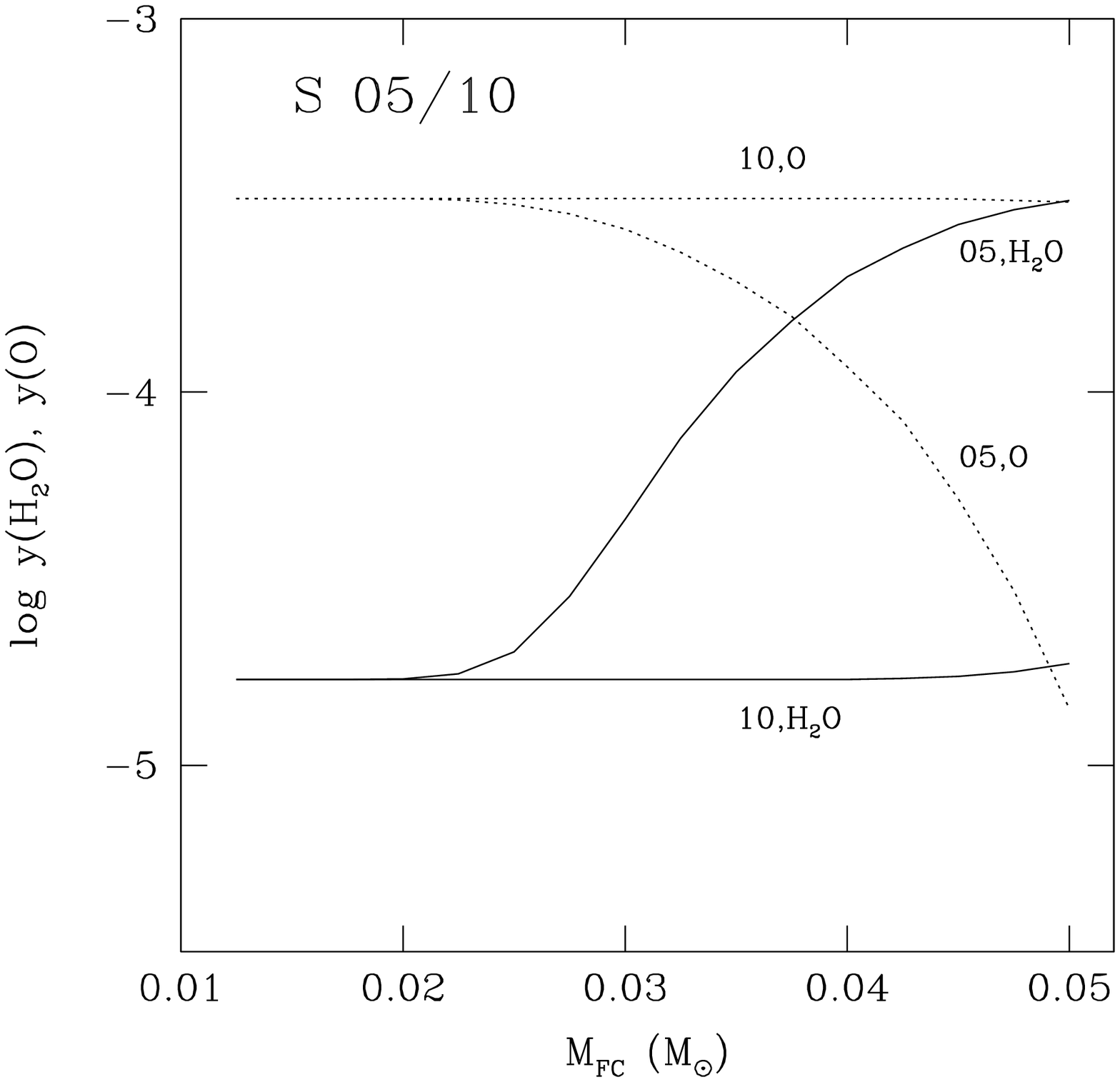}
  \end{center}
\caption{
H$_2$O (solid) and O (dashed) concentrations at the bottom of the 
postshock layer
as a function of the first core mass. 
The panels are for (a) LP05 and 10, 
(b) S05 and 10 models. }
\label{fig:chem}
\end{figure}

The value of the H$_2$O luminosity is dependent on the H$_2$O fraction 
as well as the gas temperature.
In Figure \ref{fig:chem}, we show the H$_2$O and O concentrations at the 
bottom of the postshock relaxation layer.  
In LP05 model, the postshock dust temperature is higher than the evaporation 
temperature of the ice, even in our minimum mass 
case of 0.0125$M_{\odot}$.
Thus a large amount of water vapor [ $y({\rm H_2O})=1.1 \times 10^{-4}$ ] 
is already present just below the shock. 
For the lowest mass case $M_{\rm FC} =0.0125M_{\odot}$, 
the final H$_2$O concentration remains almost the same as the initial value 
(Fig. \ref{fig:chem} a).  
In the cases with higher mass, the final H$_2$O concentration 
increases with $M_{\rm FC}$ by a chemical process in the postshock flow 
owing to higher temperature
and thus higher H$_2$O production rate (Wagner \& Graff 1987; 
Kaufman \& Neufeld 1996).
For $M_{\rm FC} \gtrsim 0.025M_{\odot}$, most of the remaining oxygen, 
which is not locked to CO, is converted to H$_2$O via OH.
This can be observed as a decrease in the OH luminosity 
(Fig. \ref{fig:L_sh} a).  
In the LP10 model, the ice mantle evaprates for $M_{\rm FC}> 0.028M_{\odot}$.
This can be seen as the discontinuous rise in the H$_2$O concentration, 
as well as the H$_2$O luminosity (Fig. \ref{fig:L_sh} b). 
Additional H$_2$O production by chemical reactions slightly increases the
H$_2$O concentration in the high-$M_{\rm FC}$ cases.
The ice mantles do not evaporate in either the S05 or S10 model.
In the S05 model, additional H$_2$O is produced by high-temperature 
gas-phase chemistry for 
$M_{\rm FC} \gtrsim 0.03M_{\odot}$, while in the S10 model H$_2$O hardly 
increases from the initial value of $y({\rm H_2O})=1.7 \times 10^{-5}$. 

In Table 1, we list the important transitions, 
their energy fluxes $F_{\rm line, sh} = \frac{1}{2} \dot{E}_{\rm line, sh}$
and the luminosities 
\begin{equation} 
L_{\rm line}= 4 \pi R_{\rm sh}^2 F_{\rm line, sh},
\label{eq:Lline}
\end{equation}
generated in the postshock relaxation layer. 
The emitted luminosities are higher and the wavelengths of the strongest 
emission lines are shorter 
for LP / small $R_{\rm sh}$ cases than for the Shu / large $R_{\rm sh}$ cases.
In general, the emitted radiation from the shock is subject to 
absorption and re-emission in the envelope (Sec. 3.2), 
and thus these values does not correspond to the observable line luminosities.
However, if cavities are present in the envelope due to outflow, 
rotation, etc., the shock emission can come out unattenuated.
In this case, the emission lines listed in Table 1 are observable 
probably as mid-infrared nebulosity.

\subsection{Processed Radiation through the Envelope}
\begin{figure}
  \begin{center}
\FigureFile(80mm,80mm){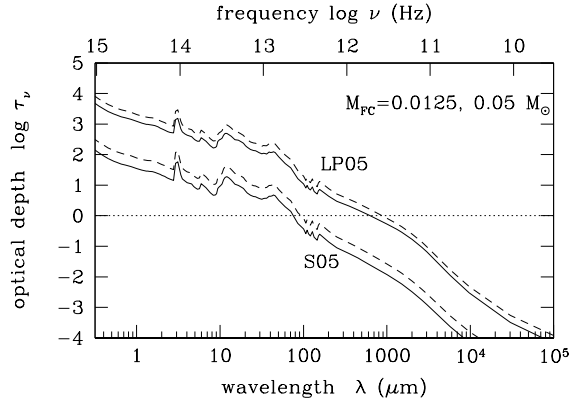}
  \end{center}
\caption{Optical depth of the envelope for models LP05 and S05
as a function of the wavelength.
The dashed and solid lines show the cases of $M_{\rm FC}=0.0125M_{\odot}$ 
and $0.05M_{\odot}$, respectively.
The dotted horizontal line indicates where the optical depth is unity.}
\label{fig:tau}
\end{figure}

A first-core object is enshrouded with an optically thick envelope, 
where radiation emitted from the accretion shock is absorbed and reemitted
as dust thermal radiation.
The total optical depths of the envelopes for each frequency 
are shown for the LP05 and S05 models in Figure \ref{fig:tau}. 
The envelope is optically thick at wavelengths shorter than 1mm 
in the LP05 model (LP10: 600$\mu$m, S05: 100$\mu$m, S10: 75$\mu$m, 
respectively).
With increasing mass of the first core, the inner portion of the envelope is 
accreted and the optical depth decreases.
This is not a large effect compared with the differences between 
the LP and Shu models. 

\begin{figure}
  \begin{center}
\FigureFile(80mm,80mm){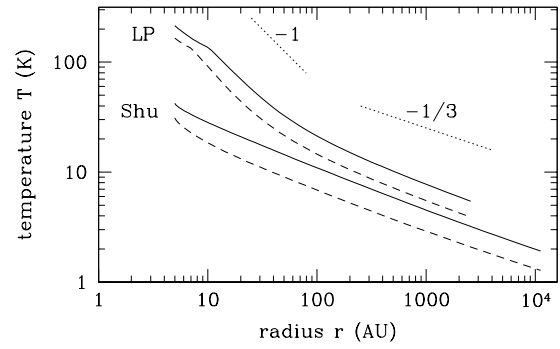}
  \end{center}
\caption{Temperature distributions in the envelope.
The dashed and solid lines show the distribution at epochs of 
$M_{\rm FC}= 0.0125 M_{\odot}$ and $0.05 M_{\odot}$, 
respectively. Lines for $T \propto r^{-1/3}$ and $\propto r^{-1}$, 
which correspond to optically thick and thin cases, respectively, 
are also shown. }
\label{fig:r_T}
\end{figure}

Figure \ref{fig:r_T} shows the temperature distribution in the envelope.
For the LP models, the dust photosphere locates at $\sim 30$ AU, 
outside of which the envelope is optically thin and reemitted photons 
escape freely.
The temperature falls approximately as $T \propto r^{-1}$ 
and $T \propto r^{-1/3}$
in optically thick and thin regimes, respectively 
(e.g., Sec. 4.2 of Hartmann 1998).
Below its vaporization temperature ($\simeq$ 150 K), 
H$_2$O ice is an important contributor to the opacity.
Therefore, for $\gtrsim 150$K, due to evaporation of ice mantle of dust, 
the dust opacity is reduced.
This causes the flattening of temperature profile 
in the innermost region observed in the LP models.
For the Shu models, where the envelope is optically thin, 
the temperature obeys $T \propto r^{-1/3}$ in the entire envelope.

Figure \ref{fig:dustSED} shows the spectrum of processed dust radiation, 
i.e. the radiation escaping from the envelope. 
The emitted dust radiation 
$L_{\nu}= 4 \pi R_{\rm sh}^2 \dot{E}_{\rm \nu, sh}$
in the shock is also shown for comparison.
However, the peak of the shock emission locates at rather short wavelength, 
around 10-100$\mu$m, due to high temperature in the postshock flow, 
while that of processed radiation appears at longer wavelength, 
100$\mu$m-1mm, which reflects the lower envelope temperature.

The envelopes of the LP models are optically thick both to 
the shock and the processed radiation, while those of the Shu models 
are optically thick to the emitted radiation, 
but optically thin to the re-emission.
Owing to this difference, the processed spectra of the LP models 
resemble the blackbody radiation $L_{\nu} \propto B_{\nu} \propto \nu^2$ 
on the Rayleigh-Jeans side, 
while those of the Shu model have flatter spectra in low frequencies 
$L_{\nu} \propto \nu^{1/2}$ approximately (Fig. \ref{fig:dustSED}).
This flat spectrum is explained by the following consideration.
The radiation spectrum from optically thin medium is 
\begin{equation}
L_{\nu} = 4 \pi \int \kappa_{\nu} B_{\nu} dM 
         \propto  \nu^{3+\beta-(3-n)/m} \simeq \nu^{1/2}, 	
\end{equation}
for density and temperature distributions $\rho \propto r^{-n}$ and 
$T \propto r^{-m}$, and an opacity of $\kappa_{\nu} \propto \nu^{\beta}$.
In the above, we used $n=3/2$, $m=1/3$ and $\beta=2$.
Comparing the LP and Shu models with the same $R_{\rm sh}$, 
we notice that, although the peak frequency of the shock emission is 
higher in the LP models, that of the processed radiation is lower 
owing to the larger optical depth of the envelope.

In Figures \ref{fig:water_LP05}-\ref{fig:water_S10}, 
we show the emitted and processed luminosities of the H$_2$O lines 
in three different epochs for all the models.
For the emitted luminosities, i.e. the unattenuated 
luminosities before processing in the envelope (eq. \ref{eq:Lline}), 
the corresponding
values of energy flux at the shock $F_{\rm line, sh}$ are also indicated 
on the left axis.
We also present in Table 2 the luminosities and the 
line flux at the distance of 150pc 
for the ten strongest H$_2$O emission lines
at the epoch of $M_{\rm FC}=0.05M_{\odot}$. 
Among H$_2$O emission lines from the shock in the LP models 
(see Figures \ref{fig:water_LP05} and  \ref{fig:water_LP10}),
those with frequencies $\sim 10^{13}$Hz (30$\mu$m) have the highest 
luminosities 
$\sim 10^{-4}-10^{-5}L_{\odot}$.
These lines are, however, heavily attenuated in the envelope 
owing to the high optical depth in these frequencies.
On the other hand, the peak in processed line radiation appears around 
$10^{12}$Hz (300$\mu$m), where the optical depth of the envelope is about a few.
With such small values of the processed luminosities 
$10^{-8}-10^{-10}L_{\odot}$, those lines are completely 
buried in the dust continuum:
$L_{\rm line}/\Delta v_{\rm D} \ll L_{\rm \nu, dust}$, where 
$\Delta v_{\rm D}$ is the Doppler width of the line.
In the Shu models, line emission in the shock peaks around $10^{13}$Hz (30$\mu$m), 
which is similar to that in the LP model.
The processed line emission, however, peaks around $3 \times 10^{12}$Hz (100$\mu$m), 
higher than in the LP models owing to the smaller optical depth in 
the Shu models. 
With maximum luminosities of $10^{-6}-10^{-7}L_{\odot}$, 
the monochromatic luminosities of these lines are higher than 
that of the dust continuum.
The final processed luminosities in the dust continuum 
and the H$_2$O lines are shown in Figure \ref{fig:L_out}.
In the LP models, the H$_2$O luminosities are only $\sim 10^{-6}$ of 
the dust luminosities, while in the Shu models, about 1\% 
of the radiation is still in H$_2$O lines after reprocessing in the envelope.
 
Profiles of brightness temperature are shown for different 
wavelength bands in Figures (\ref{fig:Tb.100G})-(\ref{fig:Tb.100micron}).
These figures show that in low frequencies where the envelopes are optically 
thin, the emission from the first cores is directly visible.
Then, a discontinuity appears in the brightness temperature distribution 
at the radius of the first core.
On the other hand, at the shorter wavelengths, where the envelope is 
optically thick, only the emission from the dust photosphere in 
the envelope is visible. 
With increasing observing frequency, the brightness temperature 
at the center decreases and the discontinuity at the first-core surface 
disappears.

\section{Summary and Discussion}
We calculated the radiation spectrum of young stellar objects in 
their first-core phase.
As envelope models, we studied cases of the Larson-Penston (LP) and 
the Shu self-similar solutions.
In the LP model, the emitted spectrum has a peak at $\sim 10 \mu$m
both for the dust continuum and H$_2$O lines.
The optical depth of the envelope at this frequency is $100-1000$, such that 
virtually all of the shock emission is absorbed in the envelope and 
reemitted as the dust thermal emission at lower frequencies.
The amount of total luminosity remains unchanged, 
which is $0.01-0.1 L_{\odot}$,
but the emission peak is shifted to lower frequency owing to the 
lower temperature in the envelope.
The processed spectrum can be approximated by the blackbody radiation with
temperature 30K, which peaks at 200$\mu$m.
Our result for the continuum spectrum is consistent with a previous study
by Masunaga \& Inutsuka (2000).
In the Shu models, the amount of total luminosity is 
$10^{-4}-10^{-3} L_{\odot}$, which is smaller by almost two 
orders of magnitude than in the LP model owing to 
the smaller accretion rate by a similar factor.
The dust temperature in the postshock layer 
is lower ($\lesssim 100$K), and thus the wavelength of the emission peak is 
longer ($\sim 40\mu$m) than that in the LP case ($\sim 10 \mu$m).
However, owing to the lower optical depth in the envelope, 
the peak of the processed spectrum is at a shorter wavelength ($100-200\mu$m) 
than that in the LP case.

Strong H$_2$O emission from accreting protostars has already 
been pointed out by Neufeld \& Hollenbach (1994), who 
considered the shock on the surface of circumstellar disks 
of more evolved objects. 
They carried out a parameter study for shocks with 
a preshock density of $n_{1}=10^{7.5}-10^{12}{\rm cm^{-3}}$ and a
velocity $v_{\rm s}=5-100{\rm km/s}$, and calculated the
emission from them.
Although, in their calculations, protostars tend to be more
massive, and shocks are thus stronger than ours 
($n_{1}=10^{9}-2 \times 10^{11}{\rm cm^{-3}}$ and 
$v_{\rm s}=1-4{\rm km/s}$ in our models), 
our shock calculations and results
are very similar to their weakest shock cases.
Provided that shock conditions are similar, 
it is a difficult task to distinguish observationally
whether the shock is on the disk surface or on the stellar surface 
without spatially resolving the emission region.
In the case of accretion onto the disk, the amont of dust extinction
can be smaller than that in our case.
Using this fact, we might be able to tell the geometry of the shock.
We need more studies in the future to confirm this possibility.

So far, we have discussed the case of spherical symmetry, where the envelope
absorbs the shock emission and reemits it at the longer wavelengths.
In practice, however, outflows are observed almost ubiquitously 
in very young stellar objects, such as Class 0 objects.
Theoretically, it is known that outflows can be launched as early 
as the first-core phase (Machida et al. 2006).
Even without outflow, a flattened disk-like density distribution is realized 
by the presence of angular momentum (Terebey, Shu, \& Cassen 1984). 
Thus, it is highly plausible that the envelope is optically thin 
in the polar direction while it is optically thick in the equatorial 
(disk) one.
If so, the shock emission escapes unattenuated 
and can be observed directly 
in the polar direction, or as a reflection nebula in other directions.
In these cases, the radiation is mostly emitted in the mid-infrared range,
and about 10\% of the emission is in H$_2$O lines at similar wavelengths 
(see Table 1).

In this paper, we demonstrated that observations in mid- to far-infrared 
and submillimeter wavelengths are important for finding the first-core
objects.
Among observational facilities at these wavelengths, we examine
observational feasibilities by currently operating {\it Spitzer 
Space Telescope (Spitzer)} and {\it Stratospheric Observatory for 
Infrared Astronomy (SOFIA)}, as well as forthcoming 
{\it Herschel Space Observatory}, 
{\it Atacama Large Millimeter/submillimeter Array (ALMA)} and 
{\it SPace Infrared telescope for Cosmology and Astrophysics (SPICA)}.  
{\it Spitzer} has a spectroscopic line sensitivity of 
$\sim 10^{-18} {\rm W/m^2}$ in mid-infrared wavelengths ($\sim 30\mu$m) 
and a photometric sensitivity of $\sim 0.1$mJy at $10\mu$m and 
$\sim 10$mJy at $100\mu$m.
Here we suppose the observation time to be one hour.  
The airborne telescope {\it SOFIA} has a spectroscopic line sensitivity of 
$\sim 10^{-17} {\rm W/m^2}$ at mid to far-infrared wavelengths 
($5-200 \mu$m) and photometric sensitivity of $\sim 1$mJy at $10\mu$m and 
$\sim 30$mJy at $100\mu$m. 
{\it Herschel} is a space telescope with its aperture 3.5m and 
oberving wavelengths $60-700\mu$m. 
Its spectroscopic line sensitivity is $10^{-18}-10^{-17} {\rm W/m^2}$
in short wavelengths ($\lambda < 200\mu$m) 
and $10^{-17}-10^{-16} {\rm W/m^2}$ in
longer wavelengths, while its photometric sensitivity is $1-10$mJy.
The millimeter and submillimeter telescope {\it ALMA} observes at
$300{\rm \mu m} - 1{\rm cm}$ (or $30-950$GHz), with its spectroscopic 
line sensitivity $\sim 10^{-20}{\rm W/m^2}$ and photospheric 
sensitivity $\sim$ 0.1mJy.
Finally, {\it SPICA} is an infrared space telescope with a 3.5m aperture, 
the same as {\it Herschel}, but its mirror is cooled to 4.5K, 
far cooler than that of {\it Herschel} (80K).
The observing wavelengths are from 5$\mu$m to 1mm, with its spectroscopic 
line sensitivity being $10^{-21}-10^{20}{\rm W/m^2}$.
Its photospheric sensitivity depends strongly on the wavelength:
it is $\sim 1\mu$Jy around 10$\mu$m and $\sim$1mJy around 100$\mu$m. 

Here, we evaluate the H$_2$O line flux from first-core objects.
After attenuation in the envelope, the maximum value of the H$_2$O-line 
luminosities is $L_{\rm line}\sim 10^{-7}L_{\odot}$ with its wavelength 
around 100$\mu$m 
in the Shu models, although it is significantly dimmer in the LP models.
If there is a cavity in the envelope owing to outflow or rotation, and
the shock radiation leaks through it without reprocessed in the envelope,
the line luminosities take the maximum value at around $30\mu$m
both for the LP and Shu solutions, with its value being 
$L_{\rm line} \sim 10^{-4}L_{\odot}$ (LP) or 
$L_{\rm line} \sim 10^{-5}L_{\odot}$ (Shu) , respectively 
(see Figures \ref{fig:water_LP05} - \ref{fig:water_S10}).
Below, as a typical distance to a first-core object, we use 1kpc, 
farther than neighboring star-forming regions, 
because, for detection of such rare objects,
the observation of a large number of young stellar objects is needed. 
The line flux can be given by
\begin{eqnarray}
F_{\rm line}&=&3.4 \times 10^{-18} {\rm W/m^{2}} 
\left( \frac{L_{\rm line}}{10^{-4} L_{\odot}} \right)
\left( \frac{D}{1{\rm kpc}} \right)^{-2}\\
&=&3.4 \times 10^{-21} {\rm W/m^{2}} 
\left( \frac{L_{\rm line}}{10^{-7} L_{\odot}} \right)
\left( \frac{D}{1{\rm kpc}} \right)^{-2},
\end{eqnarray}
where we use typical values of the luminosities for 
unattenuated ($10^{-4} L_{\odot}$) and attenuated radiation 
($10^{-7} L_{\odot}$) in the first and second expressions, respectively.
If the shock emission from a first core is 
visible without attenuation in the envelope, 
{\it Spitzer} is able to detect it in sources as far as 1kpc.
If sources are closer, {\it SOFIA} also has chance to observe them.
However, their sensitivities at far-infrared wavelengths are 
not enough for observing the attenuated emission.
On the other hand, {\it SPICA} can detect both processed line radiation 
through the envelope and direct radiation from the accretion shock 
as far as 1kpc. 
Owing to lack of the sensitivity at the mid-infrared wavelengths, 
the other future telescopes cannot observe shock radiation.
As for the attenuated line radiation, which appears in the 
submillimeter range, the sensitivity of {\it Herschel} is not sufficient. 
Although sensitivity of {\it ALMA} appears to be sufficient,
absorption by telluric water vapor prohibits the observation of 
H$_2$O lines in the submillimeter range.

\begin{figure}
  \begin{center}
\FigureFile(80mm,80mm){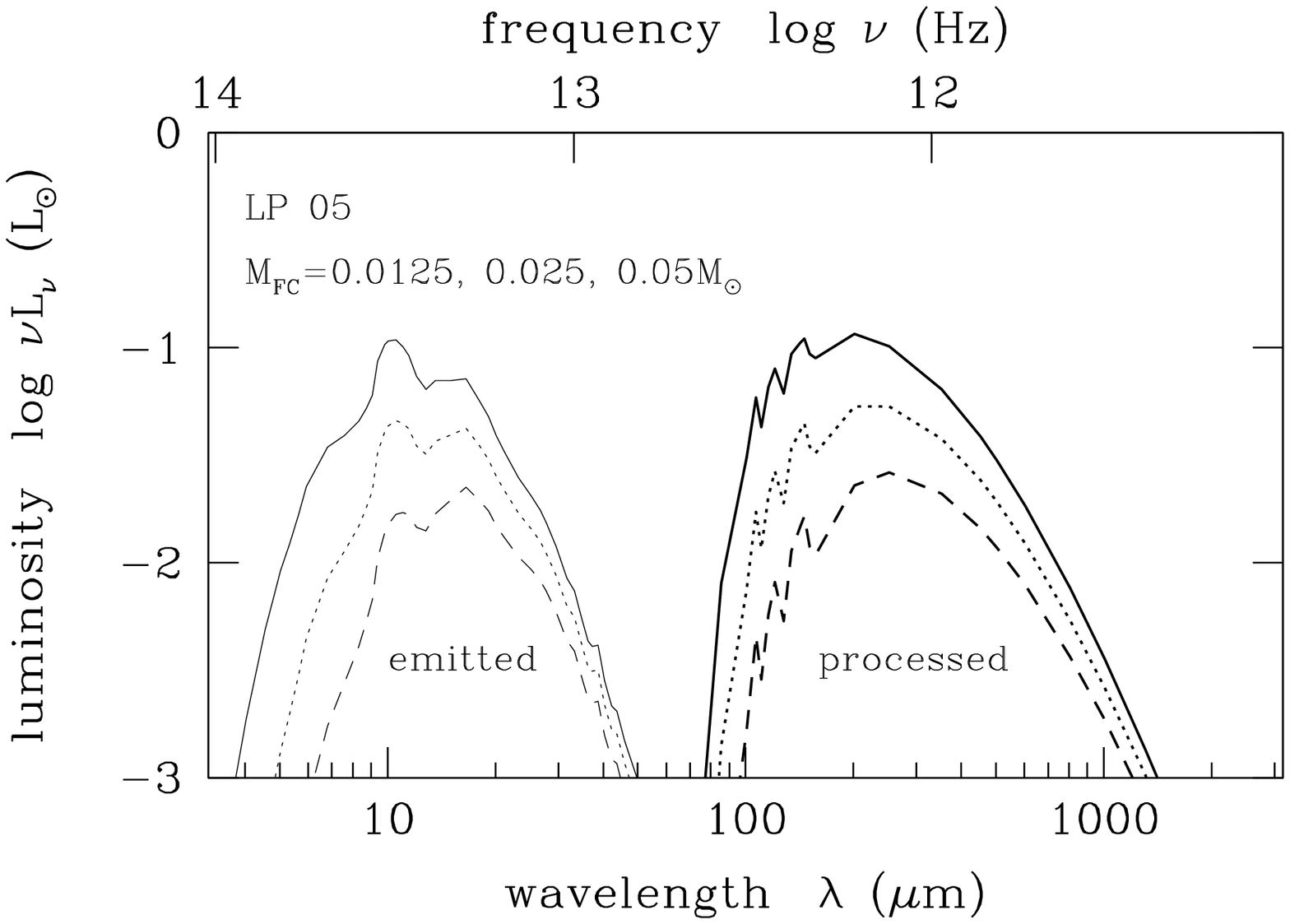}
\FigureFile(80mm,80mm){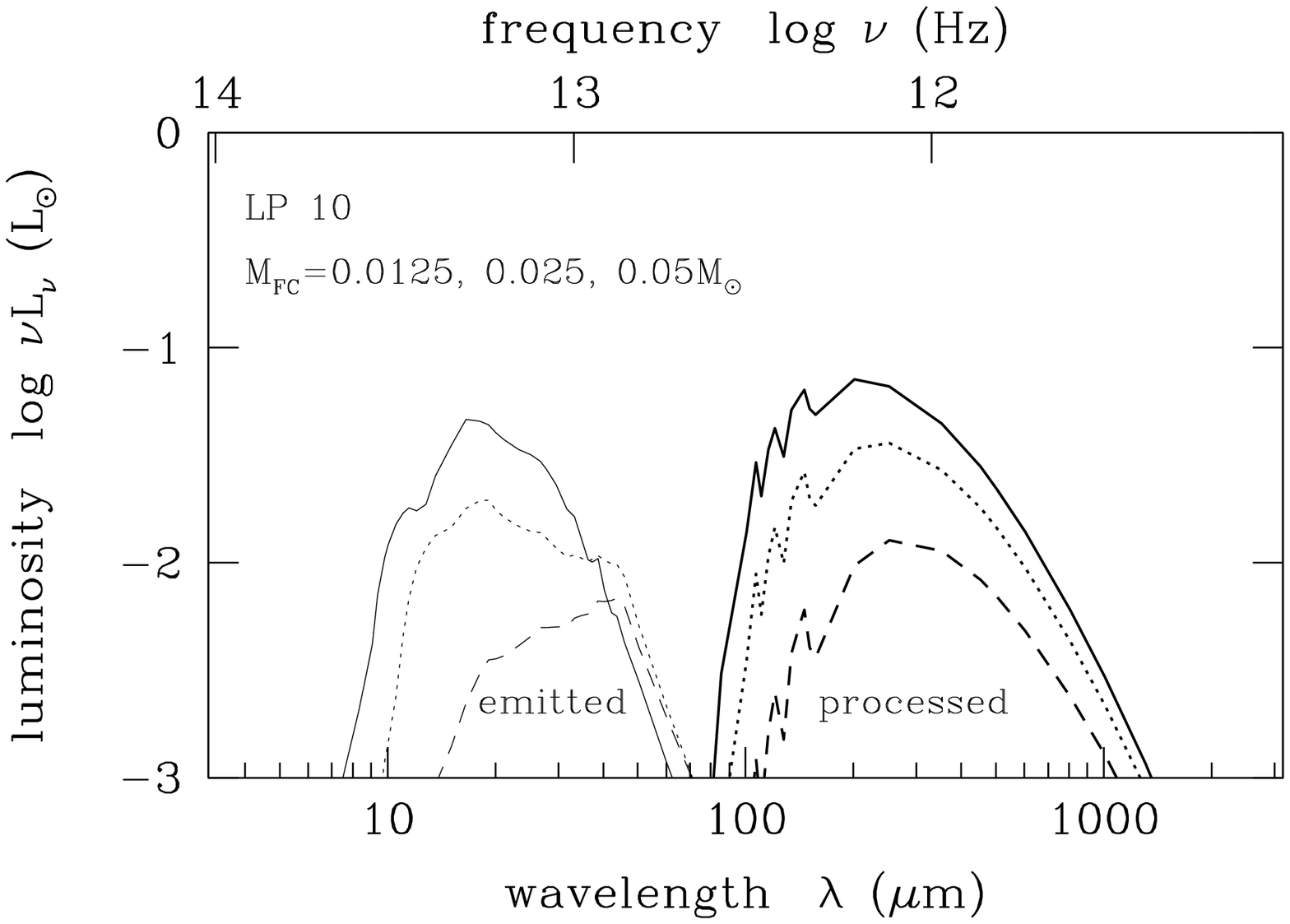}
\FigureFile(80mm,80mm){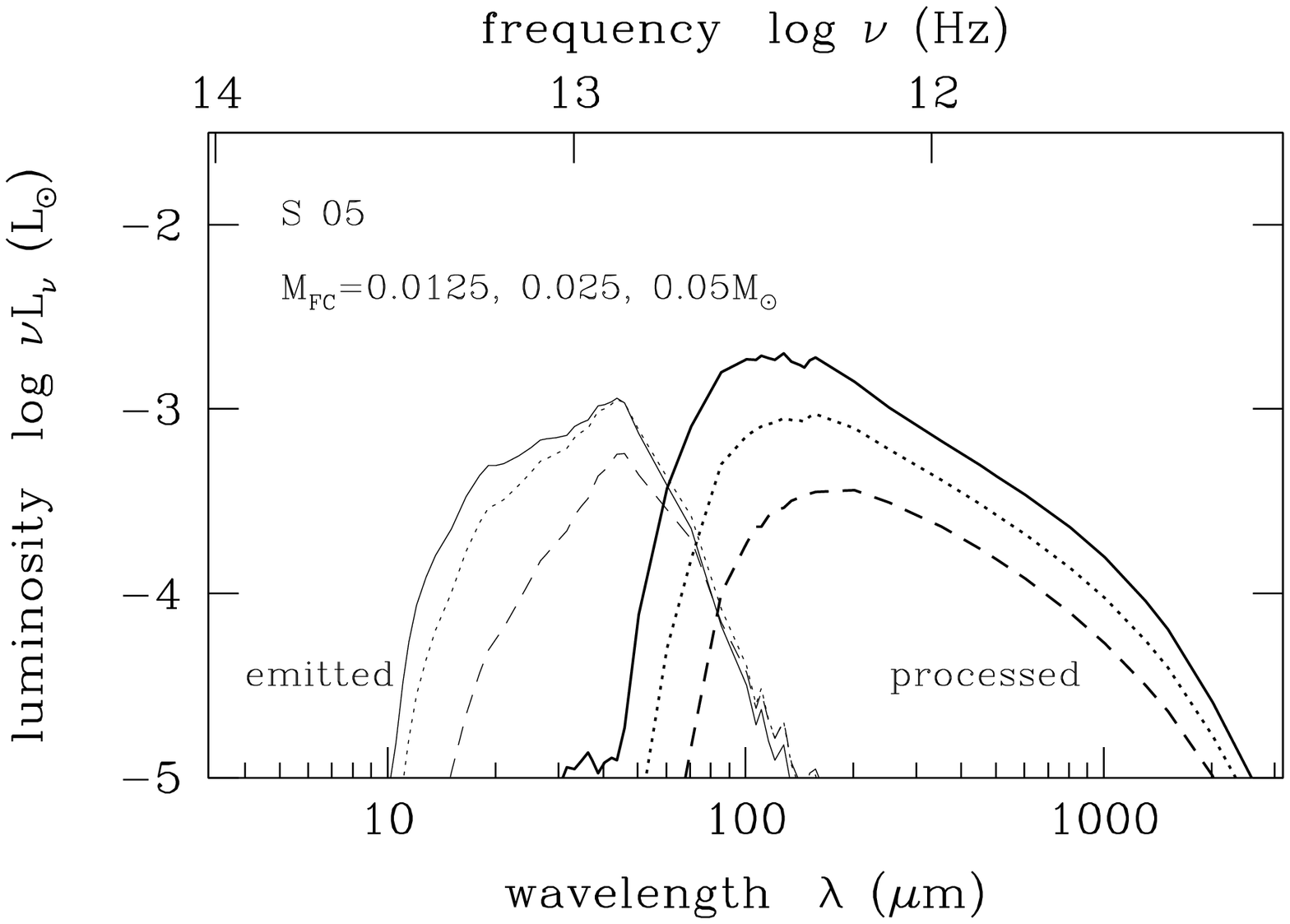}
\FigureFile(80mm,80mm){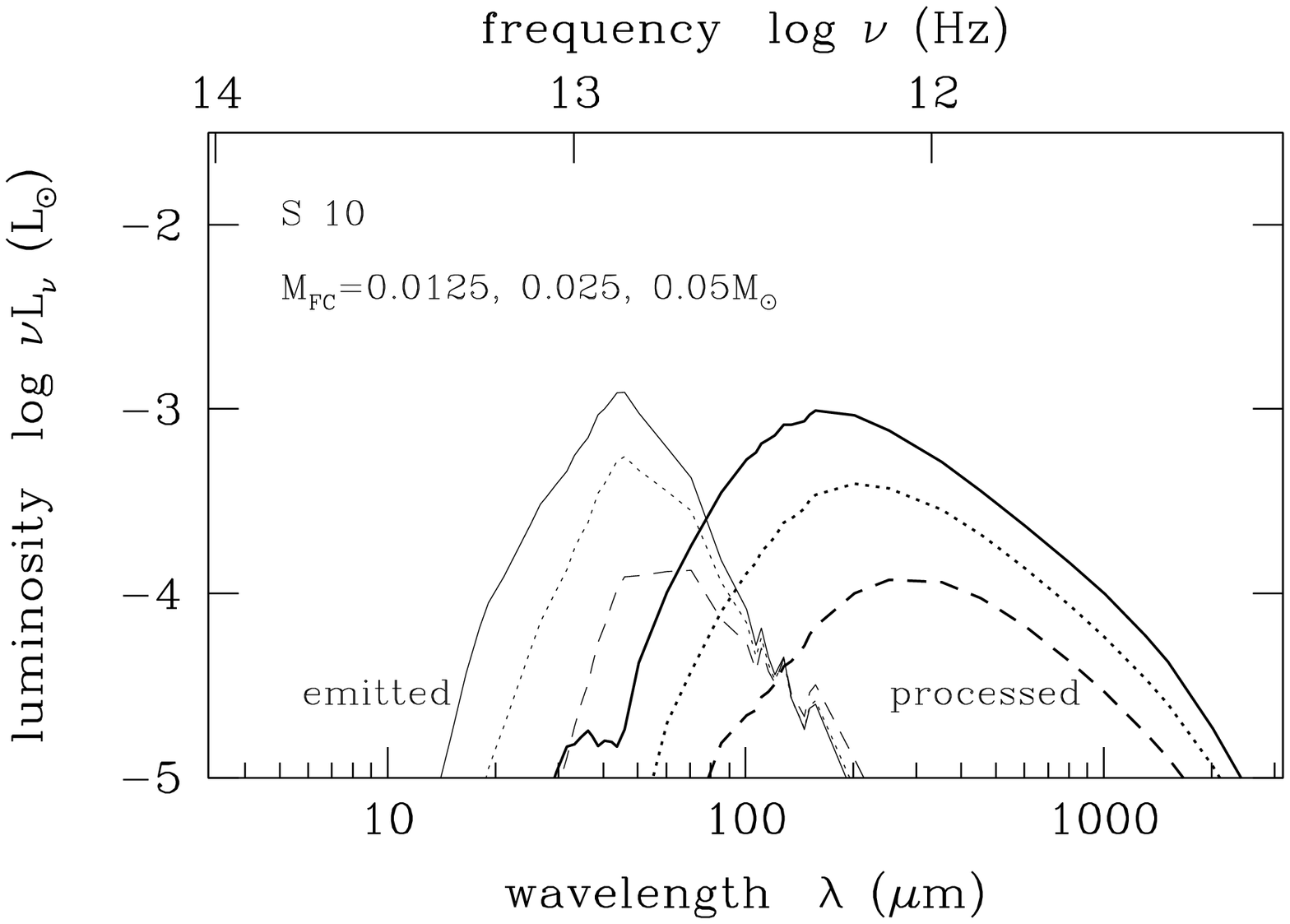}
  \end{center}
\caption{Spectral energy distributions in the dust continuum of the first-core 
objects. Both the emitted (thin) and processed (thick) spectra are shown.
The panels (a)-(d) show models LP05, LP10, S05, and S10, respectively.
In each panel, the distributions are shown for three different epochs 
of $M_{\rm FC}=0.0125M_{\odot}$ (dashed), 
$0.025M_{\odot}$ (dotted), and $0.05M_{\odot}$ (solid).}
\label{fig:dustSED}
\end{figure}

\begin{figure}
  \begin{center}
\FigureFile(80mm,80mm){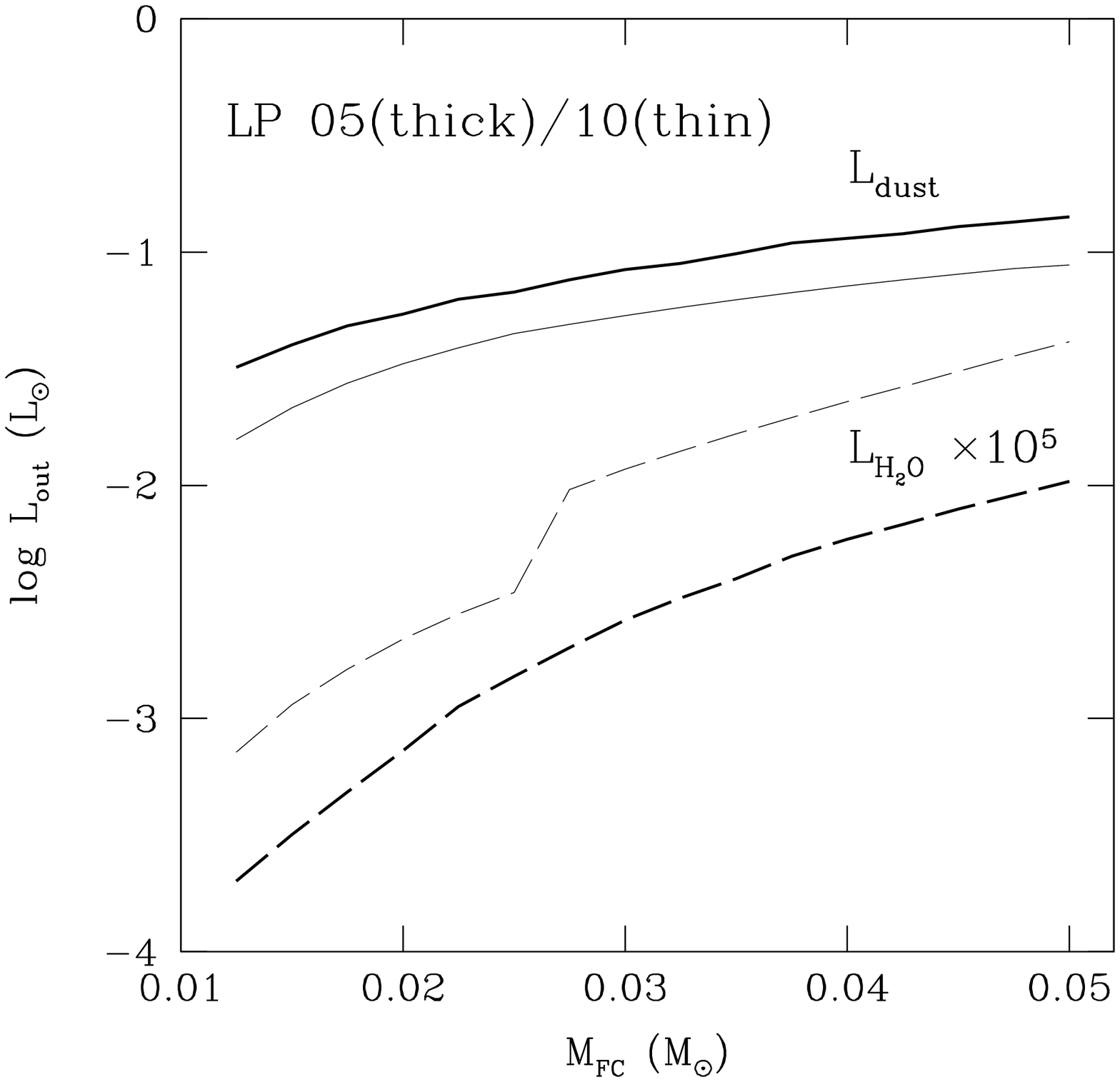}
\FigureFile(80mm,80mm){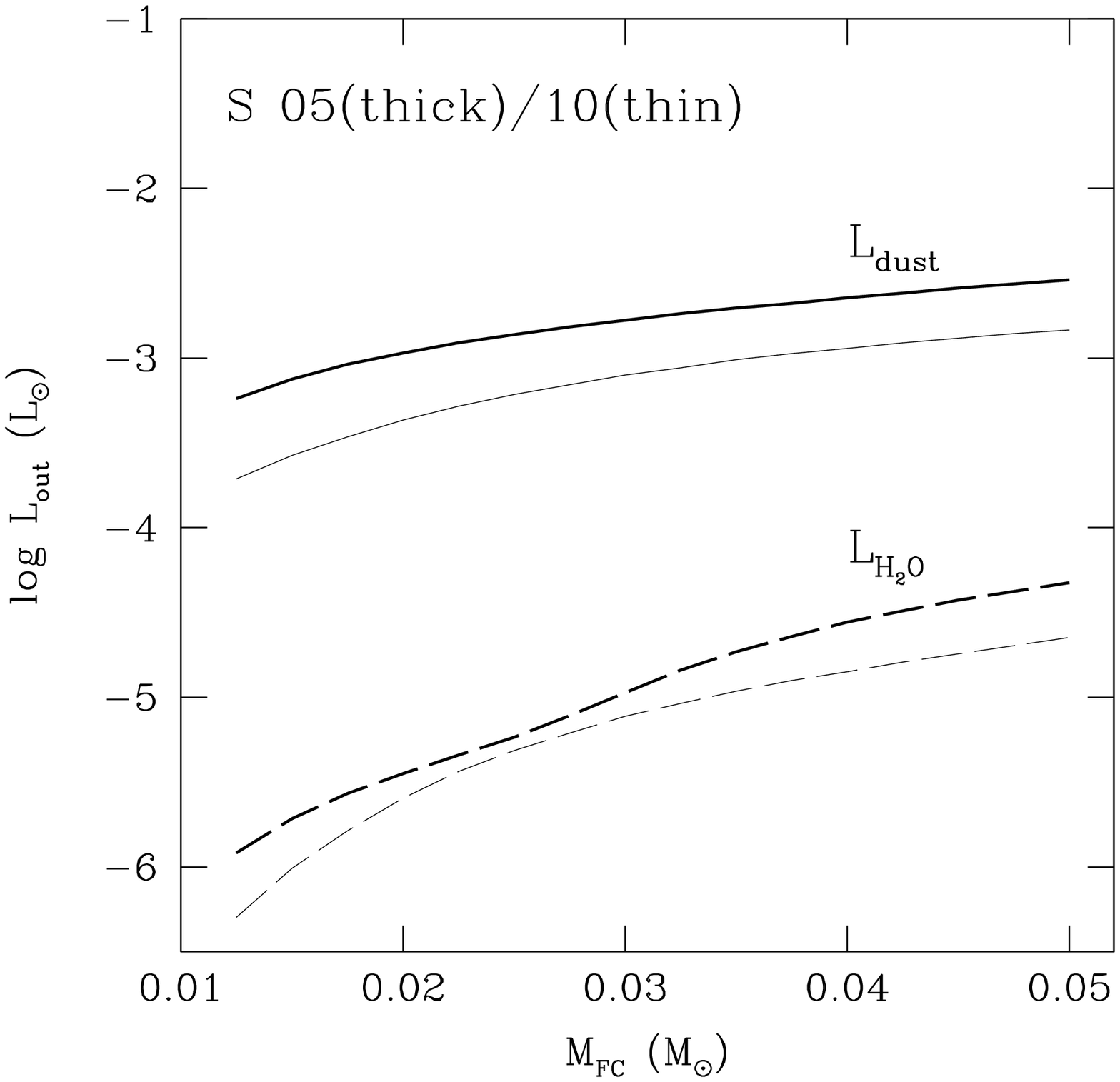}
  \end{center}
\caption{Evolution of luminosities in 
the dust continuum (solid) and H$_2$O lines (dashed) at the outer
boundary of the envelope, shown as a function of the first-core mass. 
The panels are for (a) LP05 (thick) and LP10 (thin) models, 
(b) S05 (thick) and S10 (thin) models. 
In panel (a), the luminosities in H$_2$O lines $L_{\rm H_2O}$ is multiplied
by $10^{5}$ for clarity.}
\label{fig:L_out}
\end{figure}

\begin{figure}
  \begin{center}
\FigureFile(80mm,80mm){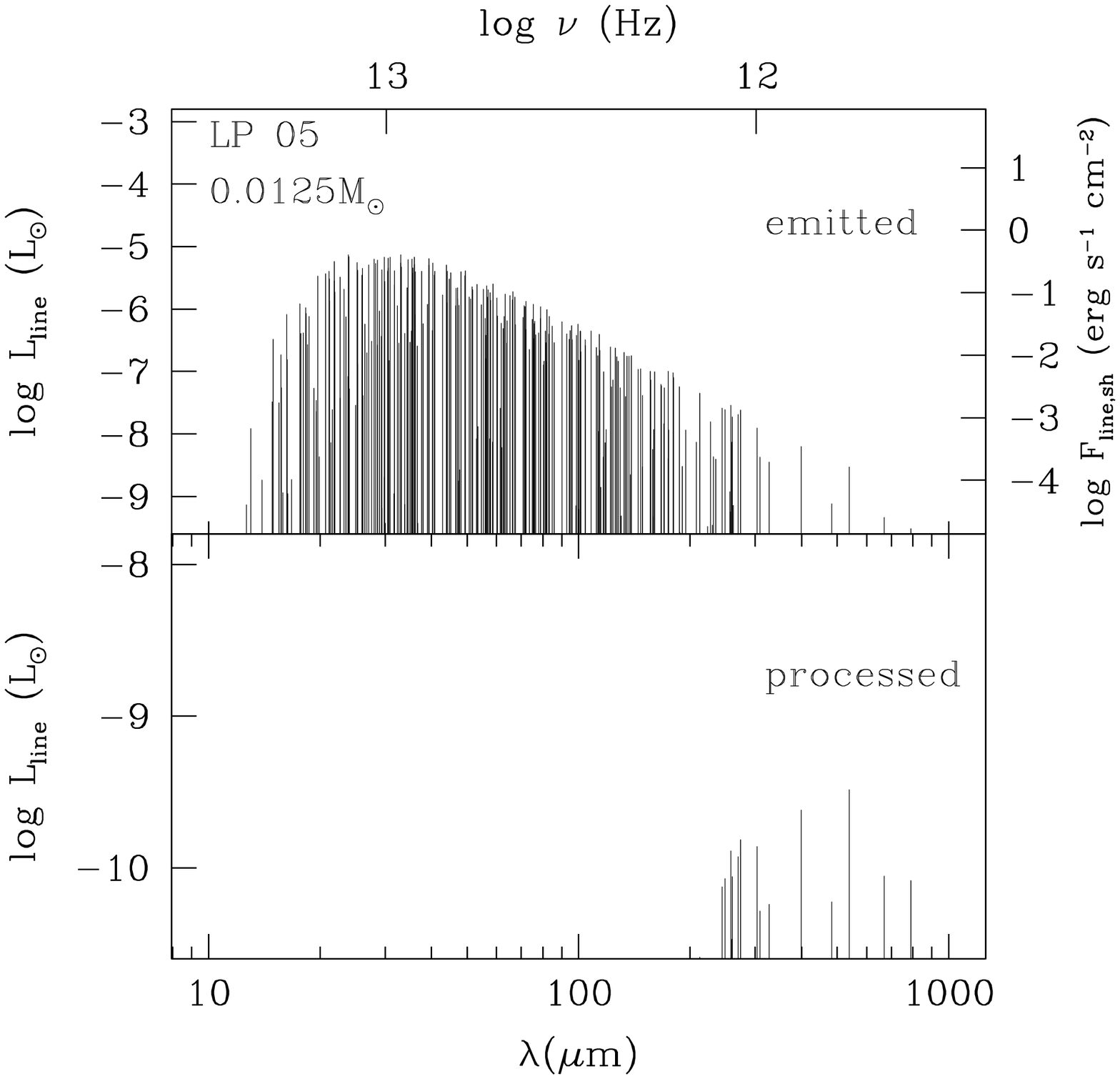}
\FigureFile(80mm,80mm){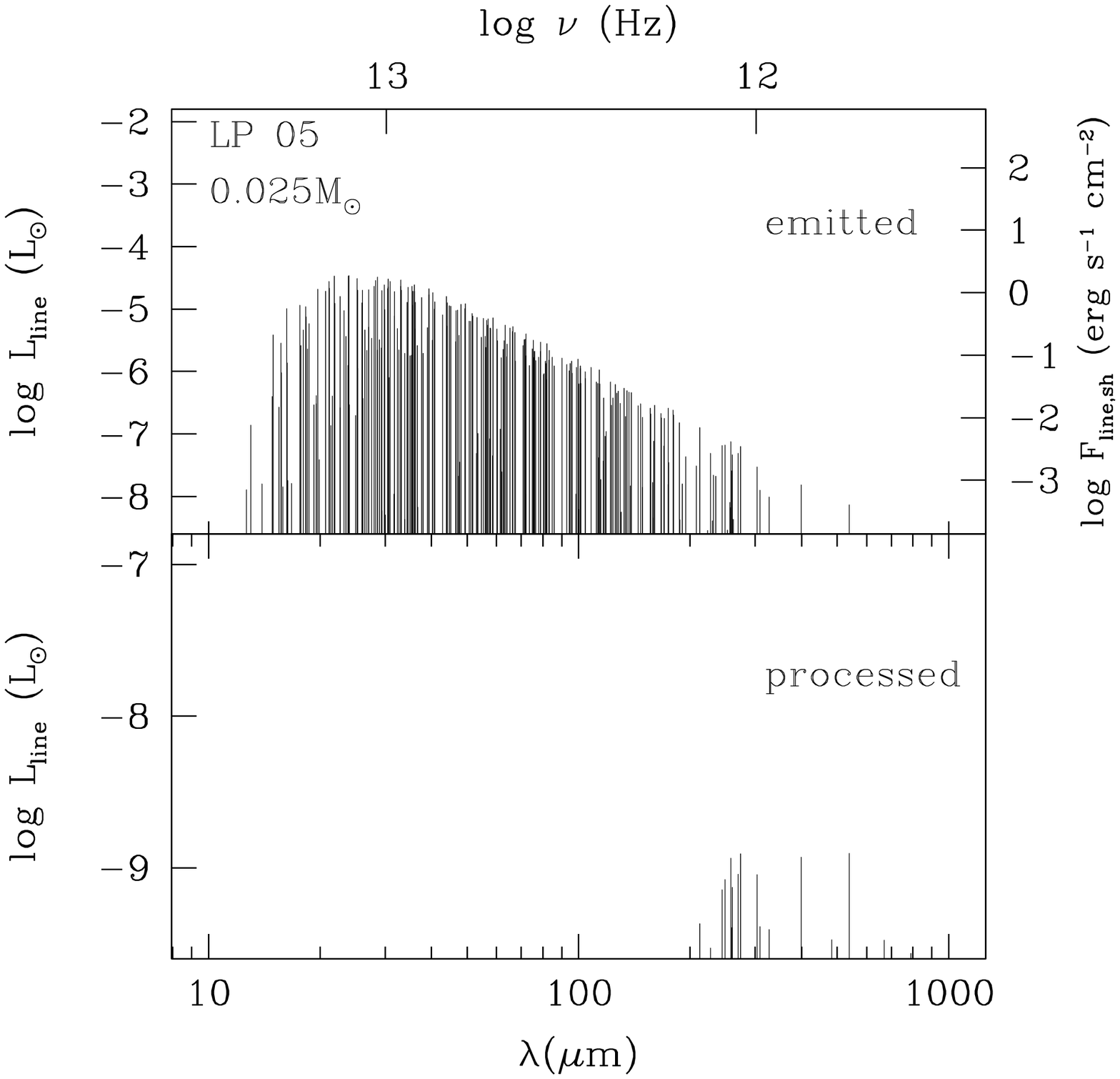}
\FigureFile(80mm,80mm){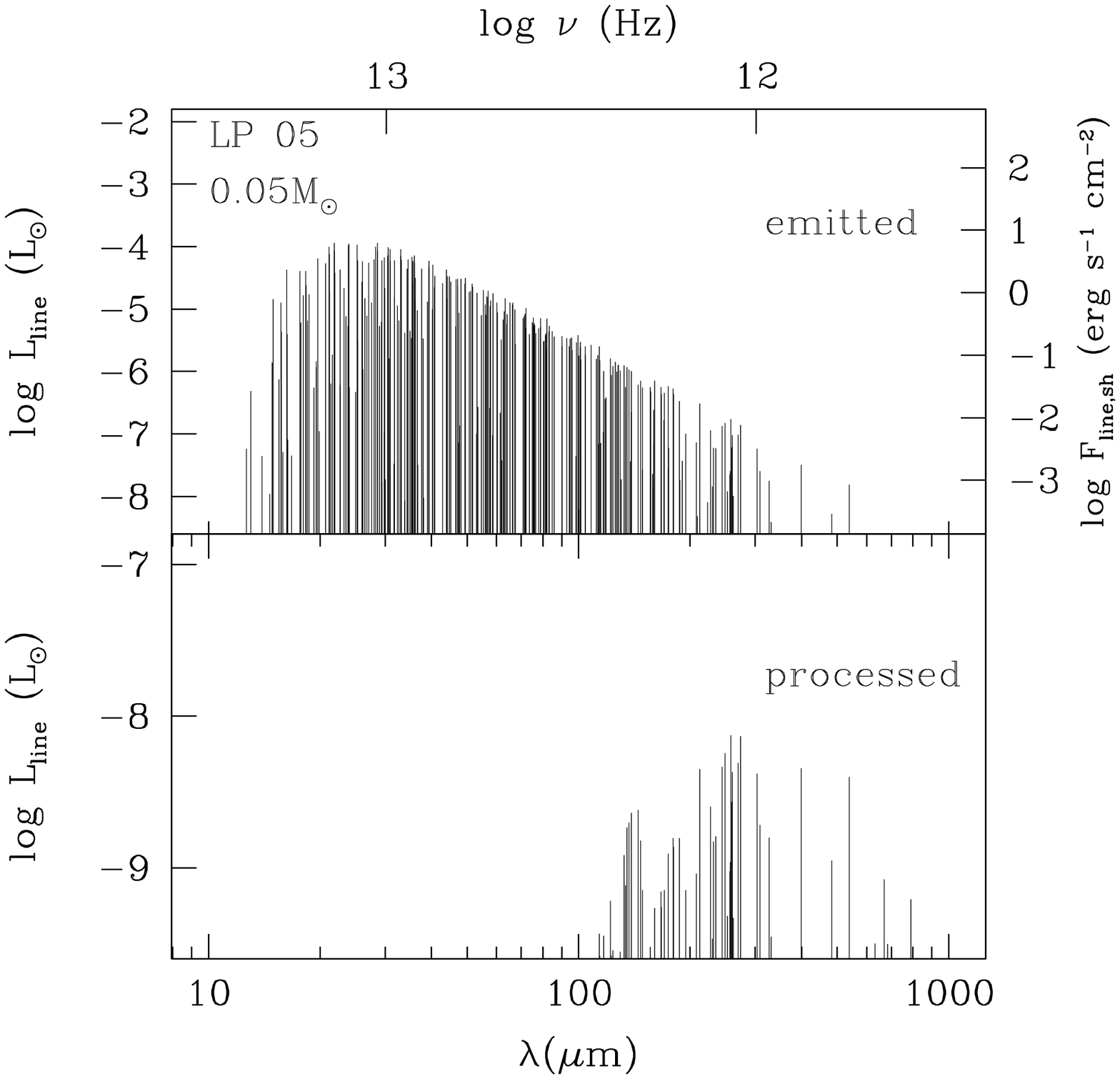}
  \end{center}
\caption{Emitted (upper halves) and processed (lower halves) 
luminosities of H$_2$O lines for the LP05 model.
Three epochs of (a) $0.0125 M_{\odot}$ (b) $0.025 M_{\odot}$ and 
(c) $0.05 M_{\odot}$ are shown.
In the upper halves, the values of the energy flux at the shock 
are also indicated.
}
\label{fig:water_LP05}
\end{figure}

\begin{figure}
  \begin{center}
\FigureFile(80mm,80mm){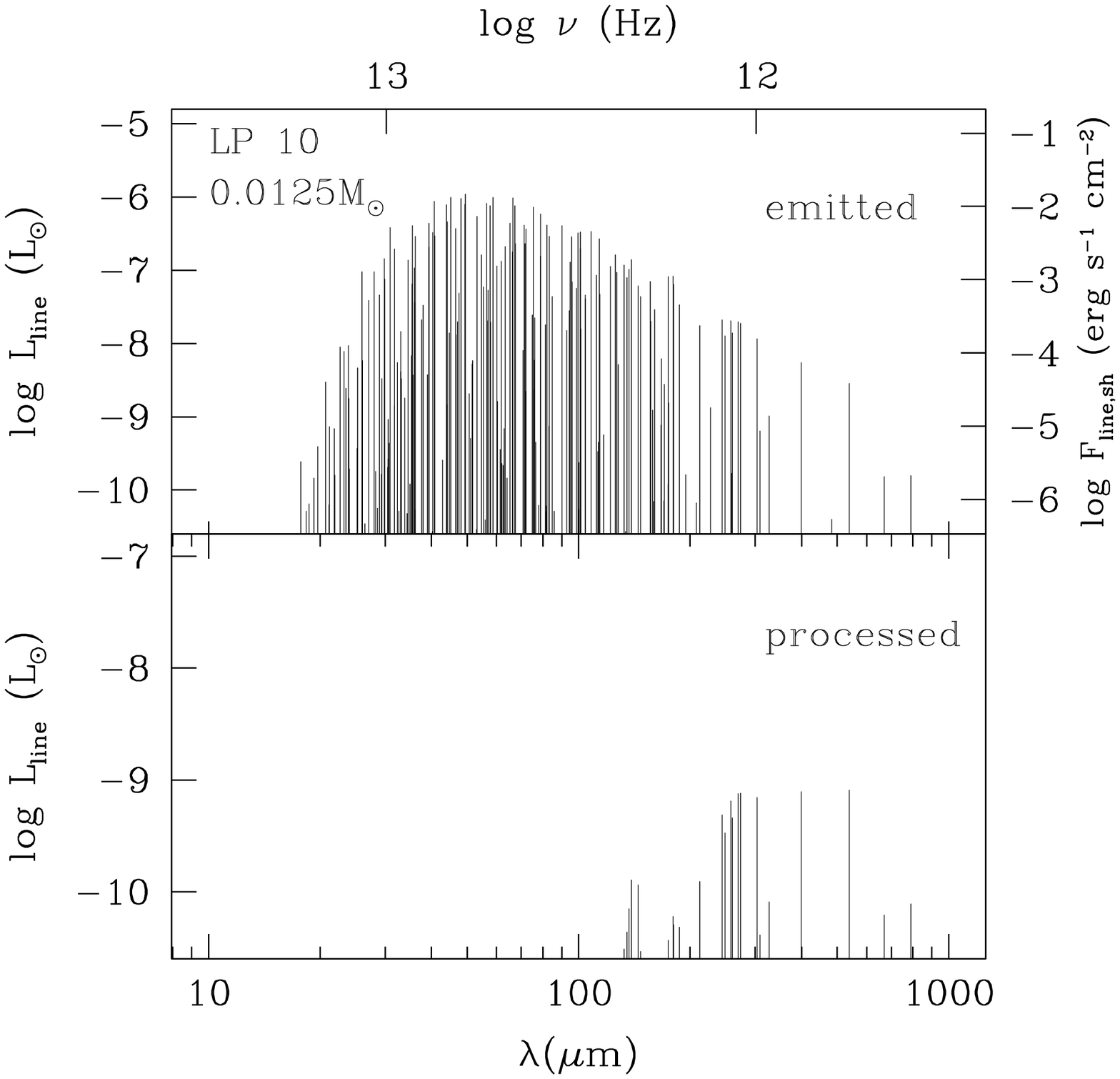}
\FigureFile(80mm,80mm){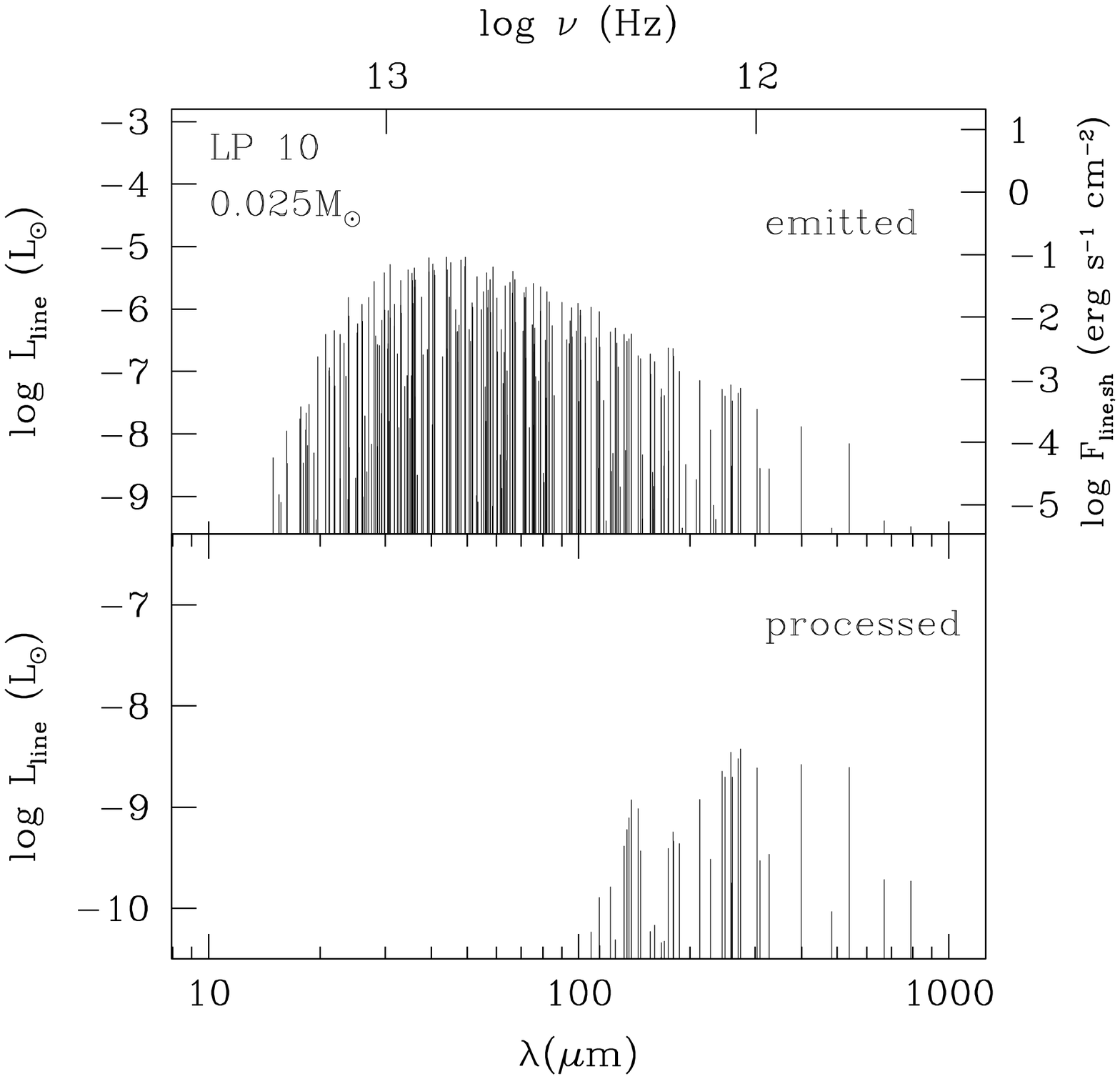}
\FigureFile(80mm,80mm){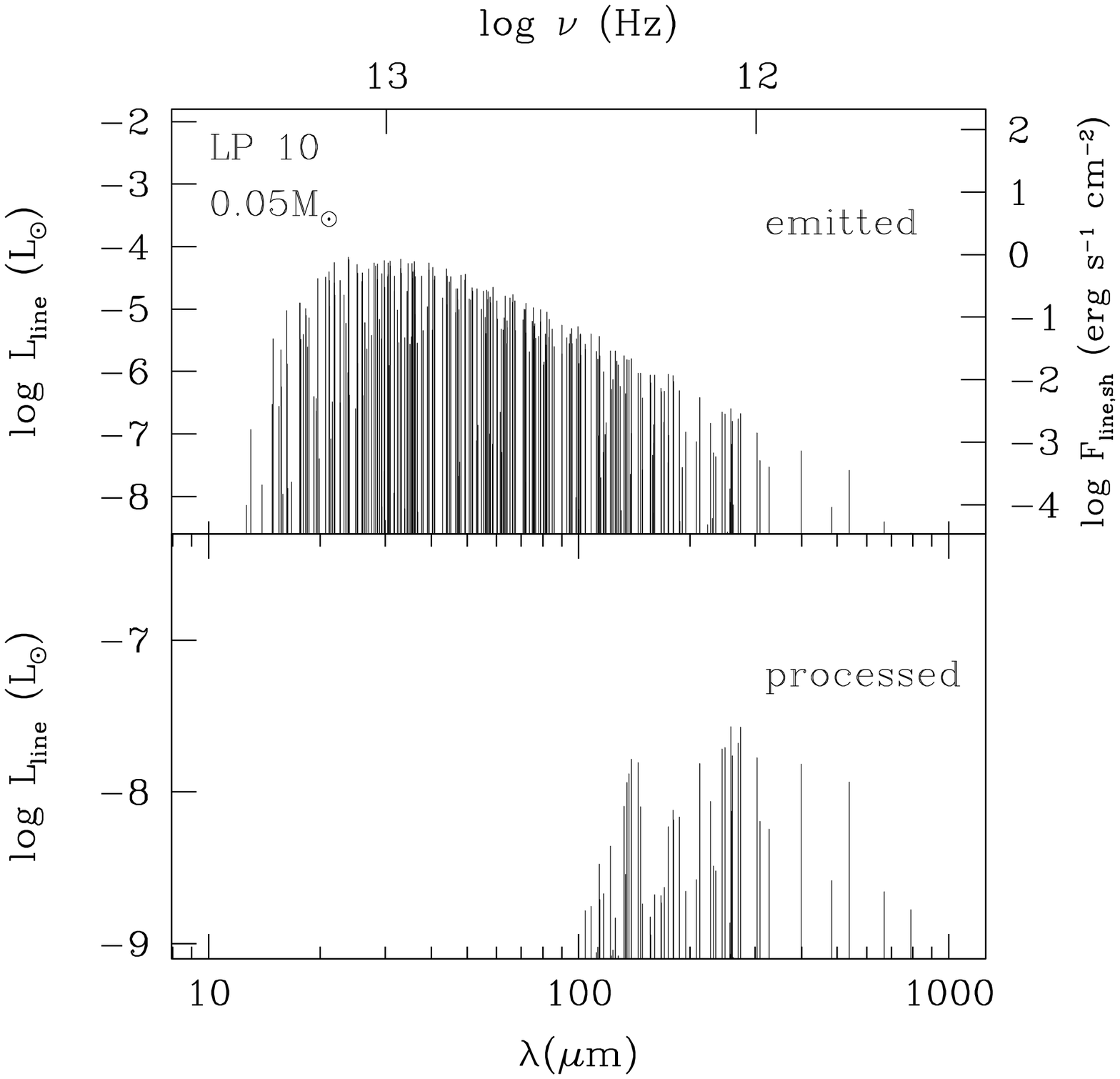}
  \end{center}
\caption{Same as Figure \ref{fig:water_LP05}, but for the LP10 model.}
 \label{fig:water_LP10}
\end{figure}

\begin{figure}
  \begin{center}
\FigureFile(80mm,80mm){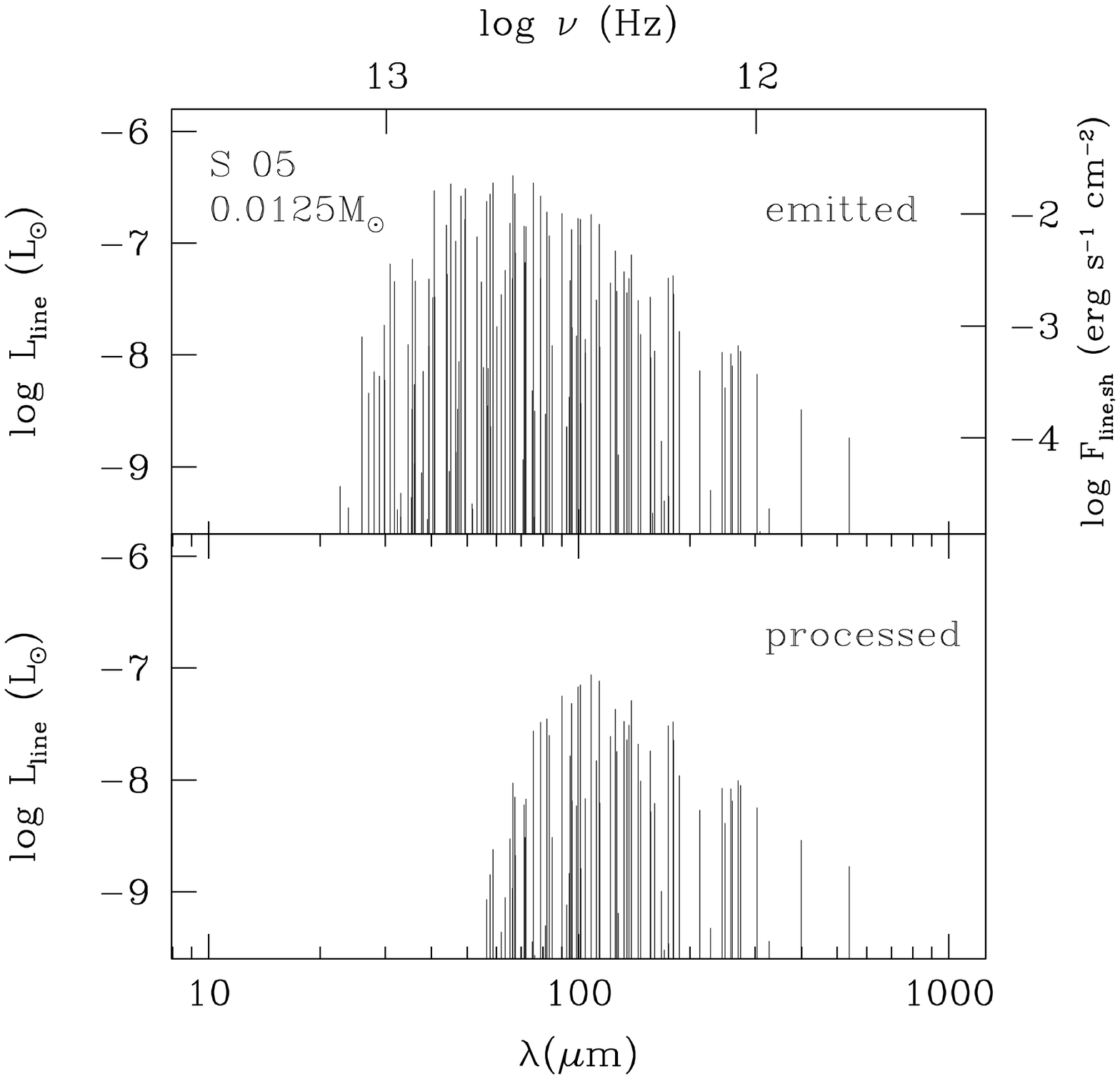}
\FigureFile(80mm,80mm){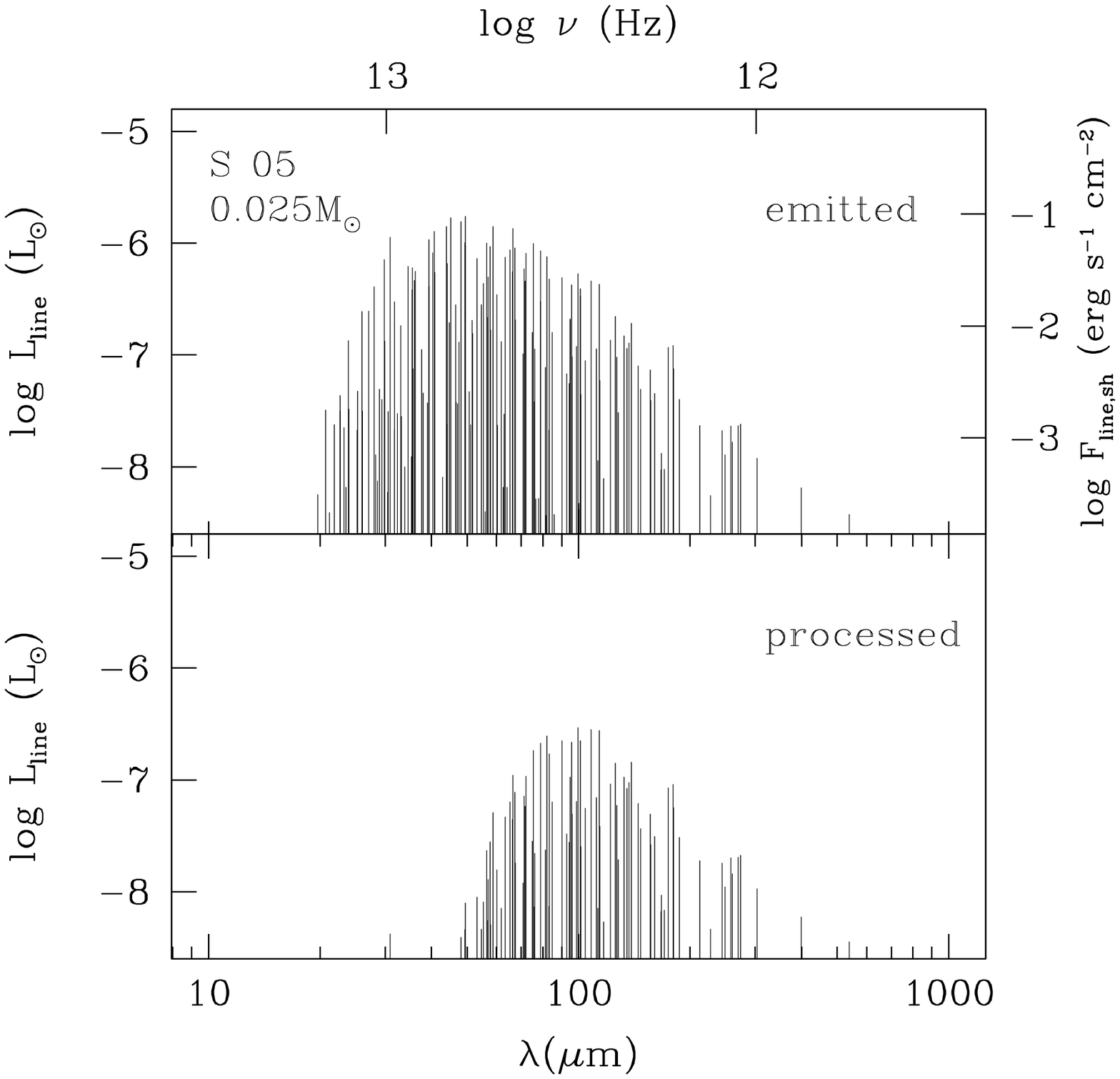}
\FigureFile(80mm,80mm){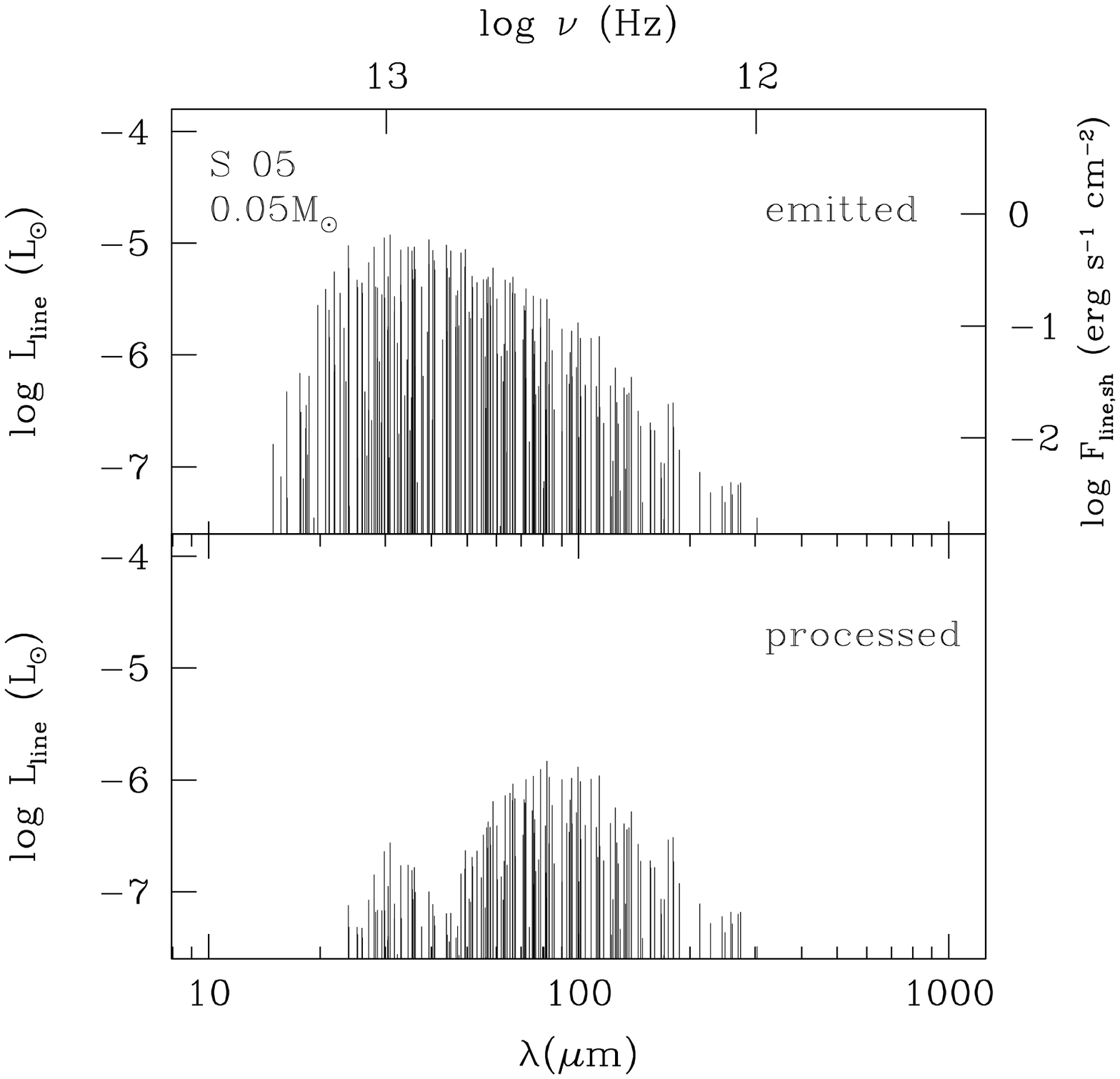}
  \end{center}
\caption{Same as Figure \ref{fig:water_LP05}, but for the S05 model.}
 \label{fig:water_S05}
\end{figure}

\begin{figure}
  \begin{center}
\FigureFile(80mm,80mm){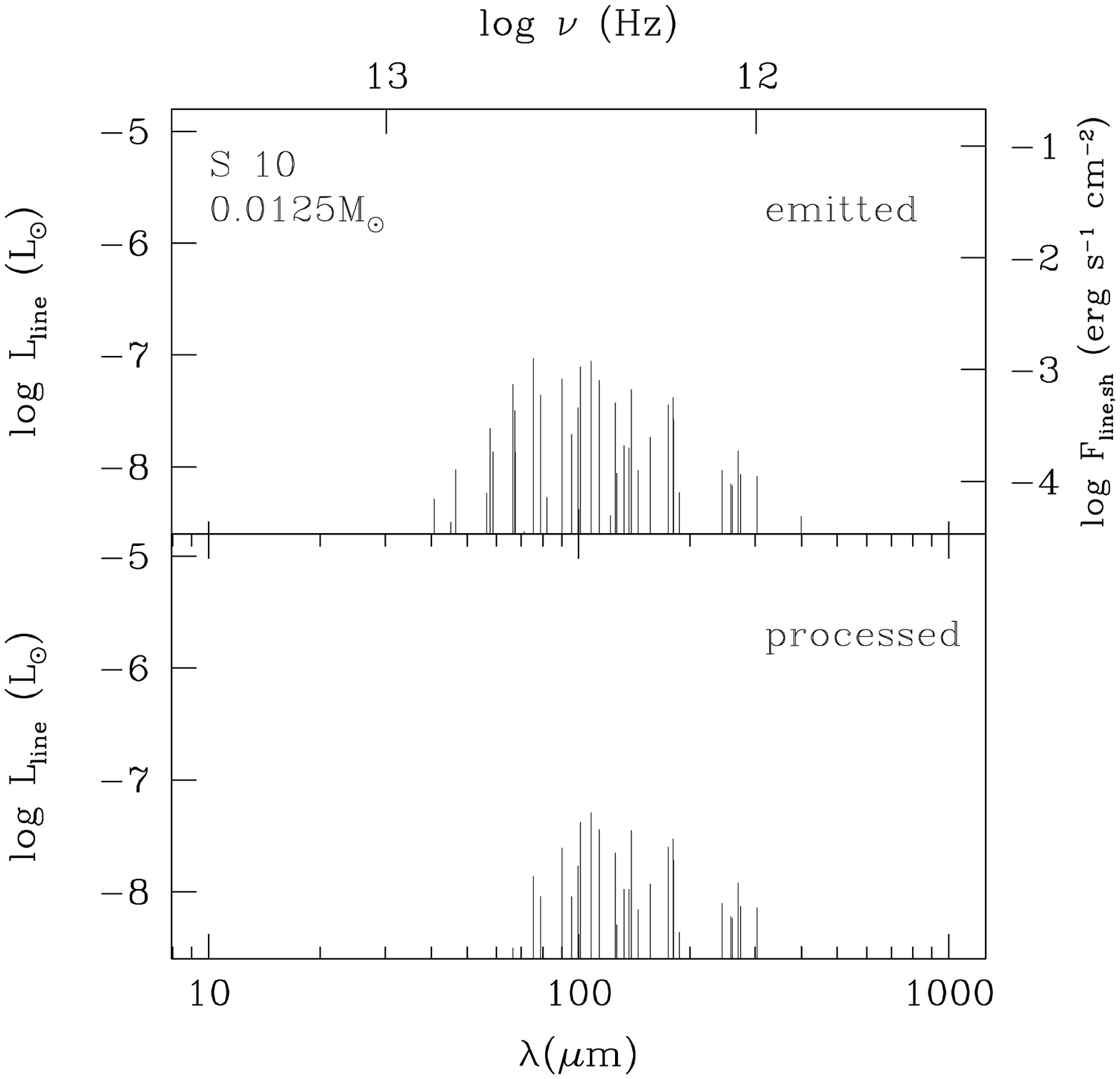}
\FigureFile(80mm,80mm){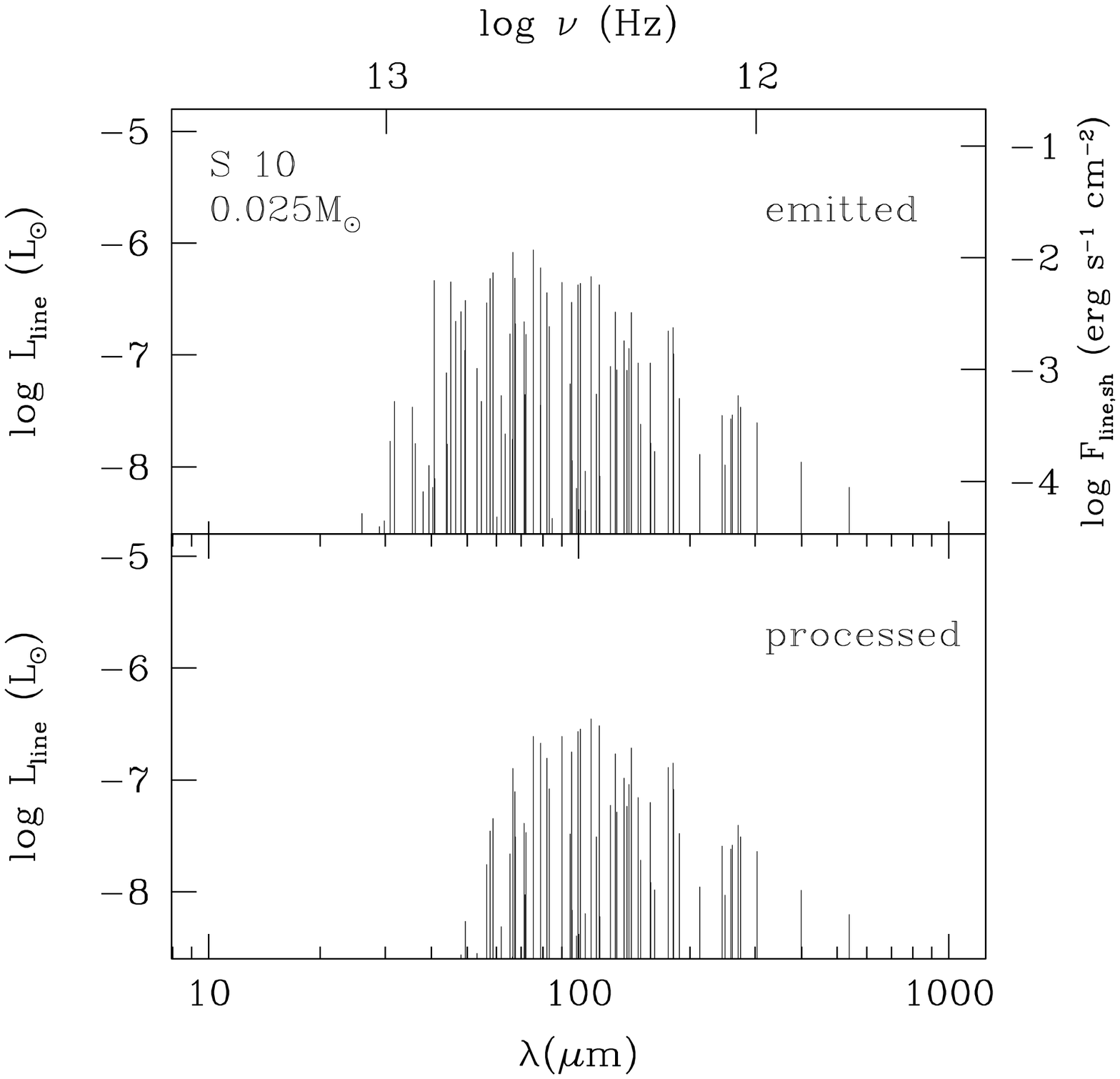}
\FigureFile(80mm,80mm){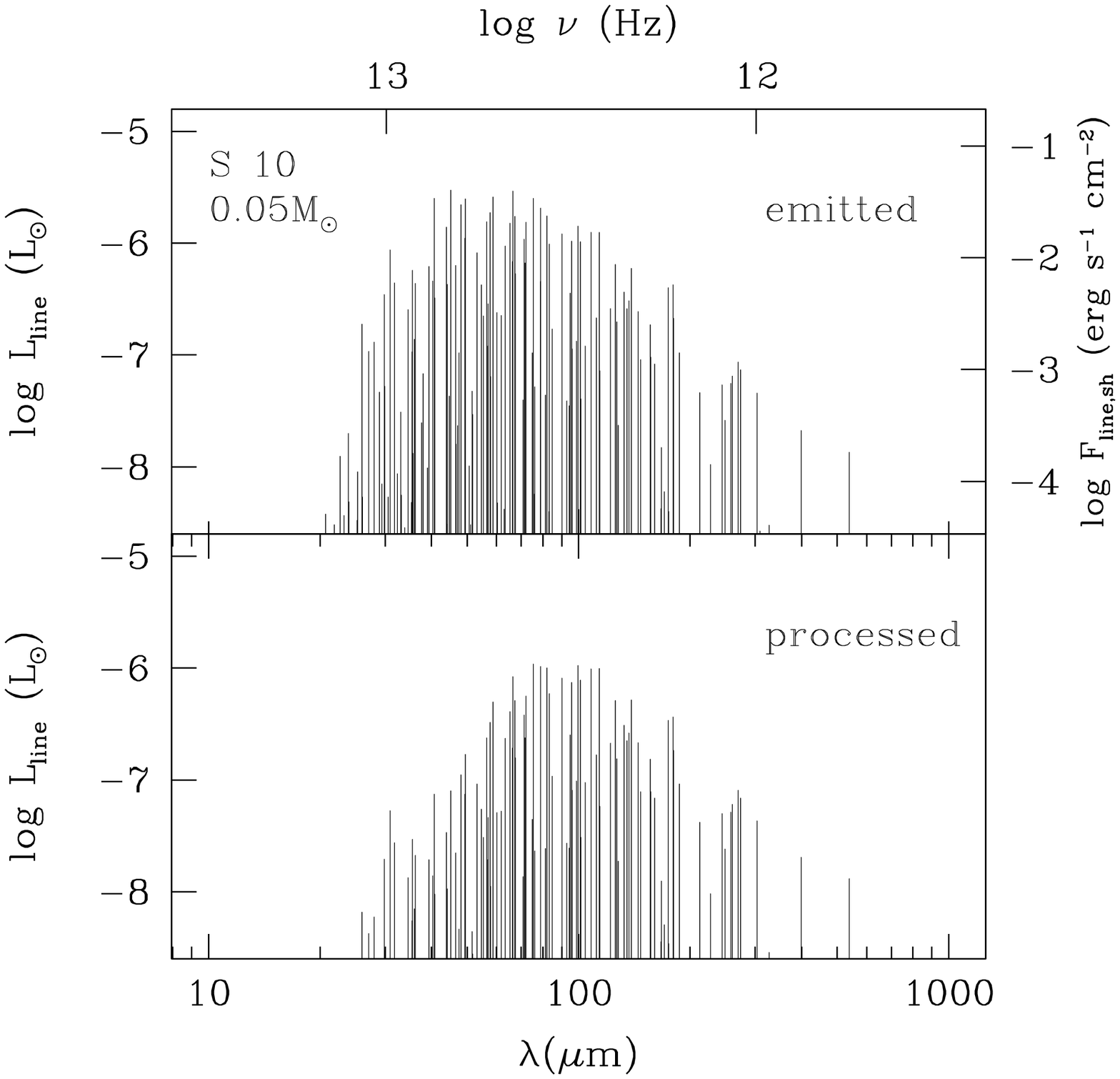}
  \end{center}
\caption{Same as Figure \ref{fig:water_LP05}, but for the S10 model.}
\label{fig:water_S10}
\end{figure}

\begin{figure}
  \begin{center}
\FigureFile(80mm,80mm){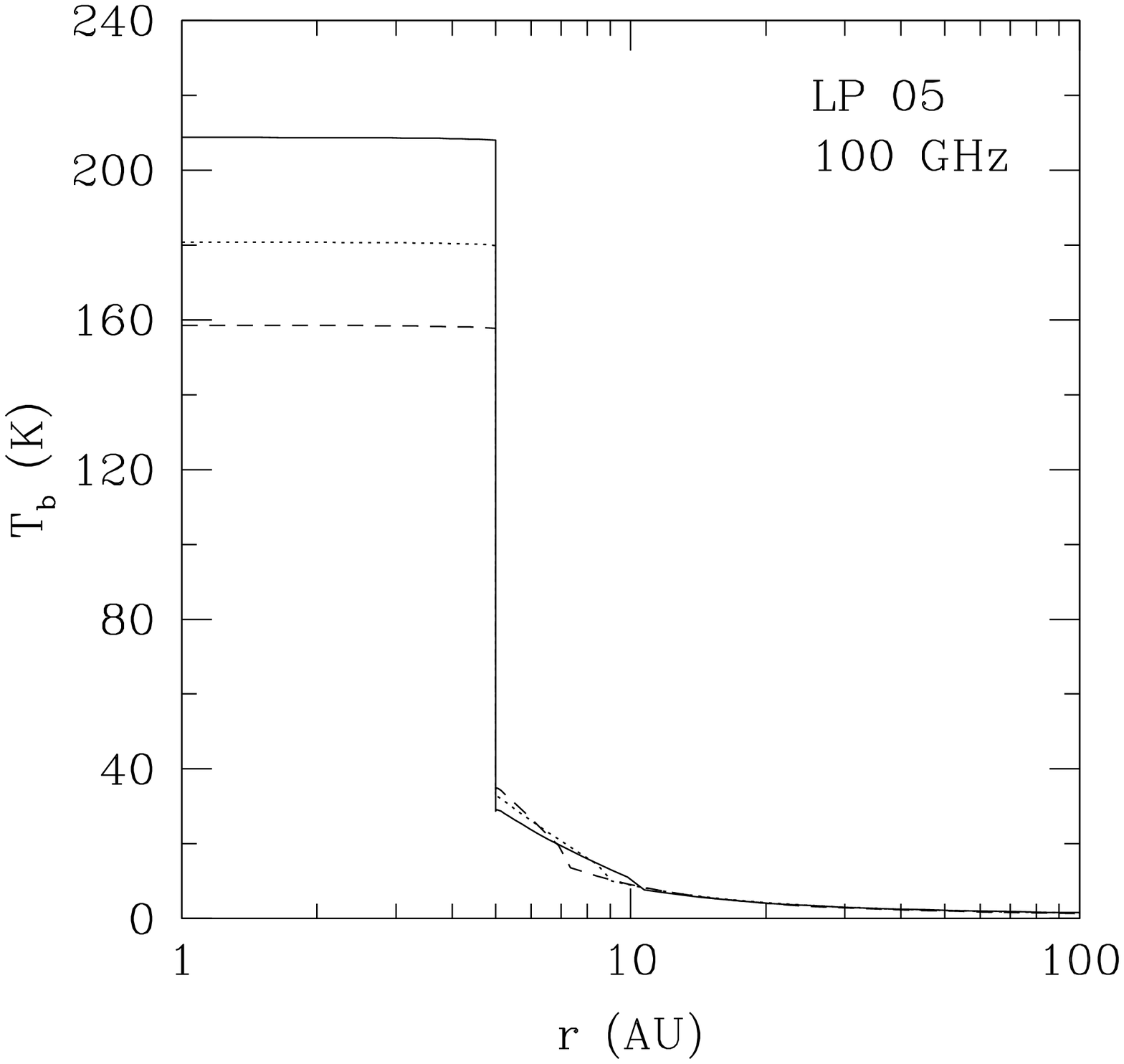}
\FigureFile(80mm,80mm){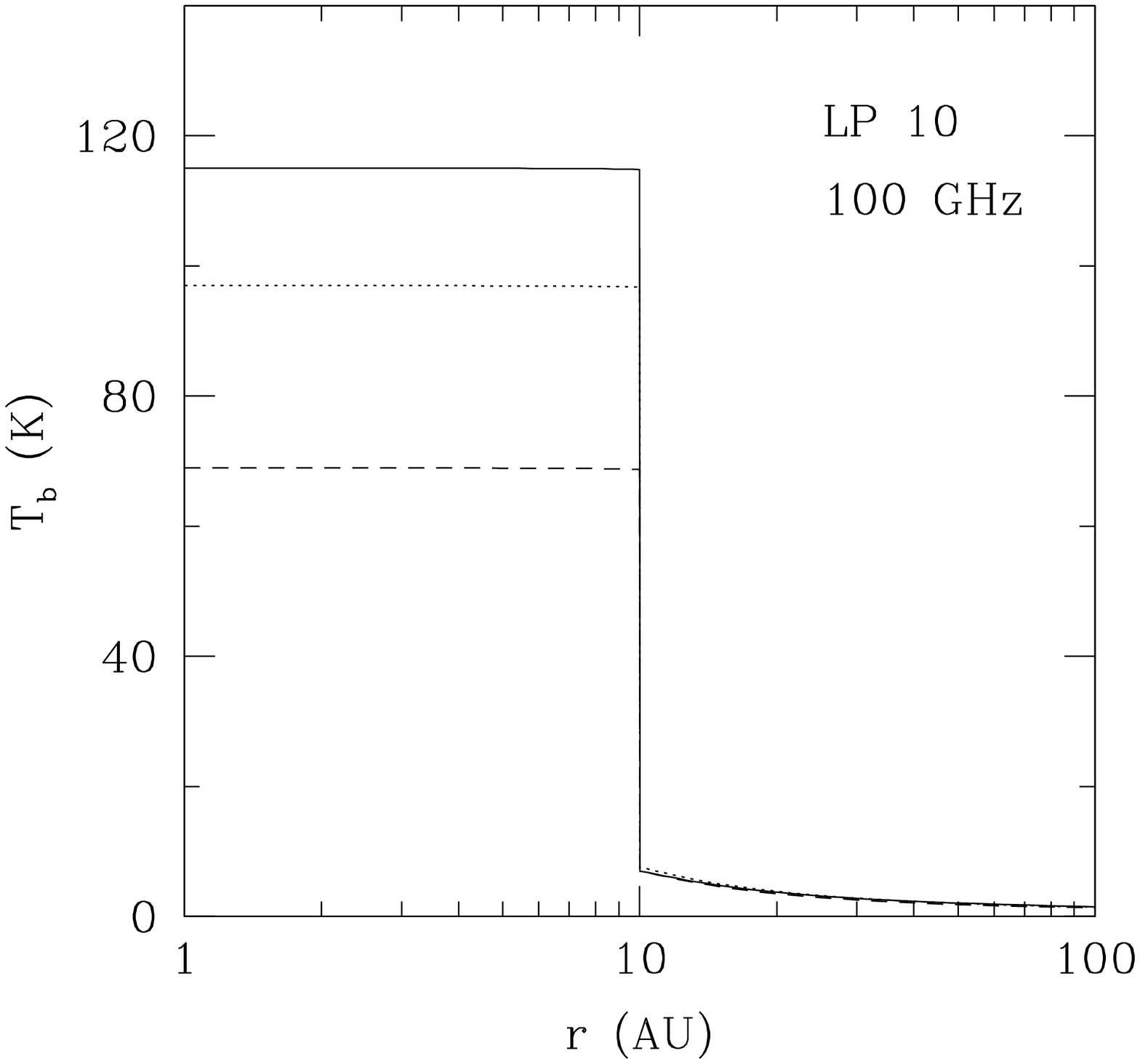}
\FigureFile(80mm,80mm){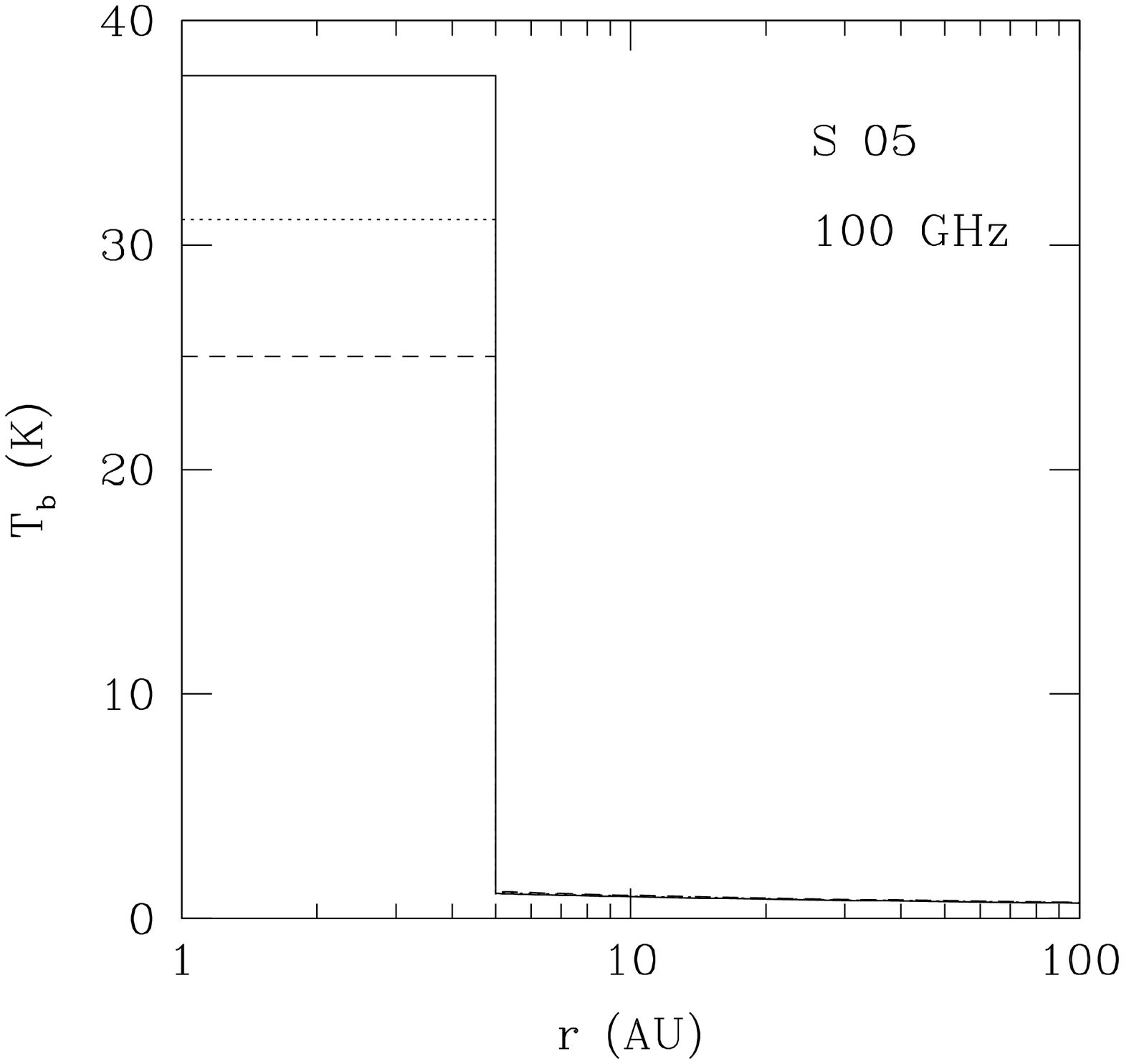}
\FigureFile(80mm,80mm){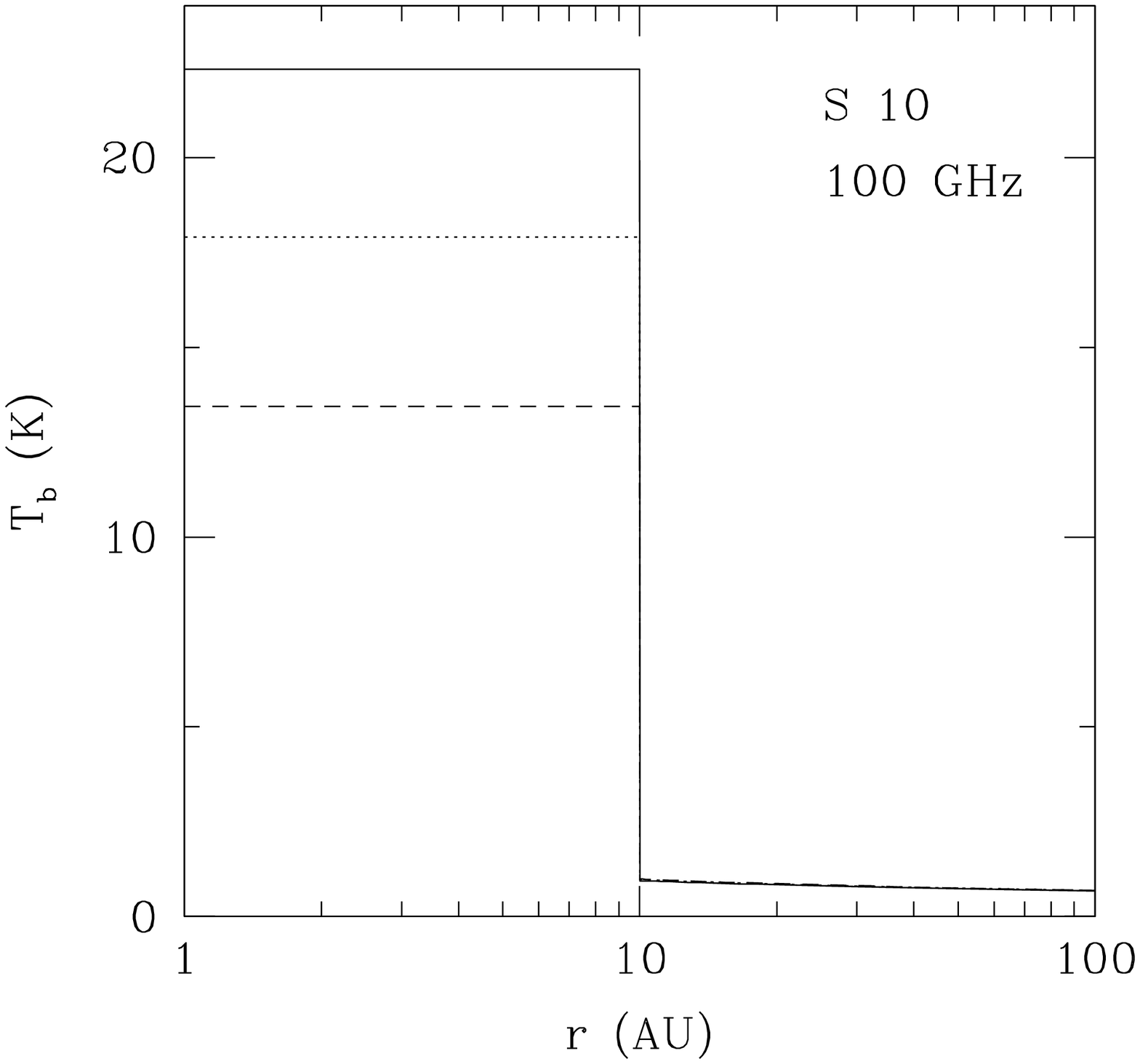}
  \end{center}
\caption{Distribution of the brightness temperature at 100GHz
as a function of the impact parameter of the line of sight. 
Panels (a)-(d) show LP05, LP10, S05, and S10 models, respectively.
In each panel, the distributions are shown in three different epochs 
$M_{\rm FC}=0.0125M_{\odot}$ (dashed), 
$0.025M_{\odot}$ (dotted), and
$0.05M_{\odot}$ (solid).} 
\label{fig:Tb.100G}
\end{figure}

\begin{figure}
  \begin{center}
\FigureFile(80mm,80mm){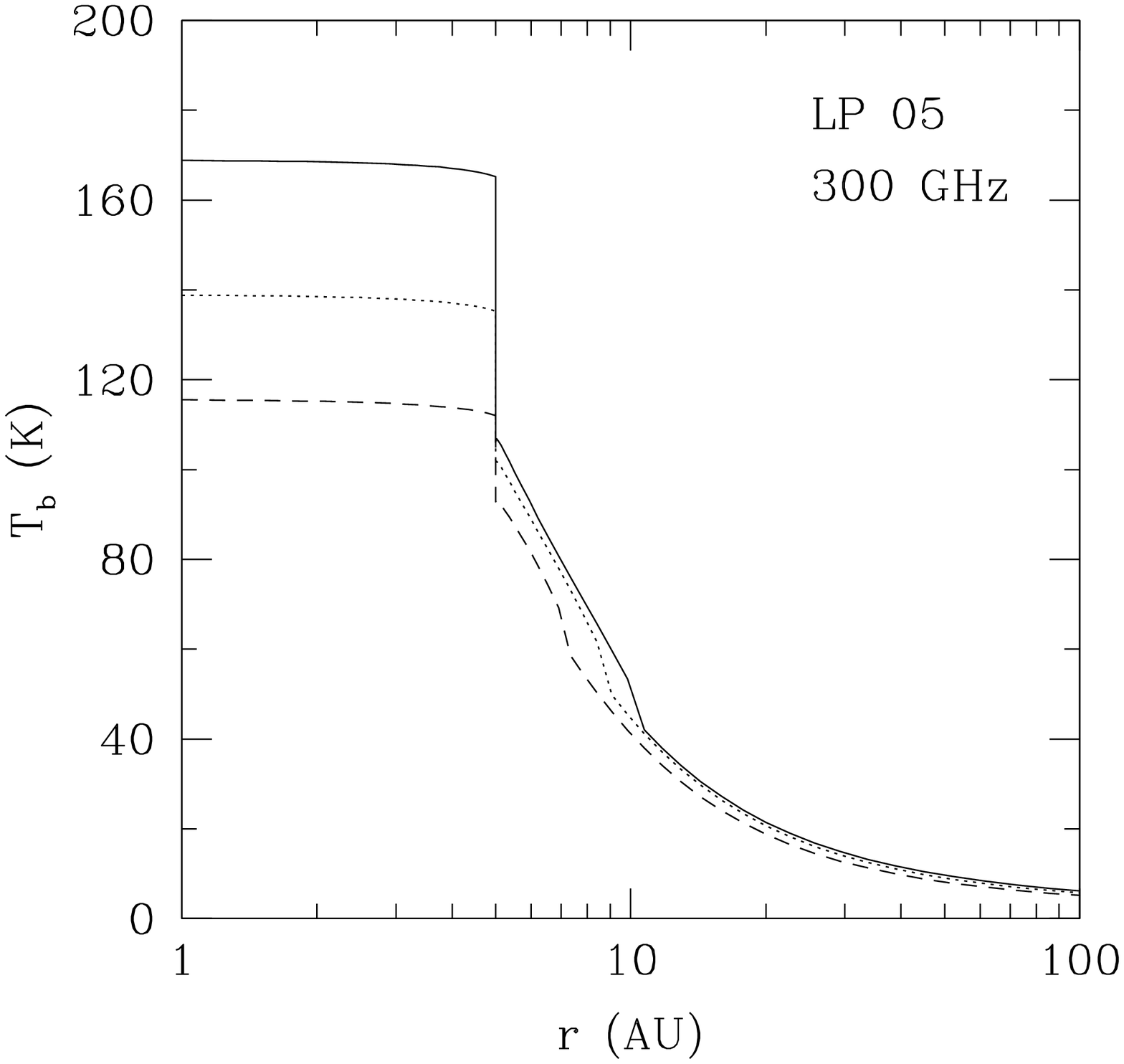}
\FigureFile(80mm,80mm){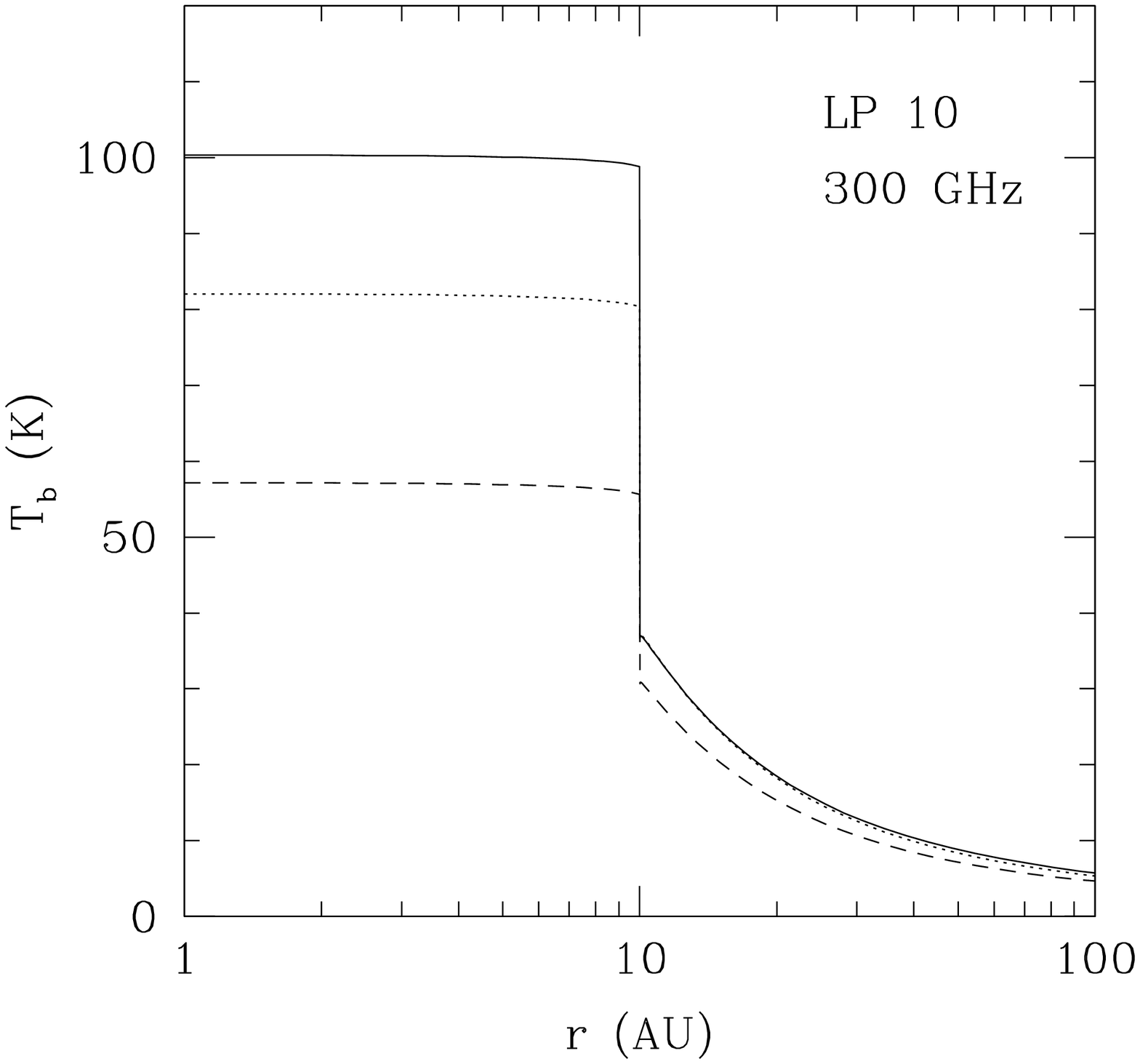}
\FigureFile(80mm,80mm){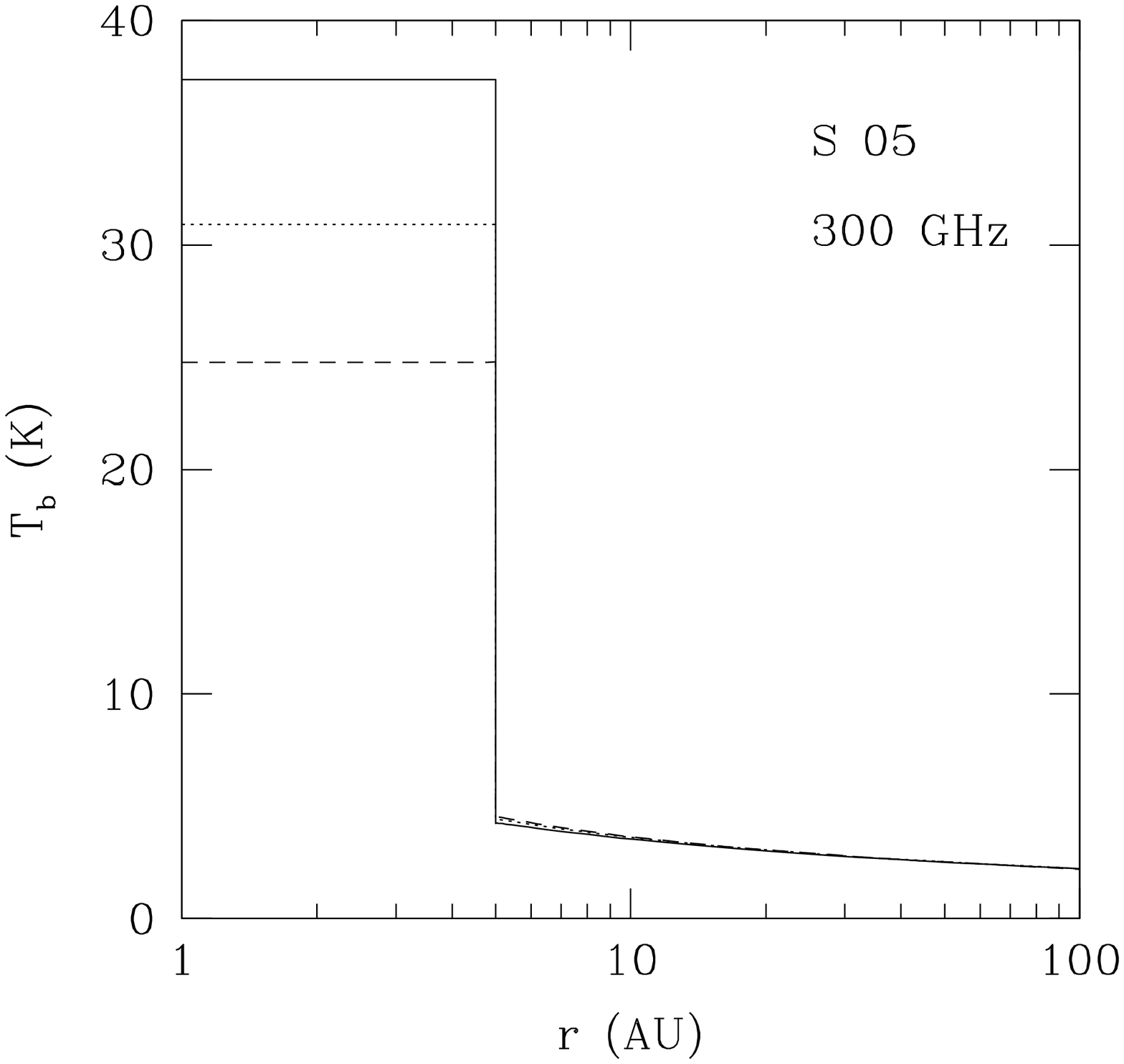}
\FigureFile(80mm,80mm){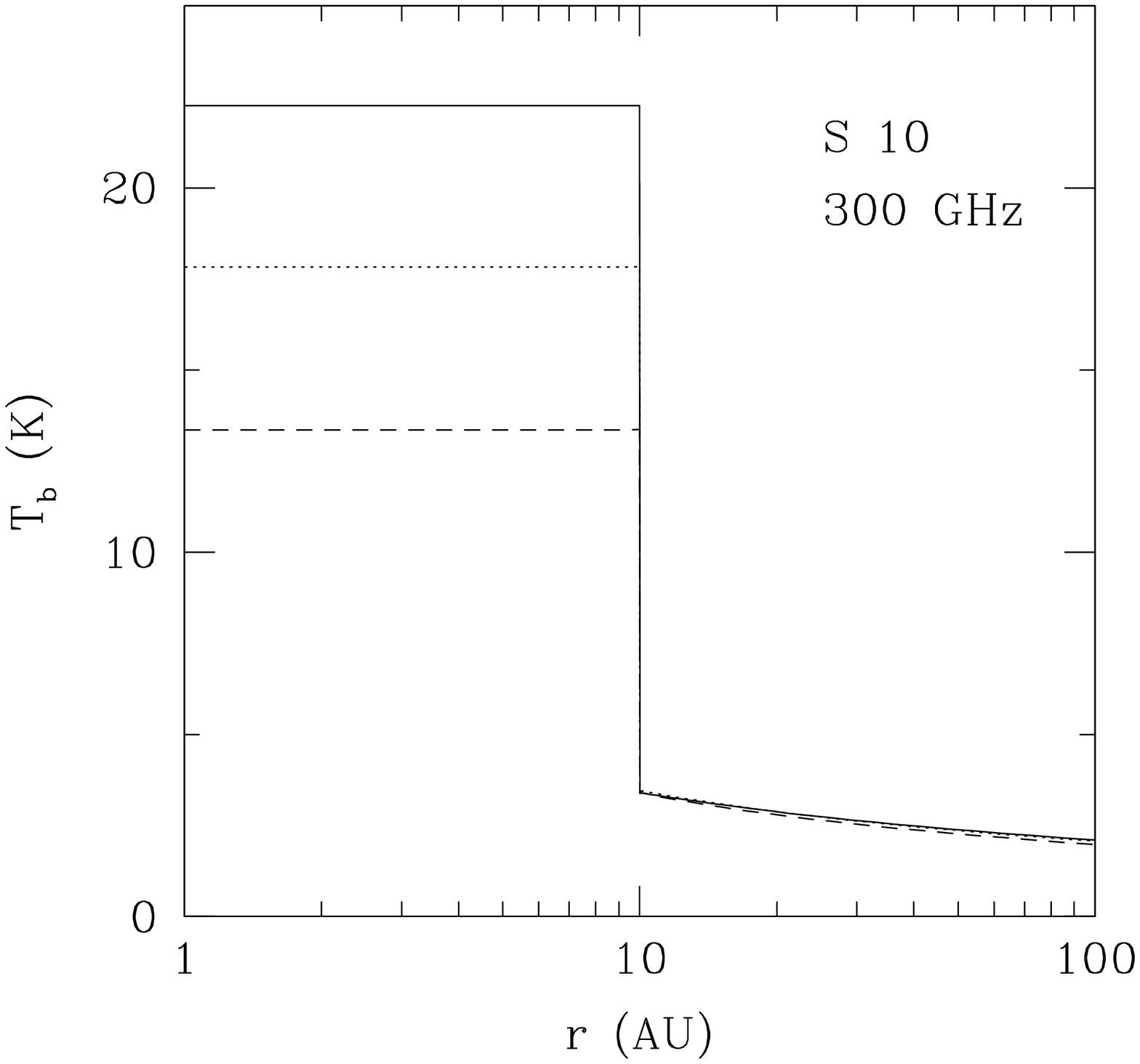}
  \end{center}
\caption{The same as Figure \ref{fig:Tb.100G} 
but for the observing frequency at 300GHz.}
\label{fig:Tb.300G}
\end{figure}

\begin{figure}
  \begin{center}
\FigureFile(80mm,80mm){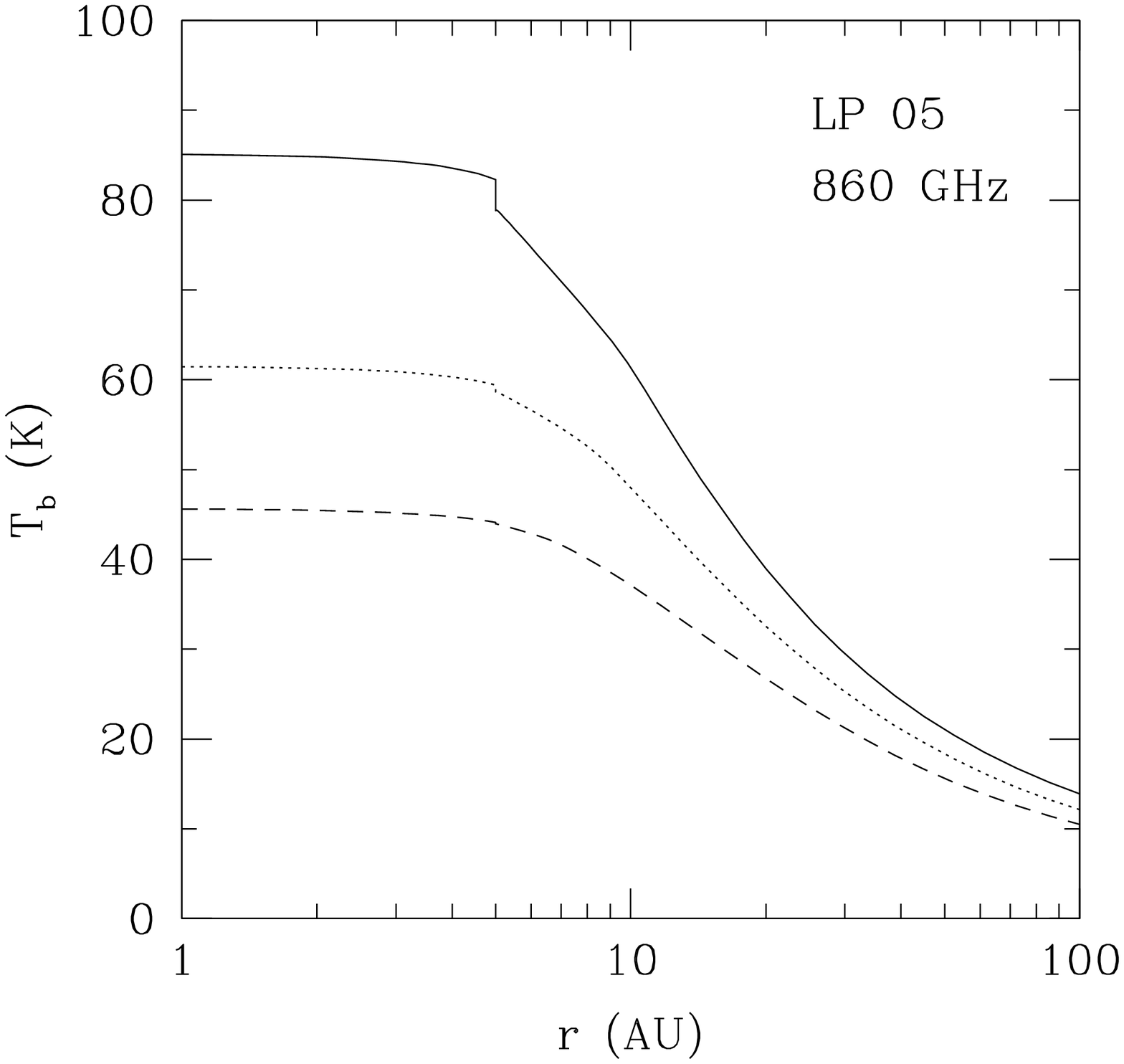}
\FigureFile(80mm,80mm){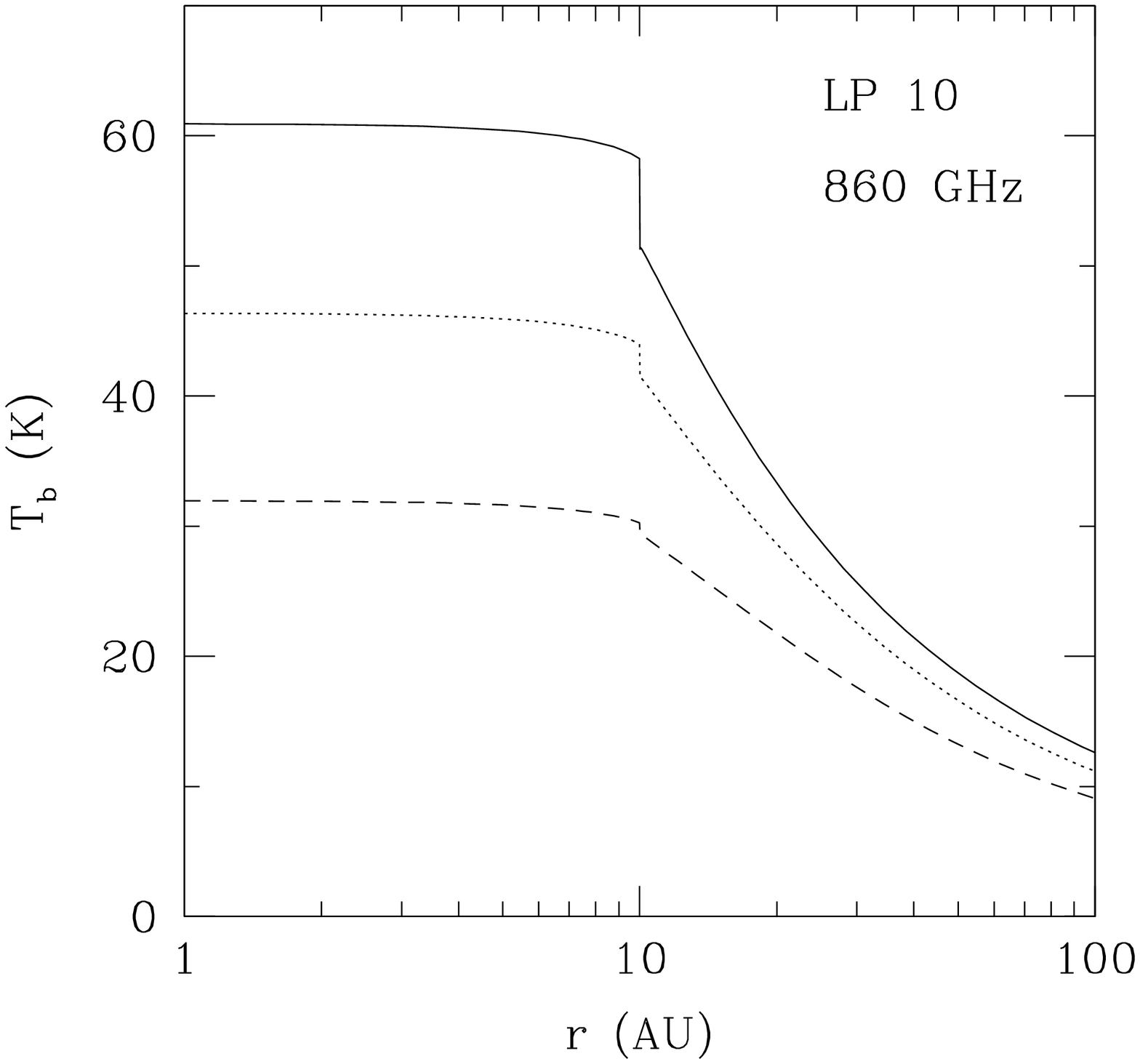}
\FigureFile(80mm,80mm){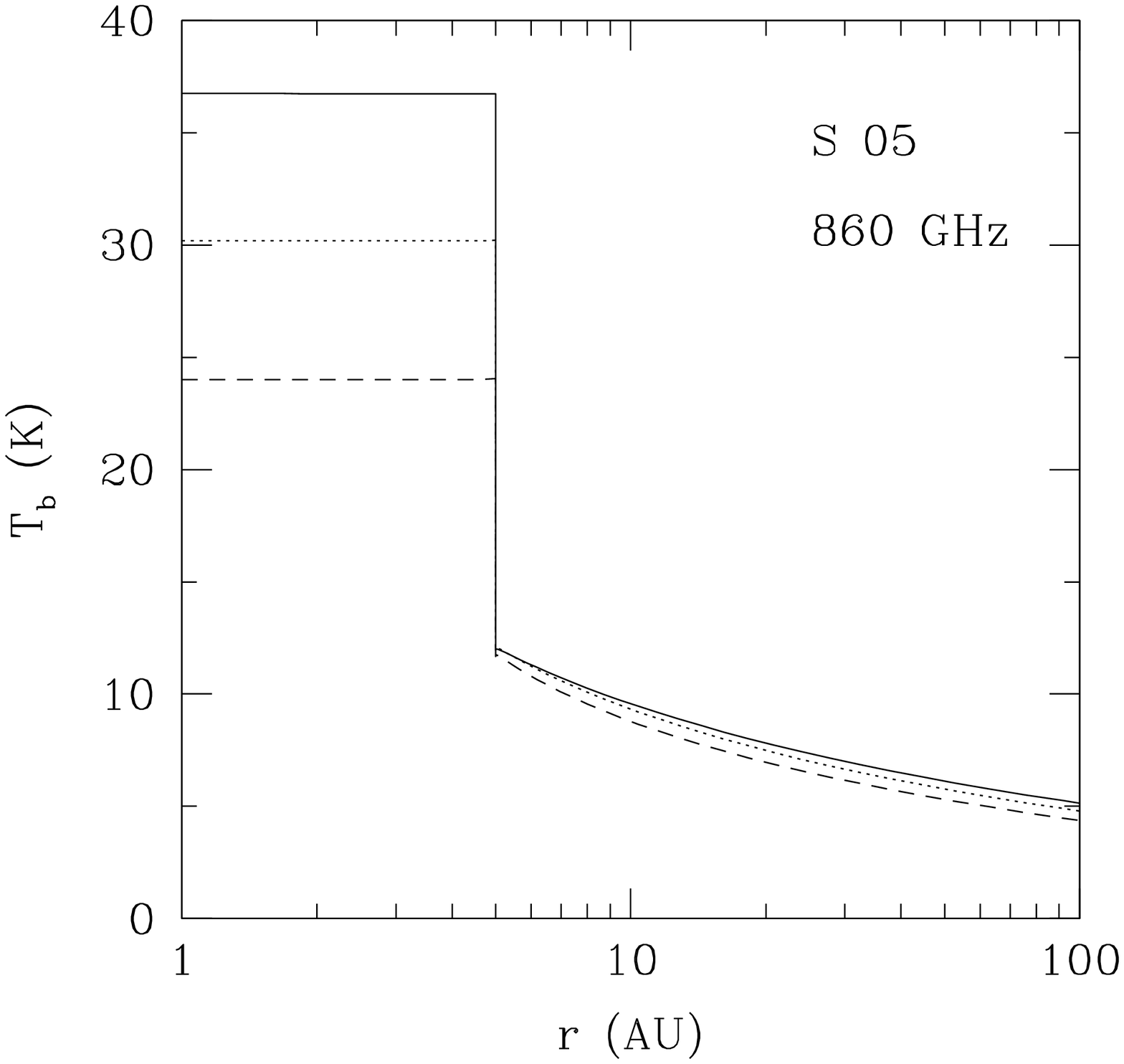}
\FigureFile(80mm,80mm){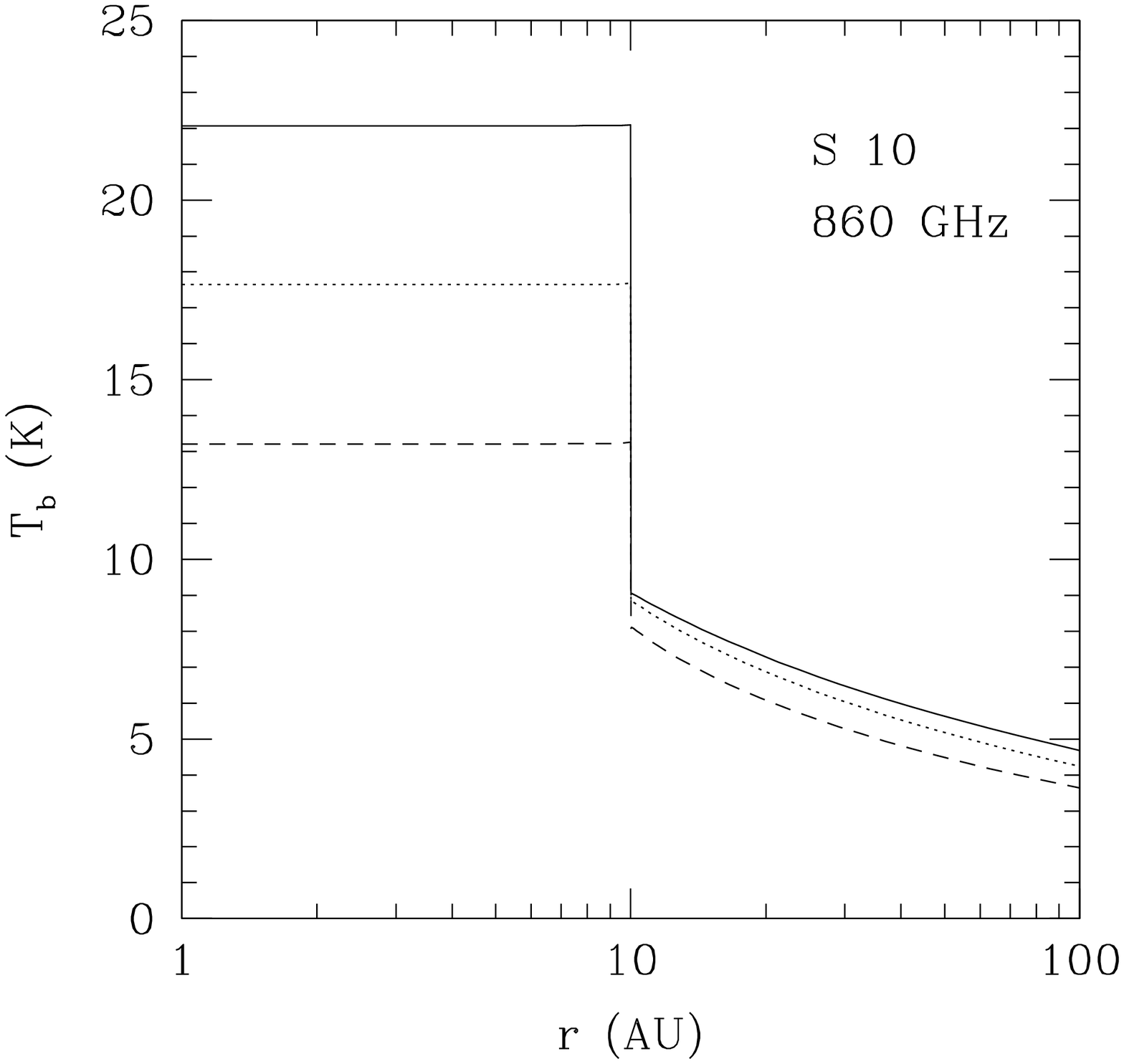}
  \end{center}
\caption{The same as Figure \ref{fig:Tb.100G} 
but for the observing frequency at 860GHz.}
\label{fig:Tb.860G}
\end{figure}

\begin{figure}
  \begin{center}
\FigureFile(80mm,80mm){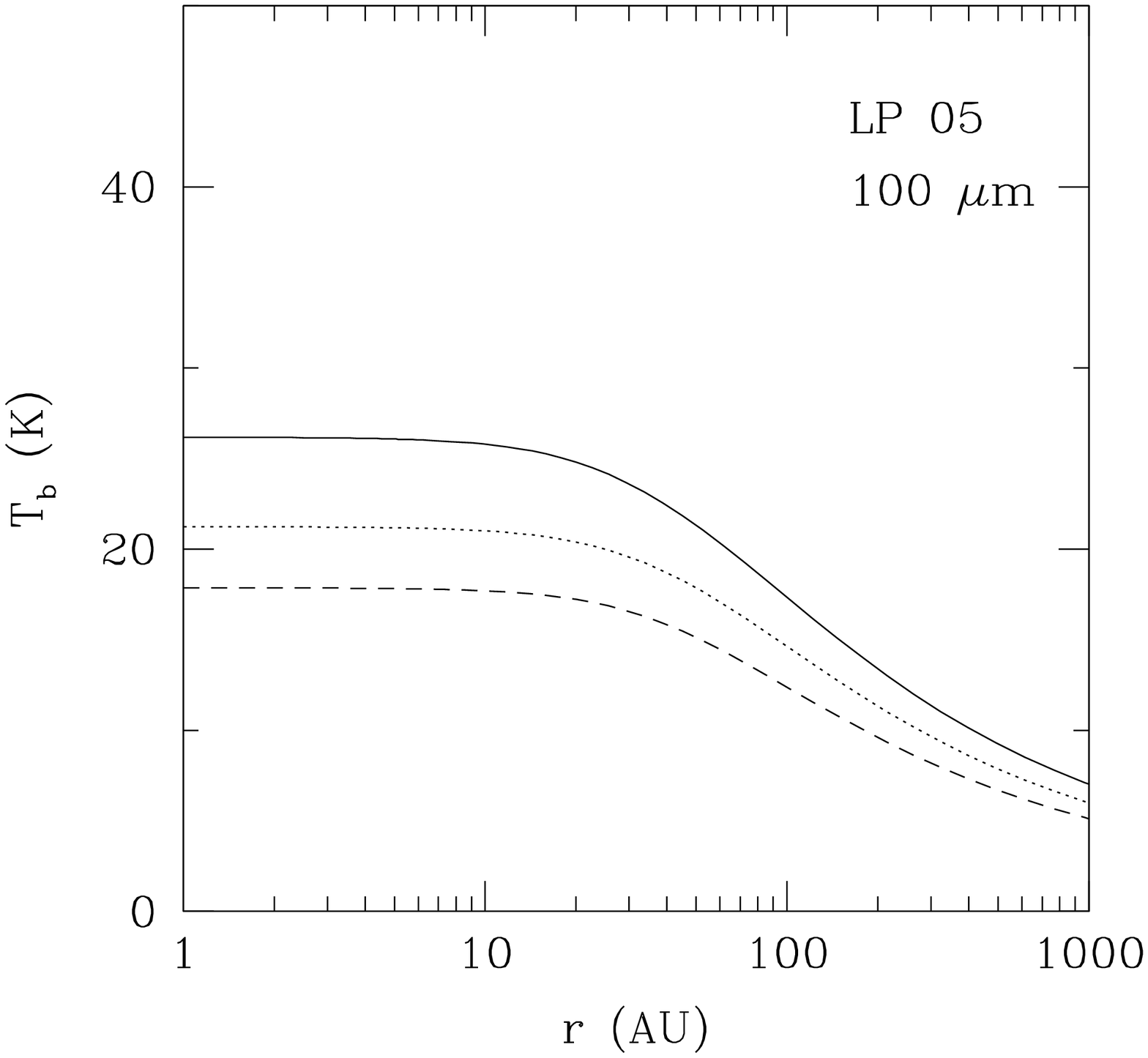}
\FigureFile(80mm,80mm){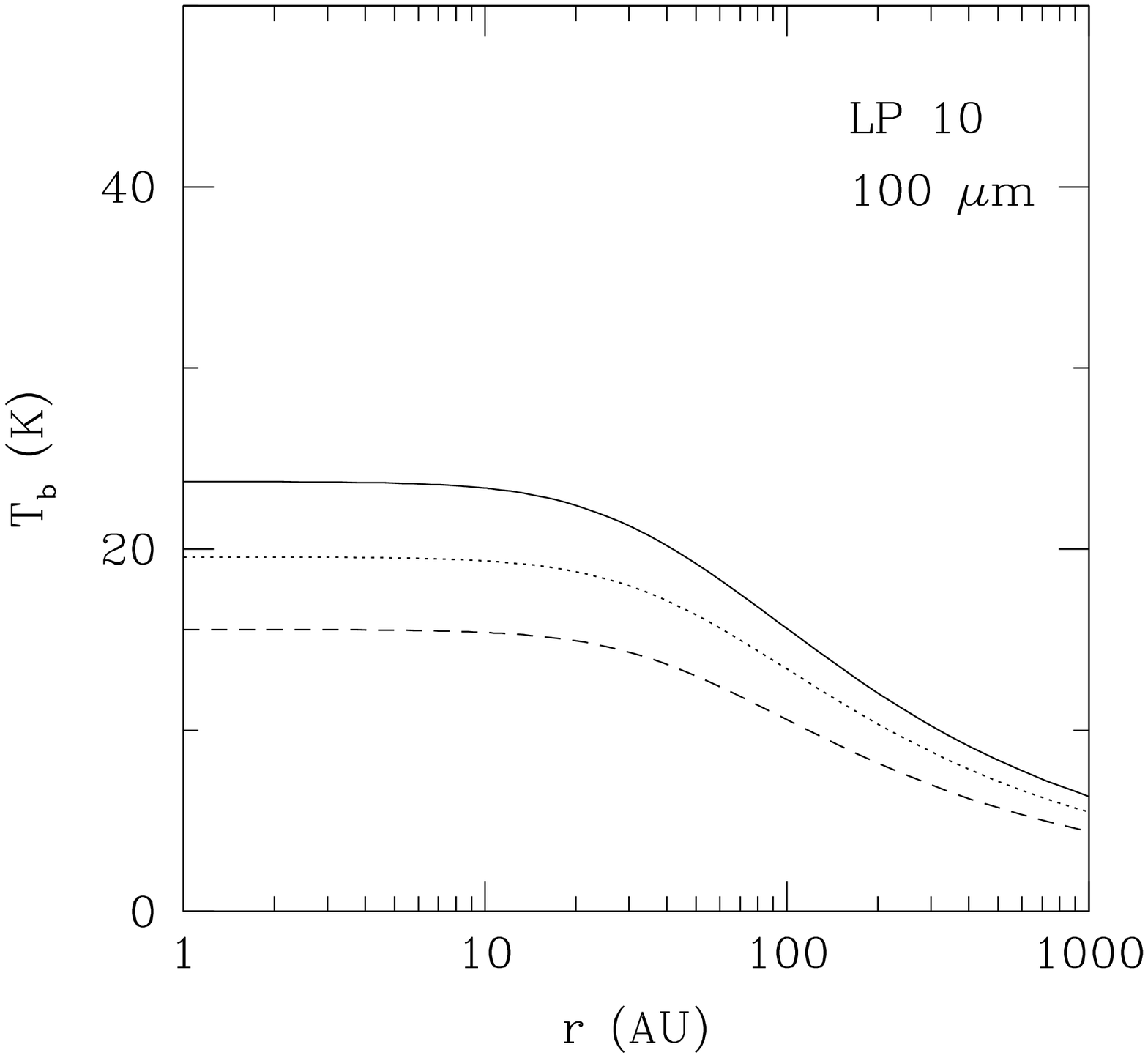}
\FigureFile(80mm,80mm){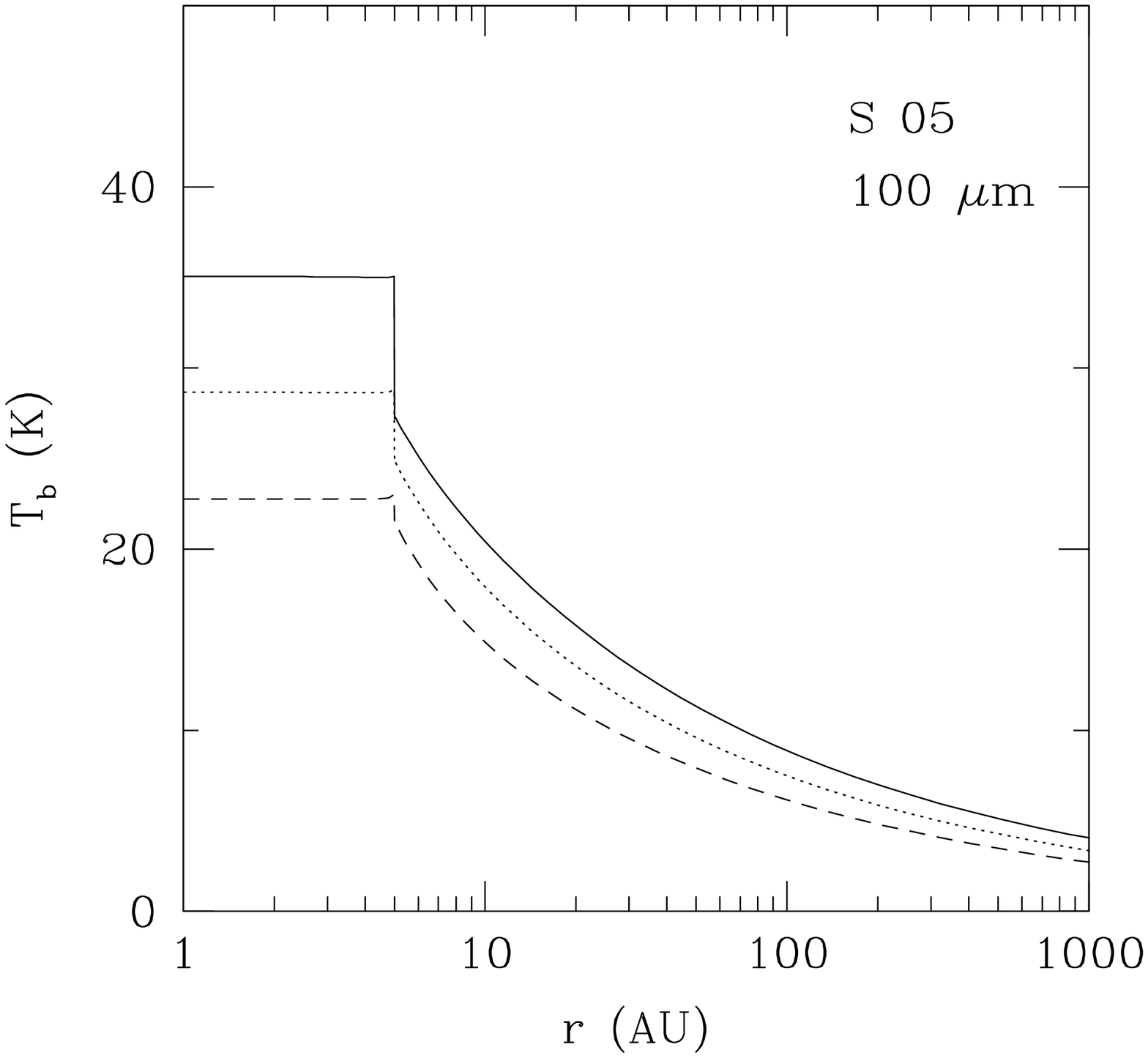}
\FigureFile(80mm,80mm){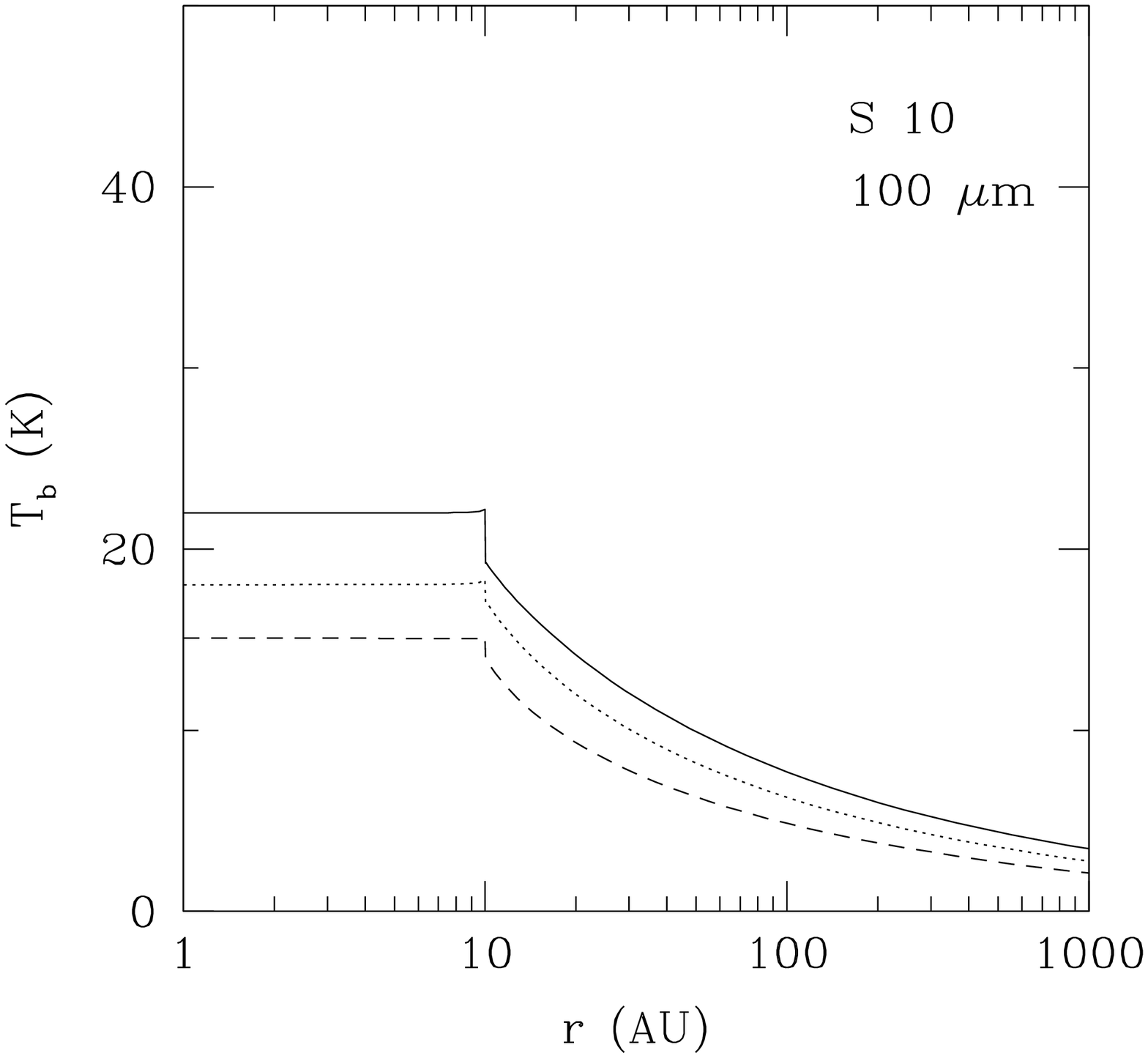}
  \end{center}
\caption{The same as Figure \ref{fig:Tb.100G} 
but for the observing wavelength
at 100$\mu$m (frequency at 3THz).}
\label{fig:Tb.100micron}
\end{figure}

Next we discuss the observability of the dust continuum.
Using a typical value $\nu L_{\nu}=10^{-3} L_{\odot}$ at 
a typical wavelength of processed radiation $\lambda=200{\rm \mu m}$ 
(see Figure \ref{fig:dustSED}),
the flux of dust continuum is  
\begin{equation}
F_{\nu}= 2.4 {\rm mJy} \left( \frac{\nu L_{\nu}}{10^{-3} L_{\odot}} \right)
\left( \frac{D}{\rm 1kpc} \right)^{-2}, 
\end{equation}
which falls below the sensitivities of currently operating 
{\it SOFIA} and {\it Spitzer}, but is above those of all 
three future telescopes discussed here.
In particular, having sensitivity both for direct and processed 
radiation, {\it SPICA} is most suitable for observing the first-core
objects not only by the H$_2$O emission, but also by the dust continuum.

In summary, {\it SPICA} can observe first-core objects as far as 1kpc 
by far-infrared dust emission and (if the collapse is Shu-like)
by the H$_2$O emission. 
In addition, if the hot dust emission and H$_2$O high-excitation-line 
emission in the mid-infrared wavelengths originating from the 
accretion shock come out unattenuated through a possible cavity, 
{\it SPICA} can detect them as direct or reflection light, depending on the
orientation between the cavity and the observer.

We thank Y. Fukui, S. Inutsuka, S. Narita, T. Onishi, K. Saigo, and
Y. Yonekura for suggestions, and the anonymous referee for improving 
the manuscript.
This study is supported in part by the Grants-in-Aid
by the Ministry of Education, Science and Culture of Japan
(16204012; 18740117; 18026008).

\newpage


\newpage


\onecolumn
\begin{table}
\begin{center}
\renewcommand{\arraystretch}{1.2}
\begin{tabular}{lccccclcccc} \hline\hline
model &  transition  &  $\lambda$  
&  $L_{\rm line}$ & $F_{\rm line, sh}$ & &
model &  transition  &  $\lambda$  
&  $L_{\rm line}$ & $F_{\rm line, sh}$ \\
 & & $({\rm \mu m})$
&  $(L_{\odot})$ & (erg s$^{-1}$ cm$^{-2}$) & &
 & &  $({\rm \mu m})$  
&  $(L_{\odot})$ & (erg s$^{-1}$ cm$^{-2}$) \\
\hline
LP05 & $9_{27}-8_{18}$ & 21.9 & 1.1E-4 & 6.2 && LP10 & $8_{36}-7_{07}$ & 23.8 & 6.8E-5 
& 9.2E-1\\
     & $7_{70}-6_{61}$ & 28.6 & 1.1E-4 & 6.1 &&      & $6_{61}-5_{50}$ & 33.0 & 6.3E-5
& 8.5E-1\\
     & $8_{45}-7_{16}$ & 23.9 & 1.1E-4 & 6.0 &&      & $8_{45}-7_{16}$ & 23.9 & 6.1E-5
& 8.4E-1\\
     & $9_{36}-8_{27}$ & 25.2 & 1.1E-4 & 5.7 &&      & $7_{25}-6_{16}$ & 29.9 & 5.9E-5
& 8.1E-1\\
     & $8_{36}-7_{07}$ & 23.8 & 1.0E-4 & 5.6 &&      & $6_{34}-5_{05}$ & 30.9 & 5.8E-5
& 8.0E-1\\
     & $8_{54}-7_{25}$ & 21.2 & 9.8E-5 & 5.3 &&      & $6_{52}-5_{41}$ & 36.0 & 5.7E-5
& 7.8E-1\\
     & $9_{45}-8_{36}$ & 28.3 & 9.8E-5 & 5.3 &&      & $9_{27}-8_{18}$ & 21.9 & 5.6E-5
& 7.6E-1\\
     & $7_{61}-6_{52}$ & 30.6 & 9.7E-5 & 5.3 &&      & $7_{52}-6_{43}$ & 33.0 & 5.5E-5
& 7.5E-1\\
     & $8_{54}-7_{43}$ & 30.9 & 9.0E-5 & 4.9 &&      & $5_{50}-4_{41}$ & 39.4 & 5.5E-5
& 7.5E-1\\
     & $6_{61}-5_{50}$ & 33.0 & 8.9E-5 & 4.9 &&      & $7_{61}-6_{52}$ & 30.6 & 5.4E-5
& 7.3E-1\\
\hline 
S05  & $6_{34}-5_{05}$ & 30.9 & 1.2E-5 & 6.5E-1 && S10  & $5_{23}-4_{14}$ & 45.2 & 3.0E-6
& 4.1E-2\\
     & $7_{25}-6_{16}$ & 29.9 & 1.1E-5 & 6.1E-1 &&      & $3_{30}-2_{21}$ & 66.5 & 2.9E-6
& 4.0E-2\\
     & $5_{50}-4_{41}$ & 39.4 & 1.1E-5 & 5.9E-1 &&      & $4_{32}-3_{21}$ & 58.8 & 2.6E-6
& 3.5E-2\\
     & $5_{41}-4_{32}$ & 43.9 & 9.6E-6 & 5.2E-1 &&      & $4_{32}-3_{03}$ & 40.7 & 2.5E-6
& 3.4E-2\\
     & $8_{36}-7_{07}$ & 23.8 & 9.6E-6 & 5.2E-1 &&      & $3_{21}-2_{12}$ & 75.5 & 2.5E-6
& 3.4E-2\\
     & $7_{34}-6_{25}$ & 34.6 & 9.3E-6 & 5.1E-1 &&      & $4_{41}-3_{30}$ & 49.4 & 2.5E-6
& 3.4E-2\\
     & $6_{52}-5_{41}$ & 36.0 & 9.2E-6 & 5.0E-1 &&      & $5_{32}-4_{23}$ & 48.0 & 2.2E-6
& 3.0E-2\\
     & $6_{43}-5_{14}$ & 28.0 & 9.2E-6 & 5.0E-1 &&      & $4_{23}-3_{12}$ & 78.8 & 2.1E-6
& 2.8E-2\\
     & $4_{41}-3_{30}$ & 49.4 & 8.8E-6 & 4.8E-1 &&      & $4_{22}-3_{13}$ & 57.7 & 1.9E-6
& 2.6E-2\\
     & $6_{61}-5_{50}$ & 33.0 & 8.7E-6 & 4.7E-1 &&      & $6_{16}-5_{05}$ & 82.1 & 1.8E-6
& 2.4E-2\\
\hline \hline
\end{tabular}
\end{center}
\caption{Most luminous H$_2$O lines emitted in the postshock layer 
before processing in the envelope at the epoch of $M_{\rm FC}=0.05M_{\odot}$}
\end{table}
 
\begin{table}
\begin{center}
\renewcommand{\arraystretch}{1.2}
\begin{tabular}{lccccclcccc} \hline\hline
model &  transition  &  $\lambda$  
&  $L_{\rm line}$ & $F_{\rm line}$ at 150pc& &
model &  transition  &  $\lambda$  
&  $L_{\rm line}$ & $F_{\rm line}$ at 150pc\\
 & &  $({\rm \mu m})$  
&  $(L_{\odot})$ & (W cm$^{-2}$) & &
 & &  $({\rm \mu m})$  
&  $(L_{\odot})$ & (W cm$^{-2}$) \\
\hline
LP05 & $3_{21}-3_{12}$ & 258 & 7.5E-9 & 1.1E-20 && LP10 & $3_{21}-3_{12}$ & 258
& 2.7E-8 & 4.1E-20\\
     & $3_{12}-3_{03}$ & 274 & 7.3E-9 & 1.1E-20 &&      & $3_{12}-3_{03}$ & 274
& 2.7E-8 & 4.0E-20\\
     & $4_{22}-4_{13}$ & 249 & 5.7E-9 & 8.5E-21 &&      & $1_{11}-0_{00}$ & 270 
& 2.1E-8 & 3.1E-20\\
     & $1_{11}-0_{00}$ & 270 & 4.9E-9 & 7.4E-21 &&      & $4_{22}-4_{13}$ & 249
& 2.0E-8 & 3.0E-20\\
     & $2_{20}-2_{11}$ & 244 & 4.6E-9 & 6.9E-21 &&      & $2_{20}-2_{11}$ & 244 
& 1.9E-8 & 2.9E-20\\
     & $2_{11}-2_{02}$ & 399 & 4.5E-9 & 6.8E-21 &&      & $3_{12}-2_{21}$ & 260 
& 1.7E-8 & 2.6E-20\\
     & $5_{23}-5_{14}$ & 213 & 4.4E-9 & 6.7E-21 &&      & $2_{02}-1_{11}$ & 304 
& 1.7E-8 & 2.5E-20\\
     & $3_{12}-2_{21}$ & 260 & 4.3E-9 & 6.4E-21 &&      & $3_{13}-2_{02}$ & 139 
& 1.7E-8 & 2.5E-20\\
     & $2_{02}-1_{11}$ & 304 & 4.1E-9 & 6.3E-21 &&      & $4_{13}-3_{22}$ & 145 
& 1.6E-8 & 2.3E-20\\
     & $1_{10}-1_{01}$ & 539 & 4.90-9 & 5.9E-21 &&      & $5_{23}-5_{14}$ & 213 
& 1.5E-8 & 2.3E-20\\
\hline 
S05  & $6_{16}-5_{05}$ & 82.1 & 1.5E-6 & 2.2E-18 && S10  & $3_{21}-2_{12}$ & 75.5 & 1.1E-6
& 1.6E-18\\
     & $5_{05}-4_{14}$ & 99.6 & 1.3E-6 & 2.0E-18 &&      & $5_{05}-4_{14}$ & 99.6 & 1.1E-6
& 1.6E-18\\
     & $4_{23}-3_{12}$ & 78.8 & 1.2E-6 & 1.9E-18 &&      & $4_{23}-3_{12}$ & 78.8 & 1.0E-6
& 1.6E-18\\
     & $4_{14}-3_{03}$ & 114  & 1.1E-6 & 1.6E-18 &&      & $6_{16}-5_{05}$ & 82.1 & 1.0E-6
& 1.5E-18\\
     & $3_{21}-2_{12}$ & 75.5 & 1.1E-6 & 1.6E-18 &&      & $4_{14}-3_{03}$ & 114  & 1.0E-6
& 1.5E-18\\
     & $6_{06}-5_{15}$ & 83.4 & 1.1E-6 & 1.6E-18 &&      & $2_{21}-1_{10}$ & 108  & 9.8E-7
& 1.5E-18\\
     & $5_{15}-4_{04}$ & 95.7 & 1.0E-6 & 1.6E-18 &&      & $3_{30}-2_{21}$ & 66.5 & 8.4E-7
& 1.3E-18\\
     & $2_{21}-1_{10}$ & 108  & 1.0E-6 & 1.5E-18 &&      & $3_{22}-2_{11}$ & 90.1 & 8.2E-7
& 1.2E-18\\
     & $7_{07}-6_{16}$ & 72.0 & 1.0E-6 & 1.5E-18 &&      & $2_{20}-1_{11}$ & 101  & 7.8E-7
& 1.2E-18\\
     & $3_{22}-2_{11}$ & 90.1 & 1.0E-6 & 1.5E-18 &&      & $5_{15}-4_{04}$ & 95.7 & 7.5E-7
& 1.1E-18\\
\hline \hline
\end{tabular}
\end{center}
\caption{Most luminous H$_2$O lines after processing in the envelope 
at the epoch of $M_{\rm FC}=0.05M_{\odot}$}
\end{table}

\end{document}